\documentclass[12pt]{article}

\usepackage[utf8]{inputenc}
\usepackage{charter}
\usepackage[xcharter,scaled=1.05]{newtxmath}
\usepackage{fullpage}
\usepackage[colorlinks=true,breaklinks=true]{hyperref}
\hypersetup{allcolors=[rgb]{0.0 0.0 0.6},linkcolor=[rgb]{0.75 0.05 0.05}}
\usepackage{graphicx}
\usepackage{xspace}
\usepackage{lipsum}
\usepackage[sort&compress,numbers]{natbib}
\usepackage{hyperref}
\usepackage[total={6.5in,9in},top=1in,headsep=0.1in,headheight=1in]{geometry}
\usepackage{enumerate}
\usepackage[shortlabels]{enumitem}
\usepackage{comment}
\usepackage{multicol}
\usepackage{wrapfig}
\usepackage{makecell}
\usepackage[usenames]{xcolor}   

\usepackage{siunitx}

\usepackage{subfigure}
\usepackage{nccmath}
\usepackage{array}
\usepackage{tabu}
\usepackage{dcolumn}
\usepackage{float}
\usepackage{url}
\usepackage{graphics}
\usepackage{hyperref}
\usepackage{ragged2e}
\usepackage[normalem]{ulem}

\usepackage{authblk}

\newcommand\snowmass{
\begin{center}
  \rule[-0.2in]{\hsize}{0.01in}\\
  \rule{\hsize}{0.01in}\\
  \vskip 0.1in
  Submitted to the Proceedings of the US Community Study\\ 
  on the Future of Particle Physics (Snowmass 2021)\\
  \rule{\hsize}{0.01in}\\
  \rule[+0.2in]{\hsize}{0.01in}\\[-2em]
\end{center}
}

\usepackage[firstpage=true]{background}
\backgroundsetup{contents={\parbox{6.5in}{\snowmass}}, scale=1,placement=top,opacity=1,color=black,position={3.25in,1.2in}}

\usepackage{fancyhdr}
\fancypagestyle{plain}{%
  \fancyhf{}%
  \fancyhead[C]{}
  \fancyfoot[C]{\thepage}
}

\fancypagestyle{empty}{%
  \fancyhf{}%
  \fancyhead[C]{{\it Snowmass 2021 White Paper Axion Dark Matter}}
  \fancyfoot[C]{\thepage}
}
\title{Snowmass 2021 White Paper\\ Axion Dark Matter}
\date{}

\author[1]{C.~B.~Adams}
\author[2]{N.~Aggarwal}
\author[3]{A.~Agrawal}
\author[ ]{R.~Balafendiev}
\author[5]{C.~Bartram}
\author[5]{M.~Baryakhtar}
\author[6,7]{H.~Bekker}
\author[ ]{P.~Belov}
\author[8]{K.~K.~Berggren}
\author[9]{A.~Berlin}
% ...

\affil[1]{Columbia University, New York, NY, USA}
\affil[2]{Northwestern University, Evanston, IL, USA}
\affil[3]{University of Chicago, Chicago, IL, USA}
\affil[5]{University of Washington, Seattle, WA, USA}
\affil[6]{Helmholtz Institute (GSI) Mainz, Mainz, Germany}
\affil[7]{Johannes Gutenberg University, Mainz, Mainz, Germany}
\affil[8]{Department of Electrical Engineering, Massachusetts Institute of Technology, Cambridge, MA, USA}
\affil[9]{Fermi National Accelerator Laboratory, Batavia, IL, USA}
% ...

\newcommand{\msfm}[1]{\ensuremath{#1}\xspace}
\newcommand{\thetai}{\msfm{\theta_\text{i}}}
\newcommand{\fa}{\msfm{f_a}}
\newcommand{\ma}{\msfm{m_a}}

\newcommand{\DMR}{DMRadio\xspace}
\newcommand{\DMRL}{\DMR-50L\xspace}
\newcommand{\DMRm}{\DMR-m$^3$\xspace}
\newcommand{\DMRG}{\DMR-GUT\xspace}
\newcommand{\DMRP}{\DMR-Pathfinder\xspace}
\newcommand{\abra}{\mbox{ABRACADABRA-10\,cm}\xspace}

\begin{document}

\maketitle

\newpage
\vspace*{7cm}
\begin{abstract}
Axions are well-motivated dark matter candidates with simple cosmological production mechanisms. They were originally introduced to solve the strong CP problem, but also arise in a wide range of extensions to the Standard Model. This Snowmass white paper summarizes axion phenomenology and outlines next-generation laboratory experiments proposed to detect axion dark matter. There are vibrant synergies with astrophysical searches and advances in instrumentation including quantum-enabled readout, high-Q resonators and cavities and large high-field magnets. This white paper outlines a clear roadmap to discovery, and shows that the US is well-positioned to be at the forefront of the search for axion dark matter in the coming decade.
\end{abstract}

\newpage

\tableofcontents 

\newpage

\section{Executive summary}
\label{sec:ex_summ}
\textbf{The Dark Matter Question} --- The nature of dark matter remains unknown, though it makes up ${\sim}25\%$ of the energy density of the universe. Its sheer quantity already makes it a clear top priority for fundamental particle physics and cosmology. Simultaneously, multiple well-motivated dark matter candidates are coming within the reach of the next generation of experiments, building on rapid technological progress. This opens the door for a breakthrough in our understanding of the universe.

\textbf{Axions} --- Among dark matter candidates, axions stand out as especially well motivated, with a clear roadmap to decisive experimental progress and potential discovery in the next decade. Axions are light pseudoscalar particles which appear in a wide variety of extensions to the Standard Model. For instance, they may arise as Goldstone bosons of a spontaneous symmetry breaking at a high energy scale $f_a$, leading to a small mass $m_a$ and weak couplings to photons, gluons, leptons, and nucleons. 

The most well-motivated example of an axion is the QCD axion, which arises as a consequence of the Peccei-Quinn solution to the strong-CP problem \cite{Peccei:1977hh,Peccei:1977ur,Weinberg:1977ma,Wilczek:1977pj}. Its coupling to gluons leads to a well-defined relationship between $m_a$ and $f_a$, which in turn leads to a narrow band of couplings to photons. The QCD axion may be produced nonthermally in the early universe by the misalignment mechanism, which, in the simplest scenarios, motivates masses $10^{-6} \, \mathrm{eV} \lesssim m_a \lesssim 10^{-4} \, \mathrm{eV}$. 

By contrast, axion-like particles (ALPs) do not couple directly to gluons, leading to a broader range of possible couplings to photons. They may be also be simply produced via misalignment in the broader mass range $10^{-20} \, \mathrm{eV} \lesssim m_a \lesssim \mathrm{eV}$, which is bracketed by astrophysical and cosmological constraints. 

\textbf{Context} --- At the time of the previous Snowmass a decade ago, no experiment had probed the QCD axion below 0.1~eV. The ADMX G2 experiment has now achieved this goal, albeit in a narrow mass range. The previous Snowmass process also spawned the Dark Matter New Initiatives (DMNI) program, which has funded DMRadio-m$^3$ and ADMX-EFR to develop project execution plans to reach the QCD axion in a wider range of masses. Simultaneously, a number of other demonstrator-scale experiments are beginning to probe unexplored axion parameter space.

\begin{figure}
    \centering
    \includegraphics[width=13cm]{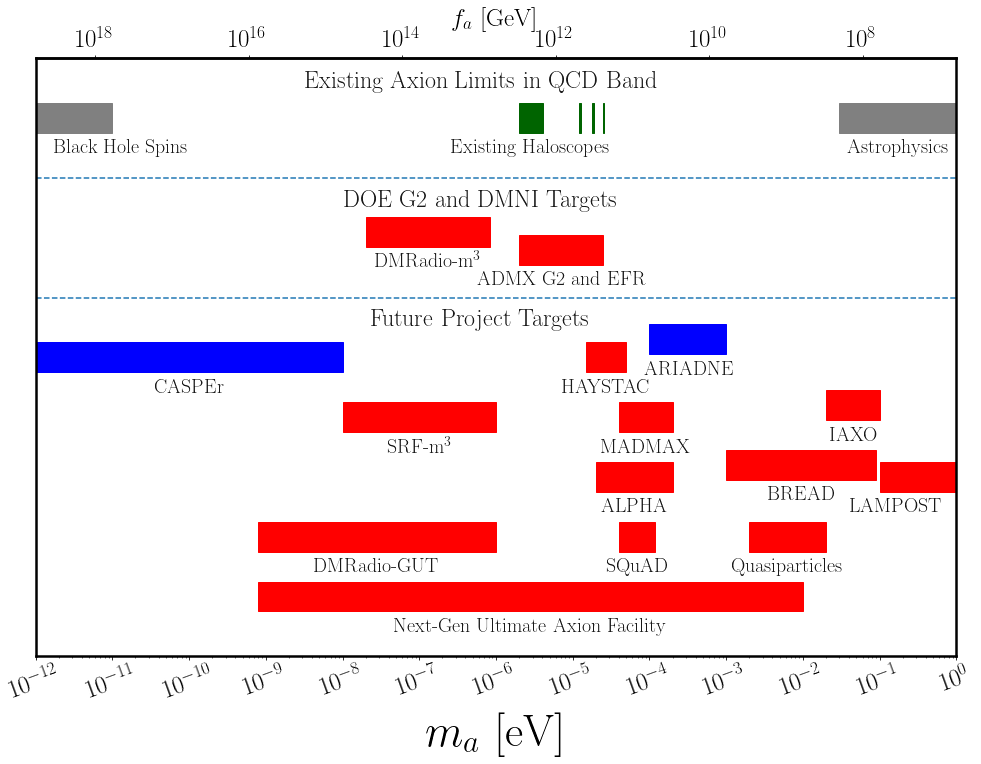}
    \caption{\label{fig:money_and_politics} Potential QCD axion mass range explored by efforts described in this white paper. The green bars indicate running experiments in the QCD region, red indicate proposals that utilize the axion-photon coupling, and blue indicates proposal utilizing alternative couplings.}
\end{figure}

This Snowmass marks the beginning of a renaissance in the search for axions. We now have a series of experiments that would allow us to definitively search the axion dark matter parameter space and a roadmap to realize them. Fig.~\ref{fig:money_and_politics} shows the potential mass reach of the proposed experiments discussed in this white paper. Several different techniques are needed across this broad mass range, but almost all require quantum-enabled readout, high-$Q$ resonators and cavities, as well as large high field magnets.

\begin{center}
{\textbf{\large{Roadmap to Discovery}}}
\end{center}

As a community we propose a phased approach that allows for discovery at all stages while nurturing emerging technologies. This plan has four components:

\begin{enumerate}
    \item
    \textbf{Pursue the QCD Axion by Executing the Current Projects:}
    The ADMX G2 effort continues to scan exciting axion dark matter parameter space and the experiments identified by the DMNI process DMRadio-m$^3$ and ADMX-EFR are prepared to start executing their project plans. 
    \item
    \textbf{Pursue the QCD Axion with a Collection of Small-Scale Experiments:} 
    The entire axion mass range cannot be explored by one or two experiments; a variety of techniques must be employed.  Most of the efforts described in this white paper would benefit from a concerted effort to foster small scale projects to either cover a limited range of the QCD axion parameter space, or bring the effort to a level of technological readiness required to reach the QCD axion paramater space. The DOE DMNI process was the right scale for many of these proposed experiments to make significant inroads to the QCD axion.
    \item
    \textbf{Support Enabling Technologies:} Many of the proposed efforts share needs in ultra-sensitive quantum measurement and quantum control, large, high-field magnets, spin ensembles, and sophisticated resonant systems. These technologies overlap strongly with other HEP efforts and synergies should be exploited. 
    \item
    \textbf{Support Theory and Astrophysics Beyond the QCD Axion:} The QCD axion is an important benchmark model, but not the only motivated one. Theoretical effort should be supported to understand the role of ALPs in dark matter cosmology, and to understand the roles axions play in astrophysical phenomena.
\end{enumerate}

\textbf{Outlook} --- In the best case scenario, the axion will have already been discovered by the time of the next Snowmass, at which point efforts would have shifted to studying its properties and determining if it makes up all, or only part of the dark matter. Alternatively, significant inroads will have been made in probing the QCD axion. A number of the projects described in this white paper will have advanced to the stage of needing mid-scale support for a final push to cover the entire range of QCD axion dark matter. The well-developed projects will have matured and merged to be able to utilize a large scale `Ultimate Axion Facility' to probe models of ALP dark matter beyond the QCD axion, or where only a fraction of dark matter is made of axions.

\newpage 

\section{Introduction}
\label{sec:intro}
Laboratory detection of dark matter (DM) is a cornerstone goal of fundamental physics today (see, e.g.~\cite{Billard:2021uyg} for a recent report focusing mostly on the European efforts).  
Astrophysical evidence for DM is abundant~\cite{Rubin:1970zza,Tyson:1998vp,Tegmark:2003ud,Clowe:2006eq,Akrami:2018vks,Bertone:2004pz}, but its microscopic properties remain mysterious. 
Many extensions of the Standard Model (SM) predict compelling DM candidates with interesting cosmological histories. 
One sought-after candidate is the axion, a pseudoscalar (spin zero, parity odd) field arising as a pseudo-Nambu--Goldstone boson of a broken Abelian symmetry~\cite{Peccei:1977hh,Wilczek:1977pj,Weinberg:1977ma}. 
This is the leading solution to the strong charge--parity (CP) problem in quantum chromodynamics (QCD) that addresses the experimental non-observation of the neutron electric dipole moment~\cite{Harris:1999jx,Baker:2006ts,Afach:2015sja,Abel:2020gbr}. 
The axion exhibits viable DM properties~\cite{Preskill:1982cy,Abbott:1982af,Dine:1982ah} and with masses below 1\,eV, its high occupation number gives rise to wave-like behaviour acting as a coherently-oscillating classical field. 
The interactions of the axion with SM forces and matter enable a diverse range of experiments to search for its direct evidence.
Axion-like particles (ALPs) are generalised pseudoscalars whose mass and couplings that allow searches across wider parameter space.

This Snowmass white paper reviews ongoing and planned experiments aiming to directly detect axion DM in the coming decade and beyond. 
Particular focus is placed on experiments located and led by the US community to align with the aims of Snowmass. A wide range of models predict different couplings and masses for axions across many orders of magnitude. 
The present most-sensitive experiments search for the axion--photon interaction around the micro-electronvolt ($\mu$eV) mass range, motivated by the predictions of a suitable DM abundance in the simplest settings. 
However, multiple decades of coupling--mass parameter space remain unexplored due to longstanding obstacles of the canonical resonant cavity haloscopes.
This motivates the burgeoning and vibrant program of experiments proposed to extend sensitivity above and below $\mu$eV masses.
These capitalize on the interdisciplinary creativity and scientific expertise of the community to enable experimental breakthroughs for decisively testing axion DM scenarios.
Realizing this program requires strengthening synergies with astrophysical probes and sustained investment in instrumentation technology crucial to opening unprecedented DM sensitivity. 

With this goal in mind this white paper proceeds along the following outline. Section~\ref{sec:motivation} reviews the particle physics motivation of the QCD axion and axion-like particles. Section~\ref{sec:cosmology} discusses the production of axions in the early universe and their late-time astrophysics. Experiments targeting axions via their couplings to photons are presented in sections~\ref{sec:above_microeV} and~\ref{sec:below_microeV}, which emphasize axion masses above and below $\mu\mathrm{eV}$, respectively. Section~\ref{sec:ultralight} discusses experiments sensitive to a wide range of other axion couplings.  Section~\ref{sec:solar_axions} considers the opportunities provided by axions produced in the sun. Synergies with astrophysical probes are presented in section~\ref{sec:synergies}. Important R\&D directions for future experiments are discussed in section~\ref{sec:randd} before section~\ref{sec:conclusion} concludes with an overview of the opportunities presented by the discussed experiments and gives an outlook to a future ultimate axion facility.

\section{Axion Motivation}
\label{sec:motivation}

\subsection{QCD Axion}
\label{sec:qcd_axion}
Yang-Mills theories, such as QCD, generically break the CP symmetry via a so-called $\bar\theta$-term:
  $\mathcal{L}_{\theta}\sim \bar{\theta} \,G \widetilde{G}$, where $G$ is the gluon field strength tensor.
Nevertheless, experimental measurements of a number of CP-violating observables in the strong sector, such as the electric dipole moment of the neutron, are compatible with the absence of such CP violation, setting stringent upper bounds on $\bar\theta \lesssim 10^{-10}$~\cite{Abel:2020gbr}.
The extreme smallness of this parameter, commonly dubbed the ``strong CP problem", constitutes a very serious fine-tuning issue that remains unresolved in the SM.
Indeed, the quest to explain small parameters has been shown in the
past to be an effective tool to gain a deeper understanding of Nature.
The strong CP problem could therefore represent a unique  guide in the
search for new physics. 

Not many solutions to the strong CP problem are known and probably the most compelling explanation to this puzzle is the Peccei-Quinn (PQ) mechanism~\cite{Peccei:1977hh,Peccei:1977ur}.
It minimally extends the SM with a new classically conserved global symmetry, the PQ symmetry $U(1)_{\text{PQ}}$, which is spontaneously broken at a scale $\sim f_a$ and explicitly broken at the quantum level by a mixed QCD anomaly. As an unavoidable low-energy consequence of this mechanism, a pseudo-Nambu-Goldstone boson (pNG) arises: the QCD axion $a$~\cite{Weinberg:1977ma,Wilczek:1977pj}. This makes this scenario extremely testable. The central ingredient of the PQ mechanism is the axion coupling to QCD,
\begin{align}\label{axglco}
    \mathcal{L}=\left(\frac{a}{f_{a}}-\bar{\theta}\right) \frac{\alpha_{s}}{8 \pi} G^{\mu \nu a} \tilde{G}_{\mu \nu}^{a}\,,
\end{align}
 that generates a non-perturbative axion potential, the minimum of which is  CP-conserving \cite{Vafa:1983tf}. Thus the axion dynamically relaxes the value of $\bar\theta_{\text{eff}}\equiv{\langle a\rangle}/{f_a}-\bar\theta$ to zero solving the strong CP problem. 

 The resulting QCD axion mass is inversely proportional to the axion decay constant $f_a$. It can be computed using chiral perturbation theory, 
\begin{align}
m^2_{a}\simeq \frac{f_\pi^2m_\pi^2}{f_a^2}\frac{m_um_d}{(m_u+m_d)^2}\,,
\label{Eq. axion mass 0}
\end{align} 
as a function of the pion mass $m_\pi$, the decay constant $f_\pi$, and the up and down quark masses $m_u$ and $m_d$. The most precise computation to date~\cite{Gorghetto:2018ocs} yields
\begin{equation}
\label{eq:massdecay}
    m_{a}=5.691(51)\mathrm{\mu eV}({10^{12} \mathrm{GeV}}/{f_{a}}).
\end{equation} 
The axion mass in Eq.~\eqref{Eq. axion mass 0} is a robust prediction of the invisible axion paradigm. Nonetheless, some extensions motivated by
the PQ quality problem, predict axions parametrically lighter \cite{Hook:2018jle,DiLuzio:2021pxd,DiLuzio:2021gos} or heavier 
\cite{Rubakov:1997vp,Berezhiani:2000gh,Gianfagna:2004je,Hsu:2004mf,Hook:2014cda,Fukuda:2015ana,Chiang:2016eav,Dimopoulos:2016lvn,Gherghetta:2016fhp,Kobakhidze:2016rwh,Agrawal:2017ksf,Agrawal:2017evu,Gaillard:2018xgk,Csaki:2019vte,Gupta:2020vxb,Gherghetta:2020ofz}.
 
The different strategies one can use to implement the PQ symmetry in a UV complete theory give rise to different axion models. The most paradigmatic ones are the DFSZ~\cite{Zhitnitsky:1980tq,Dine:1981rt}, KSVZ~\cite{Kim:1979if,Shifman:1979if}, and dynamical/composite axions~\cite{Kim:1984pt}, which are often used as benchmark examples. The DFSZ model requires at least two Higgs doublets and no additional fermions, since the fermions carrying the PQ charge are those of the SM, while hadronic models, which include KSVZ and composite axions, require new fermions to implement the PQ symmetry. Another difference among these models is that both KSVZ and DFSZ axions are fundamental/elementary scalar fields whereas the dynamical axion corresponds to a bound state, a pseudoscalar meson of a new confining force.
  
The main properties of the QCD axion, such as its lightness and feeble derivative or anomalous interactions, stem from its pseudo Nambu-Goldstone nature. This informs us about the low energy processes that would be affected by the existence of this hypothetical particle and thus which detection strategies we should pursue. Since our detectors are particularly sensitive to the electromagnetic (EM) interactions, most of the axion experiments take advantage of the axion-photon coupling, 
  \begin{equation}\label{axphotonlag}
     \mathcal{L}_{a \gamma \gamma}=-\frac{g_{a \gamma \gamma}}{4} a F_{\mu \nu} \tilde{F}^{\mu \nu}=g_{a \gamma \gamma} a \mathbf{E} \cdot \mathbf{B} \, ,
  \end{equation}
  where $F^{\mu \nu}$ is the EM field strength tensor and $\mathbf{E}$ and $\mathbf{B}$ are electric and magnetic fields. With this interaction included the Maxwell equations get modified as follows~\cite{Sikivie:1983ip},
  \begin{eqnarray} \label{m1}
      &&\pmb{\nabla}\!\cdot \! \mathbf{E} = \rho - g_{a\gamma \gamma} \mathbf{B}\!\cdot \! \pmb{\nabla}a \, , \\[3pt]
      &&\pmb{\nabla}\!\cdot \! \mathbf{B} = 0 \, , \\[3pt]
      &&\pmb{\nabla}\!\times \! \mathbf{E} = - \frac{\partial \mathbf{B}}{\partial t} \, , \\[3pt] \label{m4}
      &&\pmb{\nabla}\!\times \! \mathbf{B} = \frac{\partial \mathbf{E}}{\partial t} + \mathbf{J} - g_{a\gamma \gamma} \left( \mathbf{E}\!\times \! \pmb{\nabla} a - \frac{\partial a}{\partial t} \mathbf{B} \right) \, .
  \end{eqnarray}
  In general the axion-photon coupling has a contribution which depends on the fermionic content of the specific model and another contribution originating from the axion mixing with neutral mesons~\cite{GrillidiCortona:2015jxo},
  \begin{equation}\label{photon}
     g_{a \gamma \gamma}=\frac{\alpha}{2 \pi f_{a}}\left(\frac{E}{N}-1.92(4)\right) \, .
  \end{equation}
  Here, $E$ and $N$ are the EM and QCD anomaly coefficients and  $\alpha$ is the fine structure constant. The original KSVZ model predicts $E/N = 0$ while the original DFSZ model predicts $E/N = 8/3$. Due to their simplicity, the latter models are normally taken as benchmarks.

  While it is common to focus on the minimal benchmark models or to assume that $E/N\sim \mathcal{O}(1)$, the axion-photon coupling $g_{a \gamma \gamma} (m_a)$ is strongly model-dependent, so that its value can be changed significantly by invoking additional assumptions about the UV structure of the theory. The anomaly coefficients $E$ and $N$ can vary considerably depending on the representations of the SM group under which heavy fermions of a given axion model transform~\cite{Kim:1998va, DiLuzio:2016sbl, DiLuzio:2017pfr, DiLuzio:2020wdo,Plakkot:2021xyx}.
    If those fermions are magnetic monopoles the fine structure constant $\alpha$ entering Eq.~\eqref{photon} would be magnetic, enhancing the coupling to photons by $\alpha_m/\alpha_e\sim 4\times 10^{4} $~\cite{Sokolov:2021ydn}.
  Moreover, the dependence of the PQ scale $f_a$ on the axion mass $m_a$ can differ from the conventional relation in Eq.~\eqref{Eq. axion mass 0}~\cite{Hook:2018jle, DiLuzio:2021pxd}. Finally, the axion photon coupling can be enhanced/suppressed in models with multiple scalars where hierarchical charges are arise due to particular axion alignments in field space such as Kim-Nilles-Peloso (KNP) mechanism,~\cite{Kim:2004rp,Agrawal:2017cmd} clockwork axions \cite{Kaplan:2015fuy,Farina:2016tgd} or multi-higgs doublet DFSZ-like models~\cite{Darme:2020gyx}.

  Although the axion-photon coupling is the most popular one to search for in experiments, the Lagrangian~\eqref{axphotonlag} is actually not specific to axions. The same kind of coupling is shared by any axion-like particle (ALP), which needs not solve the strong CP problem. To discover the QCD axion, one has to probe its coupling to gluons~\eqref{axglco}. At low energies, the latter coupling induces interactions of axions with the EDMs of nucleons. The corresponding Lagrangian is,
    \begin{align}
      \mathcal{L}_{a N \gamma}=-\frac{i}{2} g_{a N \gamma} a \bar{\Psi}_{N} \sigma_{\mu \nu} \gamma_{5} \Psi_{N} F^{\mu \nu} \, ,
    \end{align}
  where $\sigma_{\mu \nu} = \frac{1}{2} [\gamma_{\mu}, \gamma_{\nu}]$ and the nucleon $N$ can be the neutron $n$ or the proton $p$. The coupling constants $g_{a N \gamma}$ corresponding to each of the nucleons depend only on the PQ scale $f_a$ and are given by the following expression,
  \begin{equation}
g_{a p \gamma}=-g_{a n \gamma}=-(3.7 \pm 1.5) \times 10^{-3}\left(\frac{1}{f_{a}}\right) \frac{1}{\mathrm{GeV}} \, . 
\label{eq:gd}
\end{equation}

Axions are also predicted to couple to leptons at tree level in a wide range of models, such as those based on Grand Unified Theories (GUTs). Even though it is challenging to probe these interactions with current experiments,  a promising avenue is to study the  influence they exhibit on various astrophysical processes, where the interactions of axions with electrons can play a crucial role (see also Sec.~\ref{sec:synergies}). The interaction Lagrangian with leptons reads, 
\begin{equation}
\mathcal{L}_{a \ell \ell}=\frac{C_{\ell}}{2 f_{a}}\, \partial_{\mu} a\, \bar{\Psi}_{\ell} \gamma^{\mu} \gamma_{5} \Psi_{\ell} \, ,
\label{eq:gann}
\end{equation}
where the coefficient $C_{\ell}$ depends on the particular lepton flavor under consideration. Note that in hadronic axion models, such as KSVZ, the couplings to leptons are suppressed, since they are generated only at the loop level through the axion-photon coupling~\eqref{photon}, whereas DFSZ constructions predict tree-level couplings to leptons.

\subsection{ALPs}
\label{sec:alps}
The designation axion-like particle (ALP) usually refers to a more general (pseudo) Nambu-Goldstone boson of a global $U(1)$ symmetry breaking. ALPs share many similarities with the QCD axions such as their similar effective theories and their potential role as DM. However, ALPs, by definition, do not interact in the same way with the gluons as the QCD axion. Since ALPs do not acquire their masses from the non-perturbative QCD effects, the mass $m_a$ and the decay constant $f_a$ are independent parameters as opposed to Eq.~(\ref{Eq. axion mass 0}) for the QCD axion. Although ALPs cannot address the strong CP problem, they are well motivated new physics beyond the Standard Model since global $U(1)$ symmetries are ubiquitous in proposed solutions to the open questions in particle physics and cosmology. The most motivated examples include the Majoron associated with the lepton symmetry~\cite{Chikashige:1980ui} that explains the neutrino masses and the familon associated with the flavor symmetry~\cite{Froggatt:1978nt} that addresses the fermion hierarchical masses and mixing. In addition, ALPs are generically predicted in the low-energy effective theories from string theory~\cite{Witten:1984dg,Conlon:2006tq,Svrcek:2006yi,Choi:2009jt,Arvanitaki:2009fg,Acharya:2010zx,Cicoli:2012sz,Halverson:2017deq}. More recently, a so-called ALP cogenesis~\cite{Co:2020xlh} scenario has been discussed where ALPs may simultaneously explain the DM abundance~\cite{Co:2019jts} as well as the observed cosmological baryon asymmetry~\cite{Chiba:2003vp,Takahashi:2003db,Co:2019wyp} using novel axion dynamics in the early universe. ALP cogenesis is of particular experimental interest because it predicts a rather precise relation between $m_a$ and $f_a$ that is experimentally accessible.

\section{Axion Cosmology}
\label{sec:cosmology}
Axion cosmology has become a broad research area and cannot be fully summarized in a few pages. This section therefore focuses on the qualities of dark matter~(DM) axions and axionlike particles~(ALPs) most crucial to their cosmological abundance and relevance to the research proposals covered in later sections.

As we will see the simplest models and mechanisms for dark matter QCD axions predict super-{\textmu}eV axion masses whose experimental search is discussed in Sec.~\ref{sec:above_microeV}. For sub-{\textmu}eV axion masses (experiments discussed in Sec.~\ref{sec:below_microeV}) non-standard assumptions for the axion models or cosmological scenarios might be required to make axions all of the DM.
Examples include entropy injection, late inflation, or naturally small initial field values.
These need not be \textit{ad hoc} modifications but can arise from solving other puzzles, such as the baryon asymmetry of the Universe.
More details on non-standard scenarios can be found in the corresponding paragraph in Sec.~\ref{sec:axion_creation_mechanisms}. It should be noted that ALPs with sub-{\textmu}eV masses are predicted abundantly in string theory~\cite[e.g.][]{Svrcek:2006yi,Mehta:2020kwu}, and that ultralight ALPs might resolve various potential cosmological issues ~\cite[e.g.][]{Hu:2000ke,Hui:2016ltb,Niemeyer:2019aqm}.

To supplement this incomplete summary, we advise the interested reader to consult some of the excellent axion reviews~\cite[e.g.][]{Marsh:2015xka,DiLuzio:2020wdo}.

\subsection{Axion creation mechanisms}\label{sec:axion_creation_mechanisms}
In the early Universe, axions can be produced both thermally and non-thermally. We focus on the non-thermal production but also comment on recent interesting developments on thermal production in Sec.~\ref{sec:thermal_production}.
The bedrock of the axion's relevance for cosmology is the realignment~(or misalignment) mechanism: it generically allows QCD axions~\cite{Preskill:1982cy,Abbott:1982af,Dine:1982ah,Turner:1983he,Turner:1985si} and ALPs~\cite{Arias:2012az} to make up significant fractions of the cold DM in the Universe.
Breaking of the PQ symmetry can also lead to the formation of topological defects, namely cosmic strings and domain walls~\cite{Kibble:1976sj,Kibble:1980mv}.
While they can emit axions during their evolution~\cite{Davis:1985pt,Davis:1986xc,Harari:1987ht,Battye:1993jv}, topological defects may be diluted away during inflation if formed too early.

\paragraph{The canonical realignment mechanism.} The scalar field equation for the \emph{homogeneous} axion field~$a(t) \equiv \fa \, \theta(t)$, where the cosmological history is described by the Hubble parameter $H(t)$, takes the form
\begin{equation}
    \ddot{\theta} + 3H(t) \, \dot{\theta} + \ma^2(t) \sin(\theta) = 0 \, , \label{eq:realignment}
\end{equation}
where the one-instanton cosine potential $V[\theta] = \fa^2\ma^2(t)\left[1 - \cos(\theta)\right]$ was assumed to describe axions and ALPs at higher temperatures. 
While the low-temperature QCD axion potential is known to be more complicated~\cite{DiVecchia:1980yfw, GrillidiCortona:2015jxo}, the relevant evolution of the field happens when the low-temperature result does not apply.
The initial conditions are usually taken to be $\theta(0) \equiv \thetai \in [-\pi,\pi)$ and $\dot{\theta}(0) \equiv 0$, where \thetai is referred to as the initial misalignment angle. The condition $\dot{\theta}(0) \equiv 0$ seems sensible as the solution diverges otherwise for $t \rightarrow 0$ or as a practical choice given that Hubble drag dampens the angle velocity $\dot{\theta} \simeq 0$ by the time the field becomes dynamical at $H \sim \ma$~\cite{Kolb:1990vq,Weinberg:2008zzc}.

The general behavior of the solutions of Eq.~\eqref{eq:realignment} in standard cosmology is as follows: at early times, when $H \gg \ma$, the field is ``stuck'' due to Hubble friction.
It is thus constant and has the same equation of state~(EOS) as dark energy or an inflaton ($w = -1$).
Around the time when $H \sim \ma$, the field becomes dynamical and starts to oscillate.
At later times, $H \ll \ma$, the field evolution eventually becomes adiabatic and the time-averaged EOS is $\langle w \rangle = 0$ and, due to the rapid oscillation, the field effectively behaves as pressureless dust or cold DM.
Under the assumption of entropy conservation, one may thus derive an estimate of the energy density today.

There exist approximate solutions to Eq.~\eqref{eq:realignment} using semi-analytical techniques, which can yield estimates of the axion DM abundance~\cite[e.g.][]{Marsh:2015xka}, as well as a small number of publicly available codes for numerical solutions~\cite[e.g.][]{Hoof:2018ieb,Karamitros:2021nxi}, including perturbation evolution~\cite{Hlozek:2014lca, Poulin:2018dzj}.
Due to the anharmonicities from the cosine potential~\cite{Turner:1985si,Lyth:1991ub,Strobl:1994wk,Bae:2008ue,Visinelli:2009zm,Kobayashi:2013nva}, at least some numerical analysis is required to get a precise prediction of the axion energy density, and thus constraints on the axion mass.

Despite the need for a partial numerical evolution of Eq.~\eqref{eq:realignment}, it is still useful compare a rough estimate for the axion abundance to the DM density today, $\Omega_\text{DM} h^2 \approx 0.12$. For QCD axions with $\fa \lesssim  \SI{e17}{\GeV}$ in a standard cosmological scenario one finds that (see e.g.\ Refs.~\cite{Preskill:1982cy,Abbott:1982af,Dine:1982ah,Turner:1983he} for early estimates, Refs.~\cite{Turner:1985si,Lyth:1991ub,Strobl:1994wk,Bae:2008ue,Visinelli:2009zm,Kobayashi:2013nva} for the inclusion of anharmonic corrections, and  Ref.~\cite{Borsanyi:2016ksw} for a recent computation)
\begin{equation}
    \Omega_a h^2 \sim \num{0.12} \, \left(\frac{\fa}{\SI{e12}{\GeV}}\right)^{7/6} \langle \thetai^2 \rangle \, , \label{eq:qcd_axion_ede}
\end{equation}
where the average misalignment angle squared, $\langle \thetai^2 \rangle$, depends on whether the PQ symmetry breaks before (see Sec.~\ref{sec:pre_infl}) or after inflation (see Sec.~\ref{sec:post_infl}).
In the former case we have a single $\thetai \sim \mathcal{O}(1)$, while in the latter case $\langle \thetai^2 \rangle \sim 4.6$~\cite{Borsanyi:2016ksw}.
Again note that Eq.~\eqref{eq:qcd_axion_ede} is only a rough estimate and also depends on the realized scenario.
That said, saturating the DM density and using the relation between $\ma$ and $\fa$ from Eq.~\eqref{eq:massdecay}, we find that QCD axion masses are naturally $\gtrsim \si{\micro\eV}$. This suggest the aforementioned separation of experiments into super-{\textmu}eV (see Sec.~\ref{sec:above_microeV}) and sub-{\textmu}eV (see Sec.~\ref{sec:below_microeV}).

Axionlike particles, which need not solve the strong CP problem, often have masses that are independent of the temperature in the early universe, or feature a very different temperature dependence than the QCD axion. This has to be taken into account when calculating their abundance. It is possible to obtain estimates for the ALP DM abundance similar to Eq.~\eqref{eq:qcd_axion_ede}, which trace out natural lines in the mass vs.\ coupling plane (cf.\ \cite[e.g.][]{Arias:2012az}).
Due to the large number of possible ALP models, there is however a wide range of conceivable models and corresponding predictions for masses  and couplings.

\paragraph{Modifications and non-standard realignment scenarios.} Note that several assumptions have been made in the canonical picture, many of which have been relaxed or modified.
As mentioned before, this is often done to connect axions to other physics or alter the preferred \fa region by diluting or enhancing the axion energy density.
Such scenarios are an active area of research, considering e.g.\ particular initial conditions set by early universe dynamics, as with a large~\cite{Co:2018mho,Takahashi:2019pqf,Arvanitaki:2019rax,Huang:2020etx} or a small~\cite{Dvali:1995ce,Banks:1996ea,Choi:1996fs,Co:2018phi} misalignment angle, or a large initial kinetic energy as in the kinetic misalignment mechanism~\cite{Co:2019jts,Chang:2019tvx}. Kinetic misalignment is intimately connected to ``axiogenesis''~\cite{Co:2019wyp}---a mechanism to explain the baryon asymmetry of the Universe; in this paradigm, the QCD axion is predicted to have a mass between $\mathcal{O}(60~\mu{\rm eV}-60~{\rm meV})$ in various axiogenesis implementations~\cite{Co:2019wyp,Domcke:2020kcp,Co:2020jtv,Harigaya:2021txz,Chakraborty:2021fkp,Kawamura:2021xpu,Co:2021qgl}, whereas for the ALP, a relation between $f_a$ and $m_a$ is predicted as $f_a \simeq 2\times10^9~{\rm GeV} (\mu{\rm eV}/m_a)^{1/2}$ in ``ALP cogenesis''~\cite{Co:2020xlh}. The radial mode dynamics of the PQ breaking field can also provide the axion dark matter abundance~\cite{Co:2017mop,Harigaya:2019qnl,Co:2020dya} via the so-called parametric resonance effect.
On the other hand, modifications of the thermal history~\cite{Visinelli:2009kt,Arias:2021rer} of the Universe, i.e.\ of $H(t)$, are conceivable, including entropy injection~\cite[e.g.][]{Dine:1982ah,Steinhardt:1983ia} or non-standard inflation scenarios~\cite{Dimopoulos:1988pw,Davoudiasl:2015vba,Hoof:2017ibo,Graham:2018jyp,Takahashi:2018tdu,Kitajima:2019ibn}.
Furthermore, extended axion models can introduce modifications to $V[\theta]$ and thus $\ma(t)$ or give additional terms from axion interactions.
These include the relaxion mechanism~\cite{Graham:2015cka}, the ``ALP miracle''~\cite{Daido:2017wwb,Daido:2017tbr}, the ``fragmentation'' scenario~\cite{Fonseca:2019ypl,Morgante:2021bks}, trapped realignment~\cite{DiLuzio:2021gos}, or many-axions in string theory~\cite{Cyncynates:2021yjw}.
As in the ``ALP miracle'' case, the initial $w = -1$ behaviour of axions has inspired models for axion inflation such as ``natural inflation''~\cite{Freese:1990rb}, ``n-flation''~\cite{Dimopoulos:2005ac}, or ``monodromy inflation''~\cite{Silverstein:2008sg,McAllister:2008hb}.

\subsubsection{The Pre-Inflation Scenario}\label{sec:pre_infl}

In one of two primary cosmological scenarios, the PQ symmetry breaks before the end of inflation (and remains broken afterwards). This occurs when the breaking scale $\sim \fa$ is greater than the Gibbons--Hawking temperature at the end of inflation, $T_\text{I} = H_\text{I}/2\pi$~\cite{Gibbons:1977mu} as well as the reheating temperature after inflation.
Two important consequences of this scenario are that
\begin{itemize}
    \item the observable universe is contained within one causally connected patch at the time of PQ breaking, where the misalignment angle is often assumed to be uniform randomly set, $\thetai \sim \mathcal{U}(-\pi,\pi)$~\cite{Turner:1985si}, and
    \item there exist isocurvature fluctuations of the field of order $\delta \theta \sim H_\text{I}/2\pi\fa$.
\end{itemize}

While the homogeneous field component dominates the energy density, field fluctuations are present -- either as vestiges from inflation or as imprints from interactions with other matter or radiation species. They remain Jeans-stable until matter-radiation equality.

As mentioned before, it is often assumed that topological defects are inflated away in this scenario and play no role for the axion energy density. However, depending on the values of $\fa$ and $H_\text{I}$, isocurvature fluctuations can impose constraints~\cite{Axenides:1983hj,Turner:1983sj,Lyth:1991ub,Hertzberg:2008wr,Visinelli:2009zm,Kobayashi:2013nva}. Furthermore, fine-tuning and possibly anthropic considerations need to be taken into account~\cite{Tegmark:2005dy,Mack:2009hv}. Despite the potential limitations, this scenario can link axions and inflationary physics, which is being actively investigated~\cite[e.g.][]{Graham:2018jyp,Co:2018phi}.

Applied to a pre-inflation axion, improved CMB polarization measurements and lensing expected from experiments like AdvACT, CMB-S4, LiteBIRD, BICEP array, Simons Observatory, and SPT-3G coupled with any local measurement of $f_a$ explored in later sections could identify at least a portion of the relic dark matter abundance as well as provide an anchor for the scale of inflation.

\subsubsection{The Post-Inflation Scenario}\label{sec:post_infl}

In the complementary scenario, the PQ symmetry breaks (for the last time) after inflation has ended. The axion field, again, takes random values $\thetai \sim \mathcal{U}(-\pi,\pi)$ in causally disconnected regions. However, the observable universe now consists of a huge number of these regions, such that the energy density of axions today can be calculated as an average over the causally disconnected regions. As a consequence, the axion DM abundance becomes independent of any one~$\thetai$ and, for QCD axions, only depends on~$\fa$. There may also be significant local DM overdensities and thus a highly non-linear evolution, giving rise to observable consequences discussed below.

Topological defects created during PQ breaking~\cite{Kibble:1976sj} are not inflated away in this scenario. They can increase the axion DM density through their evolution and decays~\cite{Davis:1985pt,Davis:1986xc,Harari:1987ht,Battye:1993jv}. While the phenomenology of cosmic strings, domain walls, and axitons has been studied extensively over the past decades~\cite{Davis:1985pt,Davis:1986xc,Harari:1987ht,Battye:1993jv,Battye:1994au,Hagmann:1998me,Hiramatsu:2010yu,Klaer:2017qhr,Klaer:2017ond,Gorghetto:2018myk,Vaquero:2018tib,Buschmann:2019icd,Hindmarsh:2019csc,Gorghetto:2020qws,Dine:2020pds,Buschmann:2021sdq,Hindmarsh:2021zkt}, no definitive consensus has been reached regarding the relevance of their contribution to the axion DM density. Recently, the contribution from string decay and wall annihilation have also been considered for the case of ALPs~\cite{Gelmini:2021yzu,OHare:2021zrq}.

The remaining large density fluctuations create local matter-dominated regions deep within the radiation era, causing gravitational collapse of modes once the Hubble scale exceeds a large fluctuation's Jeans scale~\cite{Eggemeier:2019khm,Buschmann:2019icd}. This creates mini-halos of mass up to the order of $\num{e-12}\,M_\odot$ by redshift $z = 100$, and most are expected to contain a compact soliton core with mass relation $M_\text{core} \propto M_\text{halo}^{1/3}$~\cite{Eggemeier:2019khm,Buschmann:2019icd}. The fraction of matter bound in mini-halos has been simulated to be up to 75\% at the beginning of the matter-dominated era \cite{Eggemeier:2019khm}.

At the cost of inflationary physics insight, the post-inflation axion motivates a rich late Universe phenomenology with accompanying signatures in gravitational waves and scale-dependent features in the Large Scale Structure which will be explored by expanded galaxy surveys, improved CMB measurements, and low-frequency gravitational wave detectors.

\subsubsection{Thermal Production and Evolution}\label{sec:thermal_production}

Axions can also be produced thermally in the early Universe. As dark radiation, they can modify the observed effective number of neutrino species and, as hot dark matter, influence structure formation. This leads to e.g.\ an upper limit on the QCD axion mass in the \si{\eV} range~\cite{Giare:2020vzo}. Other constraints on QCD axions and ALPs can come from CMB spectral distortions, Lyman-$\alpha$ forest, space-\ or ground-based observatories, or structure formation~\cite{Cadamuro:2011fd,Cadamuro:2010cz,Millea:2015qra,Green:2021hjh,LOI_CMBS4,CosmicProbULAloi,LOI_LightRelicsCosmo,LOI_LightRelicsCosmo2,LOI_SpectralDistortions}.

It has recently been emphasized again that the theoretical computation of the axion production rate around the QCD crossover is problematic~\cite{DiLuzio:2021vjd}. This will be relevant for the upcoming generation of CMB experiments, which will be able to probe QCD axions in this regime~\cite{Abazajian:2019eic,Giare:2021cqr,LOI_CMBS4}, with the first steps towards improvement already underway~\cite{DEramo:2021psx,DEramo:2021lgb}.

Generally, the scale of a potential thermal axion as a component of the relic dark matter stands to be constrained by any new probe of cosmological evolution reaching into new redshift ranges, complemented by local searches like those explored below to identify the axion coupling. 

\subsection{Late-Time Structure Formation and the Local Properties}

After entering the matter-dominated era, broader fluctuations begin to collapse gravitationally as their modes enter the Hubble horizon, forming e.g.\ halos and filaments, and seeding the formation of stars, galaxies, and other baryonic structures. Understanding the distribution of axions within halos and other cosmic structures, as well as their interplay with the baryons, is crucial for the efficacy of axion searches.

Several properties of the QCD axion and ALP DM distribution are important for detection: local persistent density and spectral shape, coherence qualities, and prevalence of transient structures. 

\begin{figure}
    \centering
    \includegraphics[width=\textwidth]{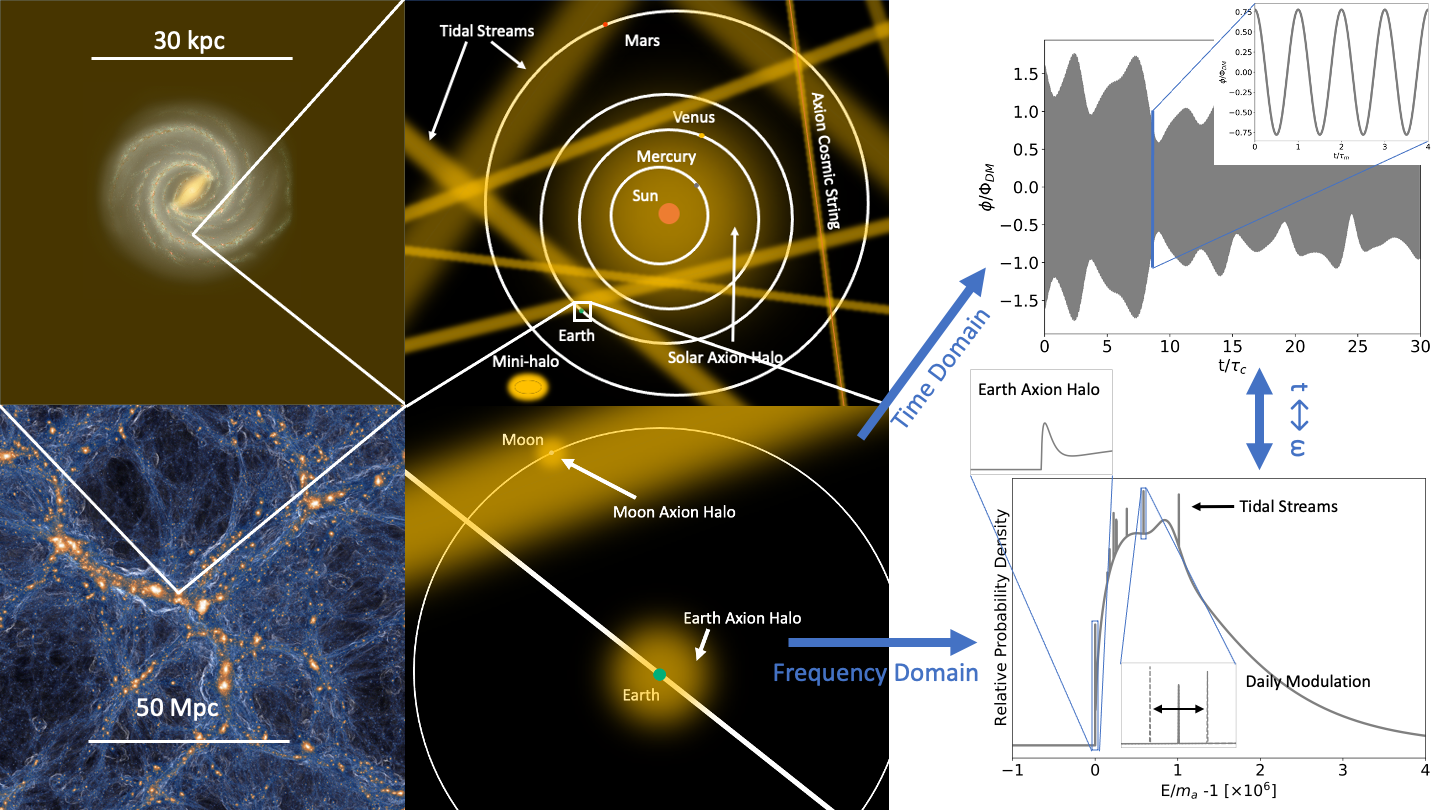}
    \caption{Illustration of Axion DM structure in the modern Universe from the scale of the cosmic web (lower left panel) to the mean axion field as may be observed by a terrestrial search (right panels). Each panel represents a view of axion DM structure at a different scale, with each of the four left panels showing the axion density in orange. The cosmic web (lower left) shows the clustering behavior of DM halos and surrounding gases. The MW (upper left) is surrounded by a halo with virial radius at least on order of magnitude larger than the Galaxy. The inner Solar System and the region immediately surrounding the Earth (middle panels) may contain axion structures beyond the diffuse component, including mini-halos and Bose stars and their tidal streams, axions bound to compact objects such as the Earth and Sun, and axion strings. The axion field as viewed by a terrestrial search (right panels) can be characterized by its coherence properties over time and the structure of its spectra. Amplitudes of sub-structures in the frequency spectra were chosen for illustrative purposes and are not necessarily indicative of a feature's expected prominence. The cosmic web image is a snapshot of gas shocks (blue) and DM density (orange) from the Illustris TNG Collaboration's TNG100-1 simulation~\cite{Springel:2017tpz}. The upper left panel contains an artist's conception of the MW's spiral structure as informed by NASA's Spitzer Space Telescope, with image credit to NASA/JPL-Caltech/R. Hurt (SSC/Caltech)~\cite{SpitzerMWimage}.}
    \label{fig:axion_abundance}
\end{figure}

See Figure~\ref{fig:axion_abundance} for an illustration of the various scales involved.

Many axion dark matter searches make use of temporal and/or spatial coherence, often on macroscopic time and length scales. Crucially, the coherence time sets the maximal usable quality factor in resonant experiments discussed in Sections~\ref{sec:above_microeV} and \ref{sec:below_microeV}. Similarly, the coherence length scale determines the volume where experiments can benefit from coherent enhancement, a fact that becomes particularly relevant in experiments aiming at large experiments searching for axions at higher masses~\cite{Brun:2019lyf,universe8010005}.

Roughly speaking on time/length scales smaller than the coherence time/length the field amplitude can be viewed as a simple plane wave with well-defined amplitude and phase. 
This only works when the energy and momentum distribution of the waves contributing to the field is sufficiently narrow and on time/length-scales small enough such that the differences in frequency and wave length can be neglected. Therefore, the coherence time is given by the energy/frequency dispersion of the field $\tau_\text{c} \sim 2 \pi / \Delta E\sim  10^{6} (2\pi/m_{a})$, and the coherence length by the momentum/wave length dispersion of the field $\lambda_\text{c} \sim 2 \pi/ (m \Delta v) \sim 10^{3} (2\pi/m_{a})$, where we have used the rough estimate ${\langle v^2\rangle}^{1/2} \sim 10^{-3}$
at Earth's location in the Milkyway. 

The improving sensitivity and resolution of searches allows us to test more detailed axion DM distribution models.
The model space of distributions is potentially very rich due to the candidates' unique nature as highly degenerate Bose fluids.
The early standard halo model, based on a non-singular or lowered isothermal sphere, produced the first conservative estimates for the halo distribution's line shape and response to daily and annual modulations due to the motion of Earth-based detectors~\cite{PhysRevD.42.3572}.
Another model of Bose DM halo was also created during this period, producing line shapes containing more fine structure and sensitivity to the path of a detector~\cite{Sikivie1998}.
Ultralight ALPs ($m \le \SI{e-20}{\eV}$)~\cite{LOI_ULADMTheory} have been simulated via a Gross--Pitaevskii~(GP) wave equation in small cosmological volumes capable of holding a cluster of field dwarf galaxies of halo mass up to $10^{9-11}\,M_{\odot}$, with semi-analytical techniques extrapolating the results to larger halo and particle masses relevant to some of the searches presented below~\cite{2014NatPh..10..496S,2020ApJ...889...88L,2020PhRvD.101h3518V,2021arXiv211009145S}. The typical QCD axion mass range, however, remains out of reach of having its wave nature resolved in simulations of MW-mass halos even with modern zoom-in techniques, pushing the use of pre-existing N-body simulations with pressureless DM~\cite{Lentz_2017}.
More recent work has found that Bose DM infall may be modified from the GP model, even above the coherence scale, with preliminary N-body simulations showing novel halo distributions and the potential for fine structure \cite{10.1093/mnras/stz488,10.1093/mnras/staa557,ManyBodyDMloi}. 

The mini-halos and Bose stars of post-inflationary axion models are also too small to be resolved in a numerical simulation of a MW-type galaxy, and their rate of survival to the present from tidal stripping due to stars, planets and other compact objects they may encounter is unsettled 
\cite{TidalStreams2016,2017JETP..125..434D,PhysRevD.104.063038,OHare:2017yze}. However, the general expectation is that there exist, in our local Galactic region, some residual number of compact mini-halos and Bose stars, and ultra-cold flows of tidally stripped matter in addition to the more diffuse component. The former will be seen as transient objects to terrestrial searches.
The later will be visible in the local DM distribution, but with the number, trajectory, and location of flows being uncertain and time-dependent \cite{TidalStreams2016,2017JETP..125..434D,PhysRevD.104.063038}.

\section{Micro-eV and Above}
\label{sec:above_microeV}

As discussed in section~\ref{sec:motivation}, simple DM production mechanisms for the QCD axion motivate searching at masses $m_a \gtrsim \mu\mathrm{eV}$. The most well-established technique in this mass range is the resonant cavity haloscope, first proposed by Pierre Sikivie~\cite{Sikivie:1983ip}, where the electromagnetic fields created by an axion in a large static magnetic field $B$ are resonantly amplified in a microwave cavity. A cavity mode of frequency $f$ is sensitive to axions of mass near
\begin{equation}
m_a = (4.1 \, \mu\mathrm{eV}) \, \frac{f}{\mathrm{GHz}}.
\end{equation}
To cover a wide range of axion masses, the mode frequency must be scanned. To maintain sensitivity to the QCD axion, the scan rate scales as 
\begin{equation}
\frac{df}{dt} \propto QV^2C^2B^4T^{-2}
\end{equation}
where $Q$ is the quality factor, $V$ is the volume, $C$ is a geometric form factor, and $T$ is the noise temperature. 

In sections~\ref{sec:admx_g2} and~\ref{sec:admx_efr}, we discuss the current and future goals of the Axion Dark Matter eXperiment (ADMX) collaboration, which is the most advanced existing experiment targeting ${\sim}\mu\mathrm{eV}$ masses. In the following sections, we discuss upcoming and proposed experiments which target progressively higher axion masses. Here the fundamental difficulty is that in order to maintain a high form factor $C$, the size of a simple cavity haloscope must scale with the axion Compton wavelength $1/m_a$. As a result, the volume falls as $V \propto 1/m_a^3$, significantly decreasing the scan rate. 

The HAYSTAC experiment (section~\ref{sec:haystac}) aims to compensate for this using quantum metrology techniques, which decrease the noise beyond the standard quantum limit. The SQuAD experiment (section~\ref{sec:squad}) boosts the signal with a high-$Q$ dielectric cavity and reduces noise using qubit-based single photon counting.  The proposed ALPHA (section~\ref{sec:alpha}) and MADMAX (section~\ref{sec:madmax}) experiments instead use alternative detectors, which break the scaling relation between the volume and the axion mass. Finally, the proposed BREAD experiment (section~\ref{sec:bread}) does not rely upon resonance at all, leading to broadband sensitivity up to the highest axion masses. There is vibrant global interest in searching for axions, and we conclude in section~\ref{sec:above_ueV_global_context} by describing some efforts based outside the US. 

\subsection{ADMX G2}
\label{sec:admx_g2}
\noindent {\bf Physics goals:}
The ADMX collaboration is operating the ``Generation-2'' ADMX haloscope project in order to cover the frequency range from 0.6{\textendash}2 GHz (corresponding to axion masses between 2.5{\textendash}8.3 \SI{}{\micro\eV}) at DFSZ coupling sensitivity by 2025. The ADMX-G2 project began collecting data in late 2016 and, as of this whitepaper, has published three data-sets~\cite{Braine:2019fqb,Du:2018uak,ADMX:2021abc}. It is currently the only experiment that has reached DFSZ axion coupling sensitivity. See Fig.~\ref{fig:4:madmax:sensitivities} for recent exclusion limits. 

\begin{figure}
\centering
\includegraphics[width=\linewidth]{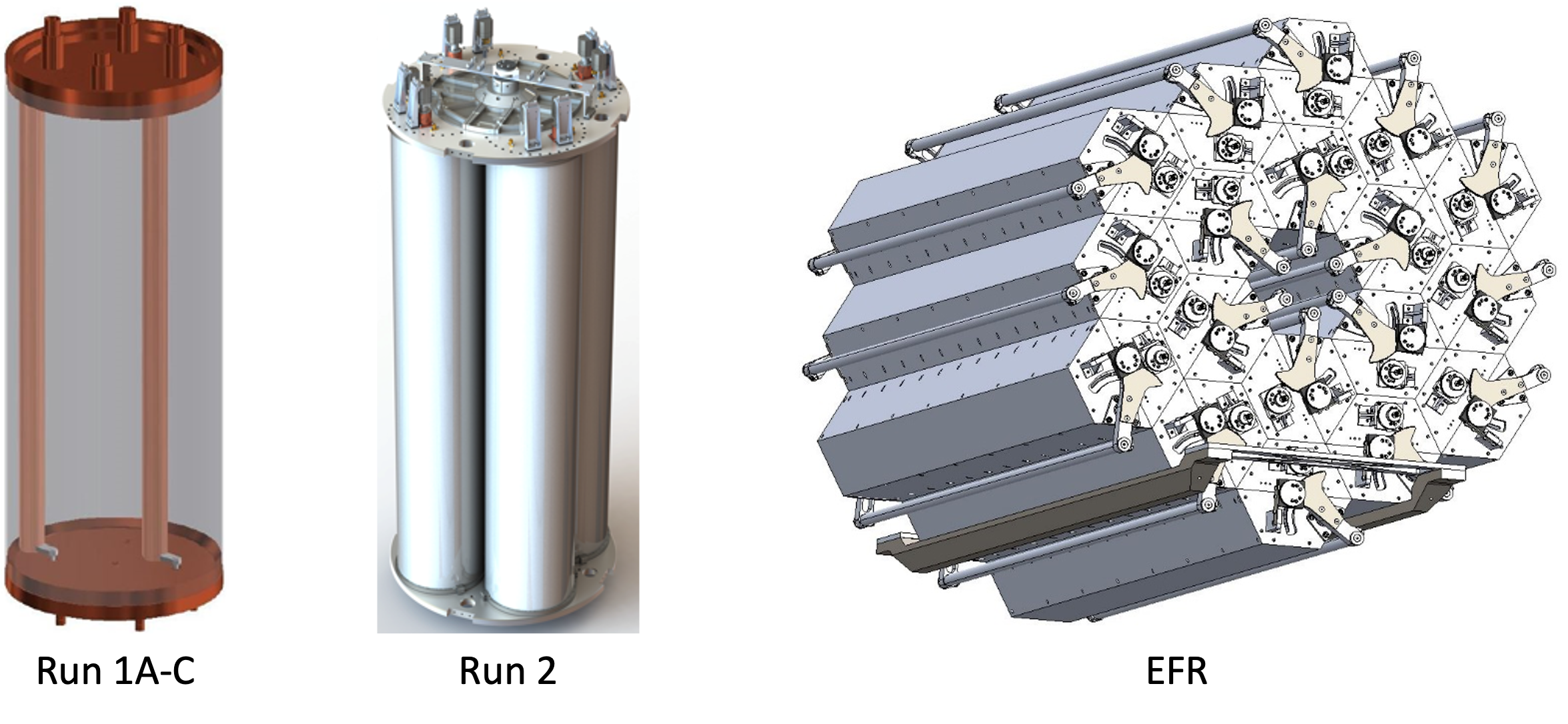}
\caption{\label{fig:admx_diagrams} Cavity diagrams for Run 1A-C (Run 1D will use one tuning rod in a single cavity), Run 2A, and ADMX-EFR.}
\end{figure}

\noindent {\bf Concept:} The ADMX-G2 experiment is built around a large (8.5 Tesla) high inductance (540 Henry) NbTi Superconducting Solenoid with a 60 cm inner diameter and 112 cm long placed at the bottom of a re-entrant bore cryostat that allows for an experimental package (cavity system and cold electronics) to be installed. The ADMX collaboration deployed various experimental packages of increasing sophistication. Since 1996 it operated at LLNL at pumped LHe temperatures (1.2 K) to reach KSVZ axion couplings. The magnet system was moved to the U. of Washington in 2010 and the ADMX-G2 project was designed around connecting a Janis dilution refrigerator directly onto the large ($\sim$ 140 liter) microwave cavity to reduce the physical temperature to $\sim$ 100 mK. A second ``bucking'' magnet is contained in a separate LHe reservoir $\sim$ 1 meter above the main cavity which reduces the field to $\sim$ 10s of Gauss in the region where the cold-electronics (circulators, switches and quantum amplifiers) provide the first stage of amplification. RF power is then routed to low-noise transistor amplifiers for a 2nd stage of amplification before the signals are routed out of the fridge where they are processed by digitizing electronics. Motion control is provided by room-temp stepper motors connected through vacuum feedthroughs to cryogenic gearboxes that move tuning rods of various size to scan frequencies. The runs are split into frequency bands to match the tunable range of the quantum amplifiers and circulators. Run 1A used a Microstrip SQUID Amplifier but subsequent runs used Josephson Parametric Amplifiers (JPAs) for amplification. The employed cavitiess are shown in Fig.~\ref{fig:admx_diagrams}. Runs 1A and 1B used two 2 inch OD copper tuning rods that ran the length of the cavity, Run 1C used two 4.4 inch OD rods and Run 1D will use a single 8.1 inch OD rod to sweep to 1.4 GHz. After this Run 2A will combine 4 frequency locked cavities coherently to take advantage of the axions coherent signal to make up for the volume lost by operating a single cavity. It is anticipated that this system will complete the scan at DFSZ sensitivity by 2025.

\subsection{ADMX EFR}
\label{sec:admx_efr}
\noindent {\bf Physics goals:}
The ADMX collaboration aims to cover the frequency range from 2{\textendash}4 GHz with the Extended Frequency Range (EFR) experiment in the period 2024{\textendash}2027. The goal is to achieve DFSZ sensitivity across this frequency range, which corresponds to an axion-mass range of 8.3{\textendash}16.5 \SI{}{\micro\eV} (see also Figs.~\ref{fig:4:madmax:sensitivities}).

\noindent {\bf Concept:} 
The cavity array inserted into the. magnet is shown in Figure~\ref{bread_and_admx} (right).
As ADMX transitions to the higher frequency parameter space, sensitivity loss due to the decreasing volume of the cavity will ensue. The ADMX collaboration will circumvent the reduction in scan rate with frequency in several ways: 

\begin{enumerate}
    \setlength\itemsep{0em} 
    \item Combine the signal from eighteen cavities to maintain the total volume (cf.~Figure~\ref{fig:admx_diagrams} (right));
    \item Increase the magnetic-field strength through the use of high-field superconducting magnet materials such as $\mathrm{Nb_3Sn}$ and rare earth superconductors (REBCO);
    \item Increase the quality factor by coating cavities with superconducting films or low-loss dielectrics;
    \item Reduce the amplifier noise level by leveraging the properties of squeezed states or developing single photon counting techniques that evade the Standard Quantum Limit.
\end{enumerate}

The magnetic field will be provided by a 9.4 T magnet, currently located at UIC. The magnet bore is horizontal, simplifying access to the cavity array and enabling a new approach to the cryogenic system and to the implementation of RF electronics. Mechanical and thermal design is underway for the horizontal cryostat that minimizes dead volume in the bore and can support the cavity weight. 

ADMX-EFR will use an array of eighteen cavities, shown on the right in Figure~\ref{fig:admx_diagrams}. The cavities use a `clamshell' design that enables the tuning rods to be inserted with ease. Each cavity has a single tuning rod operated by a rotary piezo motor.

The design of the EFR signal chain is motivated by the need to achieve low system noise.  Each cavity will have an independent signal chain, from the cavity to digitizer.  The front-end amplification will be provided by a flux-pumped JPA for each cavity. Flux-pumped JPAs reduce the risk of contamination from the pump tone within the signal band and remove the need for a directional coupler. Signals will be digitized by phase-locked digitizers and combined after calibration accounting for phase and amplitude differences between channels.

The cryogenic electronics , not visible in Figure~\ref{bread_and_admx} (right), will be physically separated from the cavities and magnet.  The cavities and cryogenic electronics will use independent dilution refrigerators to maintain their own ideal system temperatures.  The benefits of this design are two-fold: (1) the electronics can be maintained at a lower temperature than the cavities, thereby reducing the system noise, and (2) JPAs are highly susceptible to magnetic fields, and operating them away from the magnet will protect them from magnetic flux penetration that can diminish their performance.

Table~\ref{tab:target} shows the target performance for ADMX-EFR and various system parameters.

\begin{table}
\begin{center}
\caption{\label{tab:target} Experimental parameters for ADMX-G2 and EFR runs. In Run 1 where parameters varied between runs, ranges or exceptions are given.}
\begin{tabular}{| l | c | c | c |}
\hline
 & Run 1A-D & Run 2A & EFR \\ 
 \hline
 Frequency Range & 650--1390 MHz & 2--4 GHz & 2--4 GHz \\  
 Number of Cavities & 1 & 4 & 18 \\
 Volume & 106--136 $\ell$ & 85 $\ell$ & 80 $\ell$ \\
 Q & 58,000 (avg) & 60,000 & 90,000 \\
 B Field & 7.7 T (A: 6.85 T) & 7.7 T & 9.6 T \\
 Avg Form Factor & 0.45 (D: 0.35) & 0.45 & 0.4 \\
 Noise Temperature & 300--350 mK (A: 700 mK) & 350 mK & 325 mK \\
 Operations Days & 1000 & 300 & 1000 \\
  \hline
\end{tabular}
\end{center}
\end{table}

\subsection{HAYSTAC}
\label{sec:haystac}
\noindent {\bf Physics goals:} Predictions of the axion mass have tightened up dramatically, with the higher masses $m_{a} \approx$ 15-50 $\mu$eV (4-12 GHz) now being a particularly well motivated region~\cite{Klaer2017,Buschmann:2019icd}. The Haloscope at Yale Sensitive to Axion cold dark matter (HAYSTAC)\cite{Brubaker2017,Zhong2018,Backes2021} experiment, utilizes a quantum squeezed receiver which evades the Standard Quantum Limit (SQL) entirely, and thus enables an acceleration of the scan rate by a factor 2-3 in the exploration of the currently open axion mass range. At the same time, the HAYSTAC experiment is also intended to be an agile technology platform where the latest advancements in quantum measurements/instrumentation for axion searches can be explored over the next decade.

\noindent {\bf Previous Results and Future Direction:} 
From its inaugural data run in 2015, HAYSTAC has been the only experiment to operate at the SQL, and since 2018, below it, the first data publication from the Squeezed-State Receiver (SSR) appearing in Nature in early 2021~\cite{Backes2021}.  It is thus one of only two experiments to employ squeezed-states in fundamental physics, the other being LIGO. In addition, optimizations in the Bayasian analysis chain~\cite{Palken2020}, have accelerated the scan rate by a further factor of $\approx$ 3 over the standard mode of operation, and our codification of the axion haloscope analysis techniques have been adopted by several other QCD axion sensitive haloscopes~\cite{Du:2018uak,Choi2020}. With these advancements in hand, the HAYSTAC experiment is transitioning to become a data production experiment that seeks to cover the $\approx$~4-6~GHz range as quickly as possible while maintaining close or better than 1.5$\times$KSVZ sensitivity.

While the current analysis efforts focused on the 4-6 GHz range, the next iterations of the HAYSTAC experiment will focus on higher mass ranges to develop the technologies and techniques for axion searches in the 10-12 GHz range. Moving forward in terms of hardware, the HAYSTAC team seeks to upgrade the current HAYSTAC magnet to a 12 Tesla Nb$_{3}$Sn magnet, demonstrate 4 dB of squeezing or better, and continue to push forward on several promising extensions of cavity designs such as a 7-rod cavity design \cite{Simanovskaia2020} or the testing of tunable resonators utilizing photonic band gap structures to eliminate the forest of intruder TE modes with all the resonator work being led by UC Berkeley. The Josephson parametric amplifiers used in the current version of HAYSTAC, have already been proven to work effectively up to 12 GHz, but further R\&D efforts will be needed to push this effective frequency range even higher. The Colorado group is simultaneously working on the Cavity Entanglement And State-swapping Experiment For Increased Readout Efficiency (CEASEFIRE) R\&D project to improve the quantum axion search enhancement even further by attacking the biggest limitation in the initial demonstration of the squeezed state receiver, namely loss in transporting the squeezed microwave fields~\cite{Wurtz2021}. In the next three years, the goal is to deploy the CEASEFIRE concept in an axion search haloscope, demonstrating a factor of 10 or more speed up beyond the quantum limit. HAYSTAC will also continue other R\&D efforts such as a quantum non-demolition Rydberg atom single-quantum detector with an eye towards higher frequency axion searches, this effort being led by Yale (elaborated on in Section \ref{sec:RAY}). As the newest member of HAYSTAC, the group at Johns Hopkins is targeting the assembly of a second haloscope and the expansion of HAYSTAC data acquisition to provide a broader range of possibilities for analysis. The John Hopkins group will also aim to contribute to the R\&D efforts of the collaboration as a whole by providing a testbed for new detector components. 

By 2023, Yale will be completing a new Physics Science and Engineering Building that includes a state-of-the-art Advanced Instrumentation Development Center which includes fabrication facilities and other vital infrastructure to support Yale's quantum science and advanced instrumentation initiatives. The building will provide new laboratory space for HAYSTAC and enable more research on quantum devices, making Yale a national center for quantum-enhanced dark matter research and providing support for upcoming HAYSTAC runs. With exciting improvements from the analysis, hardware and infrastructure, HAYSTAC lies in an excellent position to scan over a wide range well motivated unexplored axion parameter space in the $\sim$ 4-12 GHz (15-50 $\mu$eV) mass range over the coming decade.

\subsection{RAY}
\label{sec:RAY}
\noindent\textbf{Physics goals:}
The Rydberg-atoms for Axions at Yale project plans to search for DFSZ axions with masses above 40 ueV (10 GHz) using Rydberg atoms as single photon detectors.

\noindent\textbf{Concept:}
Single-photon detection offers a promising way to eliminate quantum noise by measuring only amplitude and pushing all quantum noise to the unmeasured phase \cite{lamoreaux2013analysis}. In standard cavity haloscope experiments, the signal power scales as $m_a^{-4}$ \cite{Sikivie:1983ip}. Additionally, cavity $Q$-factors decrease with increasing frequency due to the anomalous skin effect, resulting in a further reduction to the signal power. Measuring the small resulting signals requires reduction of system noise. However, efforts to reduce this noise in quadrature measurements are inevitably limited by the unavoidable presence of quantum noise \cite{Caves:1981hw, lamoreaux2013analysis}.

With single photon detection, the ability to measure the amplitude of an axion signal is limited by the shot noise on the number of detected background photons. By exciting $^{39}K$ atoms to the appropriate Rydberg level for which the transition frequency to the next state is equal to the axion mass being probed, these atoms can be used as single photon detectors \cite{zhu2021eit}. CARRACK successfully demonstrated the power of the Rydberg technique by measuring the blackbody spectrum of their system using Rydberg atoms {\it in situ} at cryogenic temperatures \cite{carrackiibbpaper}. In principle, Rydberg atoms can span the entire mass range of the axion. Other technical aspects of the experiment, such as microwave cavities, will need additional development.

The collaboration is considering ways to enhance signal fidelity. One potentially exciting opportunity is via a sequence of so-called quantum non-demolition (QND) photon-number measurements \cite{nogues1999seeing}. QND exploits off-resonant Rabi oscillations whereby the energy initially carried out by the photon oscillates repeatedly between the cavity field and a qubit whose transition frequency is far off-resonance relative to the photon frequency. This results in a rotation of the phase of the cavity field that can be then measured, providing an unambiguous signature of the presence of a single photon. QND becomes more advantageous than a simpler scheme, such as simply detecting ionized electrons, if we can achieve a cavity Q of $\sim10^7$ or higher.

\subsection{SQuAD}
\label{sec:squad}
\textbf{Physics Goals:} 
The Superconducting Qubit Advantage for Dark Matter (SQuAD) experiment plans to perform resonant searches for dark matter axions with DFSZ sensitivity in a broad range from 10-30 GHz using high quality factor dielectric cavities combined with qubit-based single photon detectors which evade the quantum zero-point noise.  The targeted mass range can be efficiently scanned over the course of a couple of years, possibly with multiple, simultaneously running experiments.

\begin{figure}
    \centering
    \includegraphics[width=0.7\textwidth]{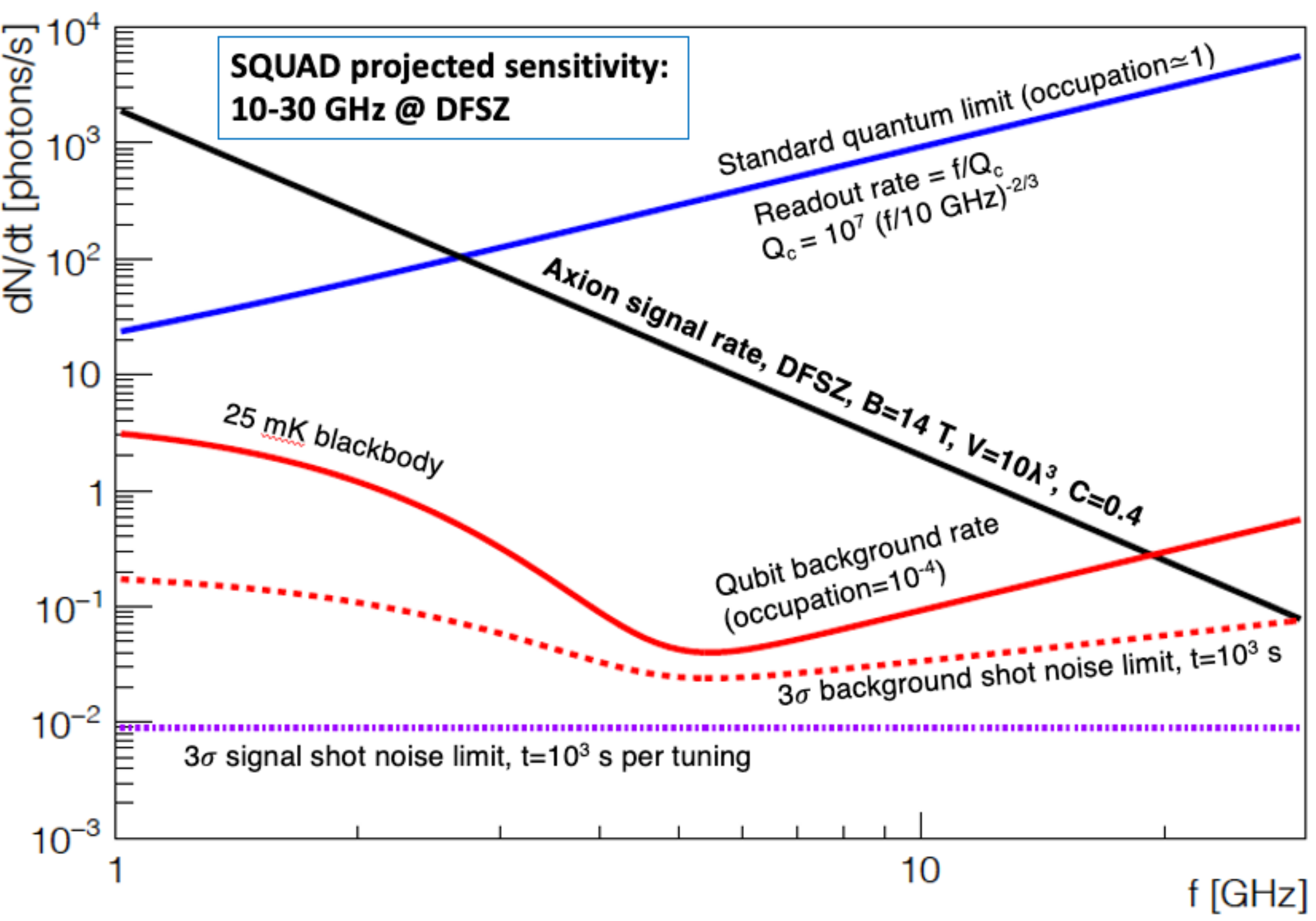}
    \caption{Predicted signal and background rates for SQuAD along with sensitivity obtained when averaging over $10^3$~s integration time per cavity tuning.  For demonstrated experimental parameters, DFSZ sensitivity is obtained for frequencies up to 30~GHz.
    }
    \label{fig:squad_sensitivity}
\end{figure}

\noindent \textbf{Concept:} 
While contemporary axion search experiments in the GHz range overcome a poor signal-to-noise ratio by extensive averaging of the noise fluctuations, this strategy cannot work for higher frequency searches at $f=10$~GHz and beyond.  As shown in Fig.~\ref{fig:squad_sensitivity}, the expected axion signal photon rate (black line) plummets with increasing frequency because the resonant cavity size must shrink to match the axion's Compton wavelength.  Meanwhile, the thermal or vacuum noise is incurred with each individual readout, and as the signal frequency increases, so does the readout rate and the resulting noise background rate (blue line).  The current techniques are thus not scalable to higher frequencies.  

The signal rate can be increased using extremely high quality factor $Q_c \approx 10^7$ dielectric cavities made by nesting a series of low loss cylindrical cavities to reflect the photon inward, keeping its electric field away from the lossy copper walls of the cavity enclosure~\cite{DiVora:2022tro}.  For high $Q_c > Q_a = 10^6$, the axion signal power saturates to its maximum value as it becomes limited only by the dark matter coherence time $Q_a/f$.  At longer time scales, the cavities can be used as incoherent photon power accumulators which store the signal energy for later readout, much as a CCD bucket would store signal electrons.  The noise rate again scales with the readout rate which can now be very slow due to the extremely long cavity storage lifetimes $Q_c/f$.  These improvements are reflected in the plot.

While higher frequency thermal photons will disappear at low temperatures, the quantum zero-point noise will be inevitable in linear amplifier readouts which measure both non-commuting quadratures of the signal photon wave.  This Heisenberg uncertainty noise can be evaded by using single photon detectors which measure only the amplitude and not the phase of the photon wave~\cite{lamoreaux2013analysis}.  Qubit-based quantum non-demolition techniques have reduced the effective dark counts by at least 4 orders of magnitude (solid red curve) relative to the quantum noise~\cite{Dixit:2020ymh}, thus reducing the required averaging time also by 4 orders of magnitude.  The resulting rate sensitivity is shown in dashed red.

\noindent\textbf{Sensitivity \& Timeline:}
Using demonstrated experimental parameters, the SNR is favorable for $f<30$~GHz and so experiments at these frequencies can be quickly done over the next couple of years.  Due to non-negligible averaging times, efficient coverage of the mass parameter space may require multiple haloscopes to be simultaneously operated, perhaps by different experimental groups.   R\&D is ongoing on developing an analogous photon counting readout based on Rydberg atoms which can be operated at the higher frequencies where qubit devices become more difficult to design and fabricate.  Background rates are expected to be similarly low for atom-based readout.

\subsection{ALPHA}
\label{sec:alpha}

\begin{figure}
    \centering
    \includegraphics[width=0.5\textwidth]{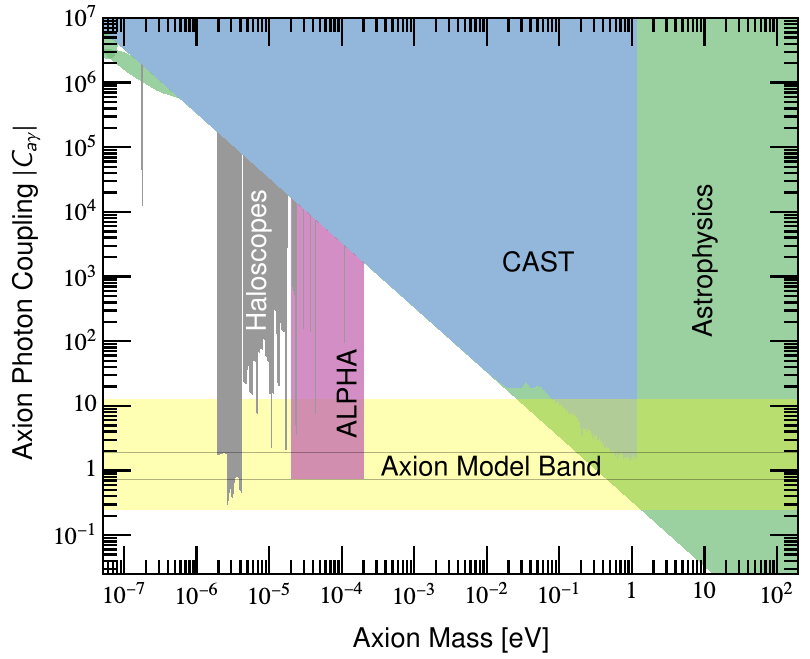}
    \caption{Axion parameter space showing the dimensionless axion-photon coupling constant $C_{a\gamma}$ as a function of the axion mass, $m_a$, with various bounds. This dimensionless coupling is defined so that it is O(1) for models which solve the strong CP problem regardless of the axion mass. The axion mass gives the frequency at which the field oscillates, typically corresponding to the microwave regime (roughly 10 GHz). Haloscope experiments looking for dark matter are shown in grey, the limit from the CAST helioscope looking for solar axions is shown in blue and astrophysical limits are shown in green. The extended QCD axion model band is shown in yellow, with traditional KSVZ and DFSZ models shown as black lines. The region to be explored by ALPHA is shown in purple, covering a significant portion of well-motivated parameter space. This projection assumes a plasma haloscope with quality factor Q=5000 inside a 13 Tesla, 50 cm bore solenoid magnet using quantum limited detection running for three years. }
    \label{fig:alphaproj}
\end{figure}

{\bf Physics goals:} The Axion Longitudinal Plasma HAloscope (ALPHA) Consortium aims to explore high mass axions (5-50 GHz) by pioneering a plasma haloscope.\\
\noindent
{\bf Concept:} At high frequencies, the small size of simple cavities leads to a dramatic loss in signal power. In order to push above a few GHz, the tyranny of volume scaling must be ameliorated. One recent proposal of how to do so is to modify the wavelength of light inside the device, giving the photon an effective mass (plasma frequency). By matching the plasma frequency to the axion mass resonant conversion can be achieved regardless of system size~\cite{Lawson:2019brd}. This ``plasma haloscope”  is capable of searching for high mass axions between 5-50 GHz, depending on detector technology. For an appropriate plasma to be found, it must be tunable, low loss and capable of operating in cryogenic environments. One remarkable option is to use a wire metamaterial,~\cite{Pendry:1998}. As the plasma frequency is simply a function of the geometry of the wires, it can be tuned through simple geometric changes. Further, by using high quality copper or superconducting wires the medium exhibits low losses, leading to a competitive device. \\
\noindent{\bf Sensitivity \& Timeline:} The ALPHA 
consortium is currently performing research in service of realizing the first plasma haloscope. ALPHA is working towards a physics capable experiment by 2026, as shown in Fig.~\ref{fig:alphaproj}, which would be sensitive to well motivated axions and dark photons~\cite{Gelmini:2020kcu} in the range of 5-50 GHz. Significant progress has been made recently in both modelling and experimentally validating~\cite{Balafendiev:2022wua,Wooten2022} the properties of wire array metamaterials, which supports projections of $Q > 10,000$. The timeline for the full experiment has also been dramatically accelerated due to the commitment of a 13~T magnet, 0.5~m diameter and 1.7~m long, by the Helmholtz Forschungszentrum Berlin to ORNL. This magnet will be be transferred and commissioned in late 2022, and made available to ALPHA.

\subsection{MADMAX}
\label{sec:madmax}
\begin{figure}
    \centering
    \includegraphics[width=0.5\textwidth]{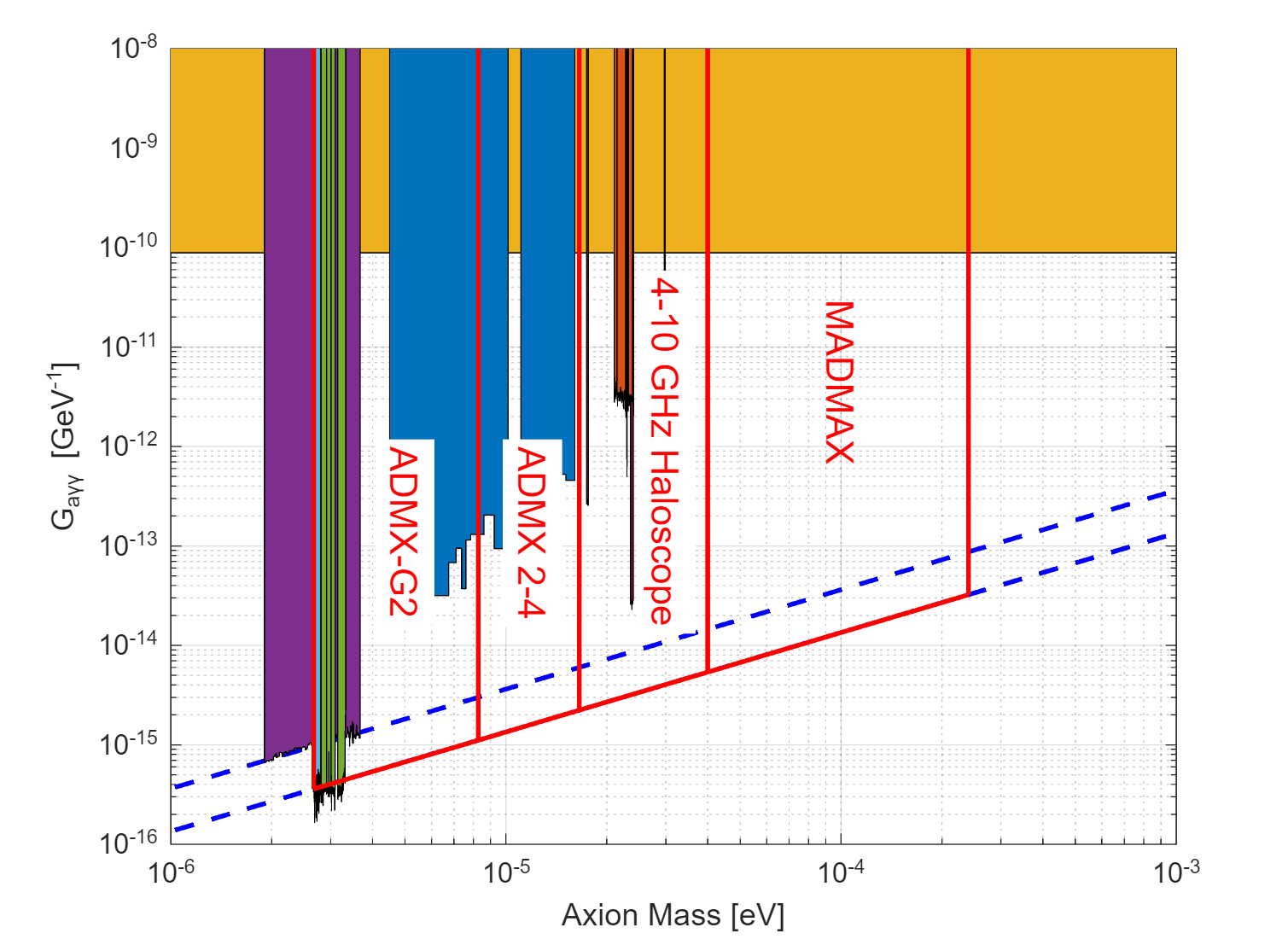}
    \caption{Expected sensitivities in the $m_a$-$g_{a\gamma\gamma}$ plane for ADMX-G2, ADMX2-4 and MADMAX (red contours), compared to existing limits (solid regions) and the KSVZ (DFSZ) QCD axion benchmark models denoted by upper (lower) blue dashed lines. Also shown is the potential reach of a generic future 4-10 GHz ADMX-like experiment.
    }
    \label{fig:4:madmax:sensitivities}
\end{figure}

\noindent \textbf{Physics Goals:} The MAgnetized Disc and Mirror Axion eXperiment (MADMAX)~\cite{TheMADMAXWorkingGroup:2016hpc,Brun:2019lyf} is designed to probe the QCD axion in the mass range motivated by the post-inflationary symmetry breaking scenario, in particular from $40$ to \SI{400}{\micro\electronvolt} ($10$ to \SI{100}{\giga\hertz}), complementary to the reach of ADMX-G2~\cite{Braine:2019fqb} and ADMX-EFR~\cite{admx24loi}, cf.~Figure~\ref{fig:4:madmax:sensitivities}.

\noindent \textbf{Concept:} 
As described above, the cavity haloscope concept suffers from diminishing volume at high frequencies. The dielectric haloscope concept~\cite{Jaeckel:2013eha,TheMADMAXWorkingGroup:2016hpc,millar2017dielectric} generalizes the cavity to a resonator with a much reduced quality factor, however with a larger volume:
Placing a dielectric disc in a dipole-magnetic field, axions can convert to electromagnetic radiation on the disc surface. By placing multiple dielectric discs in front of a metal mirror (dielectric haloscope), the emissions from the different surfaces can constructively interfere and excite resonances between the dielectrics. Both effects give rise to the so called power boost factor $\beta^2$, defined as the ratio of power emitted by only a perfect mirror in vacuum and the dielectric haloscope. 
Using $80$~lanthanum aluminate (LaAlO$_3$) discs, MADMAX is expected to reach a boost factor of $\beta^2 \sim 10^5$ over a bandwidth of \SI{20}{\mega\hertz}, giving an emitted power of
	\begin{eqnarray}
	{P_\gamma} \approx 4\times 10^{-22}\,\si{\watt} ~ \left(\frac{\beta^2}{10^5}\right) 
	\left(\frac{B_{\rm e}^2 \, A}{\SI{100}{\tesla\squared\metre\squared}}\right) \left(\frac{|C_{a\gamma}|}{1}\right)^2 \,  \left(\frac{\rho_a}{\SI{0.45}{\giga\electronvolt\per\centi\meter\cubed}}\right) ,
	\end{eqnarray}
	where $B_e$ is the external magnetic field parallel to the disc surfaces, $A$ is the disc area and $C_{a\gamma} = -\frac{2\pi f_a}{\alpha} g_{a\gamma\gamma}$ with $|C_{a\gamma}| \sim 0.7$ for DFSZ axions.

\begin{figure}
    \centering
    \includegraphics[width=0.45\textwidth]{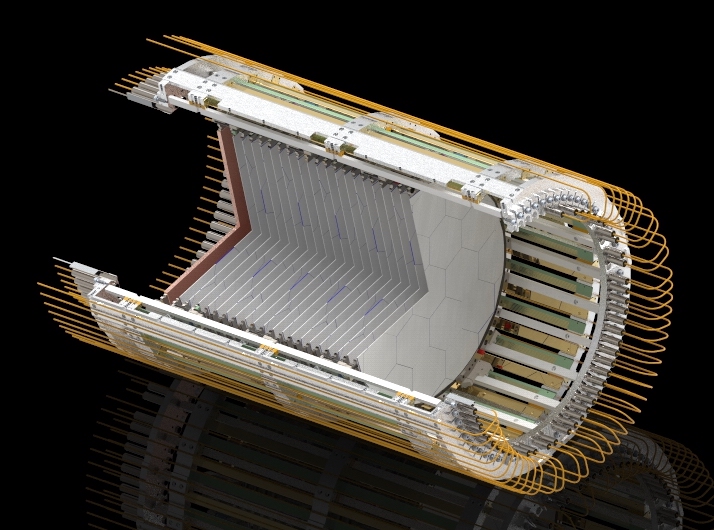}
    \includegraphics[width=0.45\textwidth]{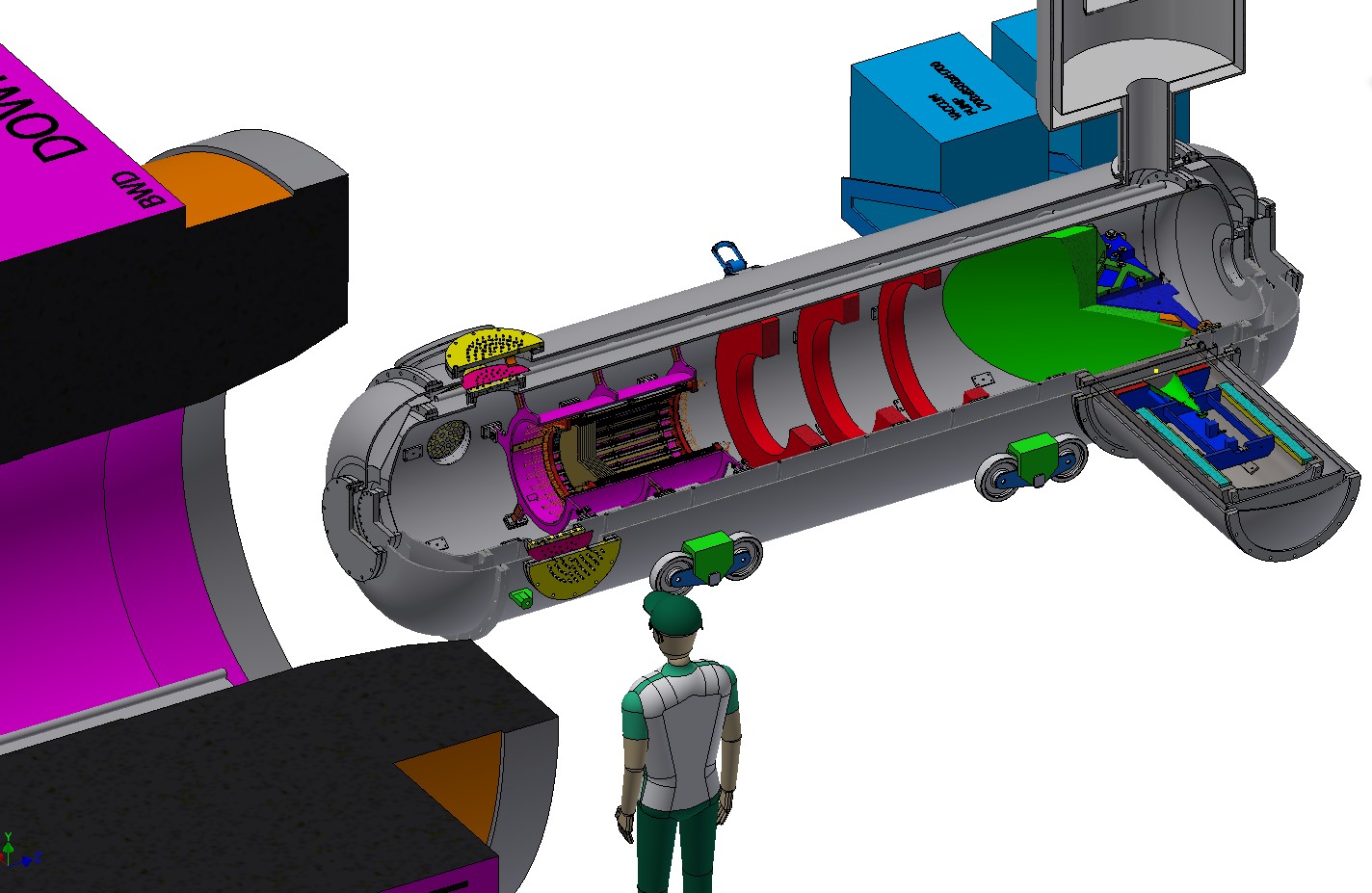}
    \caption{MADMAX Prototype design. The axion-induced radiation from the booster (left) containing $20$ discs of \SI{30}{\centi\metre} diameter is collected by the focusing mirror and antenna. It will be operated in the MORPURGO magnet at CERN, offering a \SI{1.6}{\tesla} dipole field (right).}
    \label{fig:4:madmax:prototype}
\end{figure}
\noindent \textbf{Sensitivity \& Timeline:}
The collaboration is planning the first ALP search using a proof-of-principle `closed booster system' in the 1.6-T MORPURGO magnet at CERN at room temperature. Its sensitivity is expected to exceed existing limit from CAST at around \SI{80}{\micro\electronvolt}. The collaboration is also working on tuning the system at cryogenic temperature and in a strong magnetic field. The lessons learned will be used to construct a prototype detector shown in Figure~\ref{fig:4:madmax:prototype}. It is planned to begin its operation in 2024 at CERN and place ALP DM limits from $90$ to \SI{100}{\micro\electronvolt} with $g_{a\gamma\gamma} > \SI{e-12}{\per\giga\electronvolt}$. The sensitivity of the final MADMAX will depend on outcome of the prototype booster and magnet development. 

\noindent \textbf{Opportunities for US Participation: } 
The ADMX and MADMAX collaborations have identified various opportunities to exploit synergies. These include commissioning and running a ``world's first'' large-bore, high-field dipole magnet, reliable mechanical alignment technologies at high $B$-fields and low temperatures, resonator calibration, electronics and dielectric material characterization.
These topics represent central R\&D challenges not only to MADMAX but to the whole axion community while being largely complementary in probed mass ranges. MADMAX explicitly welcomes contributions from U.S.-based groups.

\subsection{LAMPOST}
\label{sec:lampost}
\noindent \textbf{Physics Goals:} The goal of the LAMPOST(Light $a/A^{\prime}$ Multi-layer Periodic Optical SNSPD Target) experiment ~\cite{Baryakhtar:2018doz,Chiles:2021gxk} is to search for axions in the $0.1-10$~eV mass range, with eventual reach toward the 10 meV regime.

\noindent \textbf{Concept:} 
As described in Section~\ref{sec:madmax}, the dielectric haloscope generalizes the cavity to a resonator with structure on the scalar of the axion Compton wavelength in order to optimize axion to photon conversion over a larger volume ~\cite{Jaeckel:2013eha,TheMADMAXWorkingGroup:2016hpc,millar2017dielectric,Baryakhtar:2018doz}. At optical and IR frequencies, thin film deposition techniques can be used efficiently allowing a simple manufacture of the target. To cover a larger range of masses, multiple targets can be manufactured with datataking in parallel, or a single `chirped stack' can be produced with layers varying in thickness across the stack \cite{Baryakhtar:2018doz}. The photons emitted are focused with a lens onto a parametrically smaller area detector, allowing for excellent signal to noise ratios. A setup of dielectric 100-layer pair targets placed in a $10$T magnetic field will yield sensitivity to new parameter space in the sub-eV range with currently demonstrated dark count rates with superconducting nanowire single photon detectors \cite{Chiles:2021gxk}.  The ultimate goal is the development of low-noise, lower frequency single photon detection. This will extend the LAMPOST technique to the 10 meV regime and  allow access to the KSVZ and DFSZ axion parameter space.

Along with lower energy detectors, under development are background rejection techniques as well as the ability of the SNSPD detector to interface with large magnetic fields; see Section~\ref{sec:quantum_rnd}. 

\subsection{BREAD}
\label{sec:bread}
\noindent \textbf{Physics Goals:} 
The Broadband Reflector Experiment for Axion Detection (BREAD)~\cite{BREAD:2021tpx} plans to perform broadband searches for wavelike DM across multiple decades in mass from around 0.2\,THz up to astrophysical limits around optical frequencies.
BREAD was recently founded as a multi-institute collaboration with substantial US leadership.
Universities and national laboratories across the US are spearheading ongoing R\&D efforts towards proof-of-concept pilot experiments.
The experiment is planned to be located at Fermilab.
Its scientific hallmark is broadband sensitivity for axion-like DM across terahertz frequencies using a unique geometry without resonant tuning to the unknown DM mass. 

\begin{figure}
    \centering
    \includegraphics[width=0.29\textwidth]{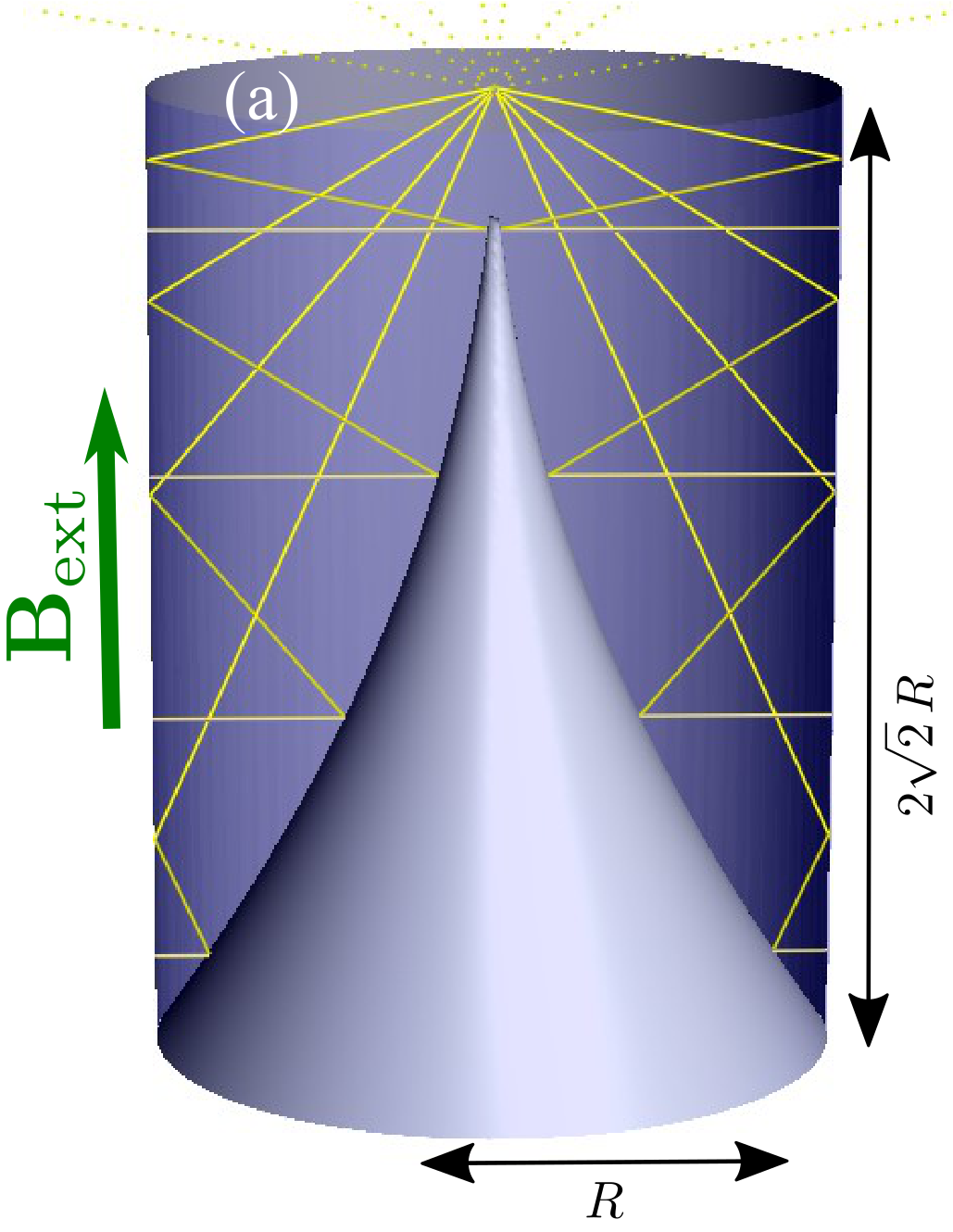}%
    \hfill%
    \includegraphics[width=0.66\textwidth]{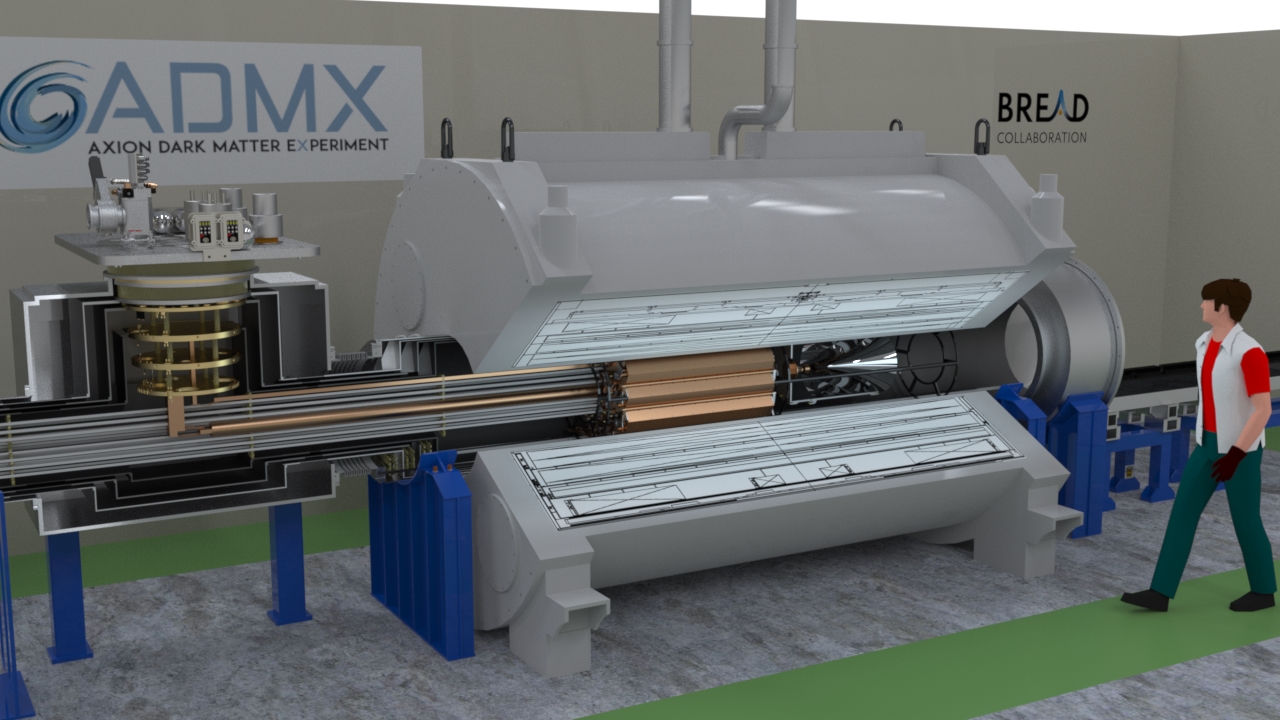}
    \caption{\label{bread_and_admx}
    Left: Ray-tracing simulation of photons (yellow lines) emitted by the BREAD cylindrical barrel reflector and focused by the parabolic surface to a focus. Reproduced from Ref.~\cite{BREAD:2021tpx}.
    Right: ADMX-EFR and the large-scale BREAD experiment will be hosted side-by-side in a former MRI magnet.
     }
    \label{fig:bread_fig1}
\end{figure}

\begin{figure}
    \centering
    \includegraphics[width=0.7\textwidth]{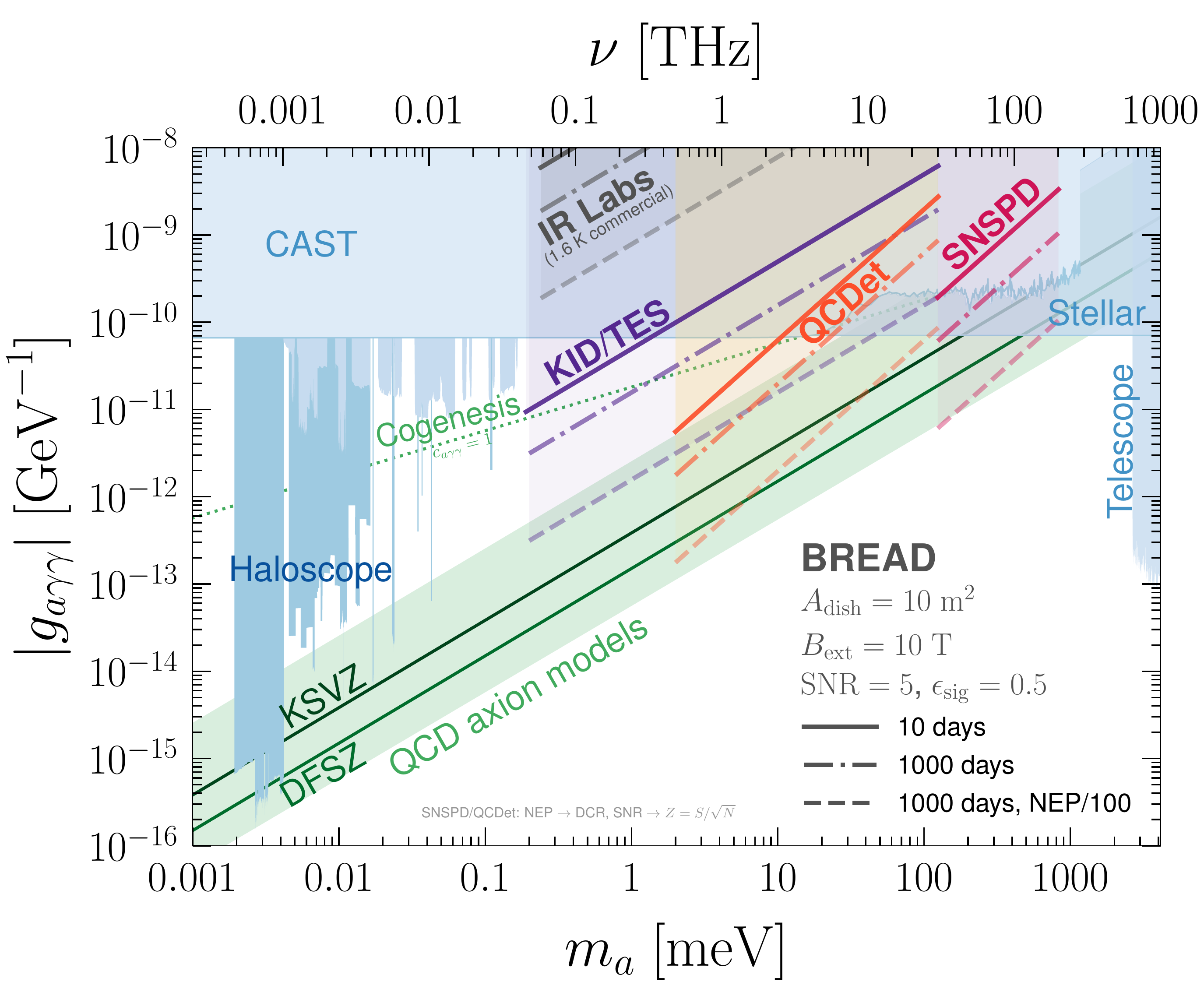}
    \caption{Projected sensitivity for BREAD by photosensor technology (thick lines) in axion coupling vs.\ mass plane. Different assumptions on exposure time and sensor noise equivalent power (NEP) are displayed in the legend. 
    Reproduced from Ref.~\cite{BREAD:2021tpx}.  
    }
    \label{fig:bread_sensitivity}
\end{figure}

\noindent \textbf{Concept:} 
BREAD realizes the dish antenna concept~\cite{Horns:2012jf} by using a cylindrical metal barrel to convert axion DM into photons, which a parabolic reflector focuses onto a low-noise photosensor. 
This geometry allows placement inside standard high-field solenoids and cryostats, which allows lower equipment costs. 
For axion-to-photon coupling $g_{a\gamma\gamma}$ and mass $m_a$, the emitted power is given by $P_a = \frac{1}{2} \rho_\text{DM} (B_\text{ext}^{\parallel} g_{a\gamma\gamma} / m_a)^2 A_\text{dish}$~\cite{Horns:2012jf}, where $\mathbf{B}_\text{ext}^\parallel$  is the external magnetic field with nonzero component parallel to the dish with area $A_\text{dish}$.

Figure~\ref{fig:bread_fig1}\,(left) shows the ray-tracing simulation in the BREAD geometry  inside a solenoid with magnetic field $B_\text{ext}$.
Photon paths are shown as yellow lines, which are emitted perpendicularly to the barrel surface with length $L = 2\sqrt{2}R$ and the parabolic surface with radius $R$ focuses the rays to a point.
Various state-of-the-art photosensor technologies for projected sensitivity studies are considered in Ref.~\cite{BREAD:2021tpx}. 
This includes commercial cryogenic thermistors (\textsc{IR Labs}) alongside established solutions widely deployed for astronomy applications, namely kinetic inductance detectors (KID) and transition edge sensors (TES). 
Meanwhile, quantum capacitance detectors (QCDet) and superconducting nanowire single photon counters (SNSPD) exemplify emerging technologies with low-noise photon counting capabilities.

\noindent \textbf{Sensitivity \& Timeline:}
Progress towards small-scale prototype experiments targeting dark photon benchmarks (detailed in Ref.~\cite{BREAD:2021tpx}) are ongoing at Fermilab.
Figure~\ref{fig:bread_sensitivity} shows the projected sensitivity of large-scale BREAD realizations with various photosensor technologies. 
Specific benchmarks are assumed for the magnetic field $B=10\,\text{T}$, dish area $A_\text{dish} = 10\,\text{m}^2$, with experimental runtimes of 10 and 1000\,days. 
This also motivates photosensor R\&D will continue to reduce noise for probing QCD axion benchmarks. 
These sensitivities could be achieved by hosting BREAD as a side experiment to ADMX-EFR within the same magnet, as shown in Fig.~\ref{fig:bread_fig1}\,(right).

\subsection{Global Context} 
\label{sec:above_ueV_global_context}
In synergy with programs in the US, wavelike dark matter programs are gaining momentum around the globe. Here we elaborate on some currently funded projects. Due to space constraints, we must omit several other projects, such as the cavity haloscopes developed at the Center for Axion and Precision Physics Research in Korea~\cite{CAPP2021}. 

\subsubsection{ORGAN}

ORGAN is a high-mass axion haloscope, targeting the 15-50 GHz region (60-200 $\mu$eV). The experiment operates in a BluFors XLD system, with the capacity to reach a base level ~7 mK, inside a 12.5 T environment. The pathfinding run placed the first limits on axions from a haloscope above 10 GHz, beating CAST at 26.6 GHz \cite{MCALLISTER2017}. After this, ORGAN entered a planning and commissioning stage for long term operations (Phase 1 and 2), which commenced in 2021 \cite{OrganPhase1a}. Phases 1a and 1b are short, targeted scans of 15-16 GHz and 26-27 GHz ranges respectively, using equipment on hand (HEMT amplifiers and established resonator designs). Phase 2 encompasses the entire 15-50 GHz region, broken down into 5 GHz sub-phases, which will re-scan the regions in Phase 1 with enhanced sensitivity, using single photon counters (SPCs), or quantum limited amplification. With efficient GHz SPCs, ORGAN aims to reach the DFSZ axion limits. Phase 2 is expected to commence in 2023. Phase 1a ran in 2021, successfully excluding ALP Cogenesis models between 15.3–16.2 GHz (63-67 $\mu$eV) \cite{OrganPhase1a}. Phase 1b is in commissioning, expected to commence in 2022, using HEMT amplification, and newly designed sapphire loaded cavity resonators with dielectric-boosted sensitivity \cite{OrganDBS2020,OrganSM2018}.

\subsubsection{RADES}
The RADES project is developing axion dark matter haloscopes
to be used in dipole magnets.
Up to the present day, RADES has collected data at CAST and SM18 magnets (CERN), and in the long run it plans to use the BabyIAXO and Canfranc magnets for new data campaigns. Most cavities of RADES are based on an arrangement commonly used in radio-frequency filters: an array of small microwave sub-cavities, whose size controls the operation frequency, connected by rectangular irises, providing high volumes. The detection principle, the theoretical framework and the haloscope details, can be found in a recent overview \cite{universe8010005}. Besides the goal of increasing the sensitivity of detection through achieving large geometric factors and high quality factors, a key feature of the system must be the tuning of the resonant frequency. RADES designed and built a mechanical tuning system, but they are also working on another tuning method: electrical tuning by ferroelectric materials, which can provide an avenue that is less prone to mechanical failures and thus complements and expands existing techniques. Also, ferroelectrics allow the ability to independently adjust the many different cavity frequencies to maintain the correct mode structure and field pattern.

\subsubsection{ALPS II}

The Any Light Particle Search II (ALPS II) probes for axion-like particles
with a coupling factor of $g_{a\gamma\gamma}\geq2\times10^{-11}\,{\rm GeV}^{-1}$
at all masses below 0.1~meV \cite{DIAZORTIZ2022100968,HALLAL2022100914}. Axion-like particles
with similar coupling factors would explain astronomical observations
such as TeV transparency and larger than expected stellar cooling
rates \cite{PhysRevD.87.035027,Giannotti:299846}. ALPS II is also sensitive to potential
scalar cousins of the pseudo-scalar axion and to hidden sector photons.

\noindent{\textbf{Method:}}
Light-shining-through-a-wall (LSW) experiments such as ALPS II are
broadband experiments which send a strong laser beam through a magnetic
dipole field. Some of the photons will be turned into axions which
pass through a photon-blocking wall behind which they enter a second
magnetic field. Inside this field, a small fraction of these axions
reconvert back into photons. Compared to the other methods described in 
this whitepaper, this method is less sensitive to weak axion couplings,
but does not require the axion to comprise dark matter. 

ALPS II uses two 120~m long strings of straightened HERA dipole magnets
(5.3~T) in series, a 70~W laser system similar to the advanced LIGO
laser systems, and two high finesse 122~m long optical cavities to
enhance the source and the signal photons on each side of the wall.
This setup increases the regenerated photon rate by 12(!) orders of
magnitude compared to earlier LSW experiments \cite{EHRET2010149}.

\noindent{\textbf{Current schedule:}}
The ALPS collaboration is an international collaboration with groups
from Germany, the UK, and the US which is currently commissioning
the ALPS experiment at DESY in Hamburg, Germany. The first science
data sets with sensitivities reaching $g_{a\gamma\gamma}<10^{-10}\,{\rm GeV}^{-1}$
for pseudo-scalar (axion-like) and for scalar particles as well as
for hidden sector photons are expected for the fall 2022. It will
take a few more years and a few improvements such as better cavity
mirrors with improved surface roughness before the final sensitivity
will be reached.

\noindent{\textbf{Future plans:}}
Unfortunately, the sensitivity does not scale well with experimental
parameters and another leap in sensitivity using the LSW method
is unlikely. However, the magnet strings and optical infrastructure
enables additional ground-breaking experiments such as measurements
of vacuum birefringence predicted by Heisenberg and Euler in 1938(!)
or high frequency gravitational waves \cite{EJLLI20201,HF_GW2019}.

\section{Below Micro-eV}
\label{sec:below_microeV}

It is also interesting to consider axions with mass well below $\mu \mathrm{eV}$, corresponding to axion frequencies below $\mathrm{GHz}$. As noted in section~\ref{sec:cosmology}, ALPs can be produced with the appropriate dark matter abundance in this range through the ordinary misalignment mechanism. The QCD axion could also comprise dark matter, given a tuning of the initial misalignment angle, anthropic arguments or modified cosmologies~\cite[e.g.][]{Tegmark:2005dy,Mack:2009hv,Visinelli:2009bg,Visinelli:2009kt,Arias:2021rer,Dine:1982ah,Steinhardt:1983ia,Dimopoulos:1988pw,Davoudiasl:2015vba,Hoof:2017ibo,Graham:2018jyp,Takahashi:2018tdu,Kitajima:2019ibn}. 

In this mass range, the cavity haloscopes introduced in section~\ref{sec:above_microeV} cannot be used, since it is impractical to build a cavity large enough to have a sufficiently low mode frequency. Instead, most experiments use the lumped element approach, where the electromagnetic fields produced by the axion drive a current through a pickup coil. In sections~\ref{sec:abracadabra} and~\ref{sec:shaft}, we discuss ABRACADABRA and SHAFT, two pioneering experiments which used this technique to achieve broadband sensitivity. In sections~\ref{sec:DMRadio-m3} and~\ref{sec:DMRadio-GUT}, we discuss the current and future plans of the DMRadio experiment, which aims to probe the QCD axion by amplifying the axion-induced current in a resonant LC circuit. 

Finally, in section~\ref{sec:heterodyne}, we discuss the recently proposed heterodyne approach with SRF cavities, which decouples the axion frequency and the mode frequency by using a superconducting cavity containing an oscillating background magnetic field. The presence of this field introduces new noise sources, but also significantly increases the signal power, especially at low axion masses. This approach thus achieves comparable sensitivity to axions without requiring large, high-field magnets or quantum sensing techniques. 

\subsection{DMRadio-m$^3$}
\label{sec:DMRadio-m3}

\begin{figure}[ht]
  \centering
  \includegraphics[width=.89\textwidth]{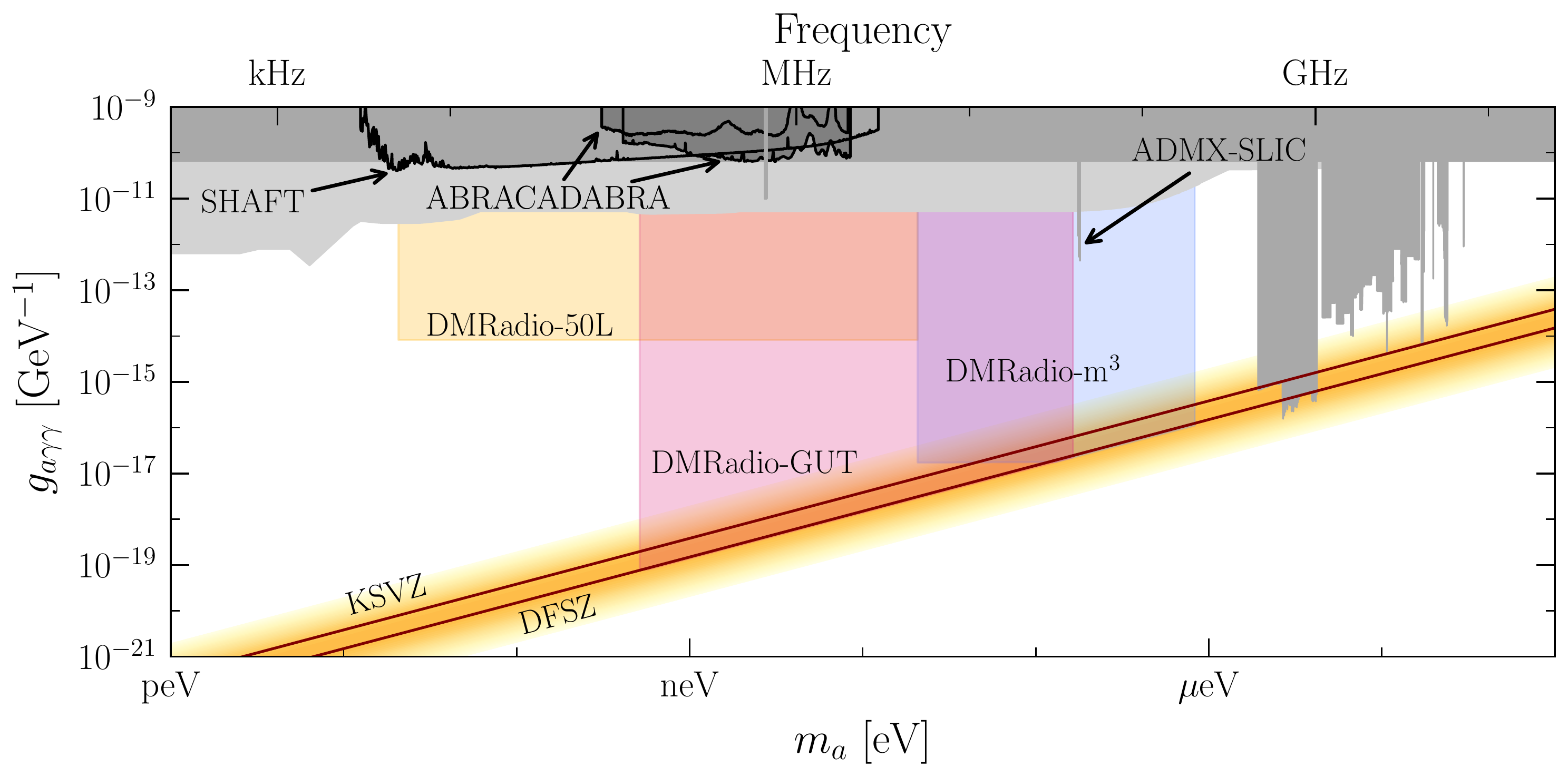}
  \caption{The projected sensitivities for \DMRL, \DMRm, and \DMRG. The      \DMR program aims to probe the QCD axion below $1\,\mu\mathrm{eV}$.     This assumes scan times of 3 years for \DMRL, 5 years for \DMRm,       and 5 years for \DMRG.
  }
  \label{fig:dmrsensitivity}
\end{figure}

\noindent {\bf Physics goals and status:}
The \DMRm experiment aims to search for the QCD axion over the mass range $20\,\mathrm{neV}\lesssim m_a\lesssim 800\,\mathrm{neV}$, achieving DFSZ sensitivity over the range $120\,\mathrm{neV}\lesssim m_a \lesssim 800\,\mathrm{neV}$ with a 5 year total scan time. This mass range begins to probe the top of the  
GUT-scale QCD axion masses, in addition to testing a wide swath of interesting ALP parameter space (see Fig.~\ref{fig:dmrsensitivity}.) \DMRm will extend the reach of the \DMRL experiment that will probe the ALP parameter space between 5\,kHz and 5\,MHz ($20\,\mathrm{peV}\lesssim m_a \lesssim 20\,\mathrm{neV}$). The \DMR program also opens the door for an ambitious and definitive search for the GUT-scale QCD axion called \DMRG (see Sec.~\ref{sec:DMRadio-GUT}).

\noindent {\bf Experimental approach:} 
\DMRm uses a lumped element approach consisting of a coaxial inductive pickup inside of a solenoidal magnet, with a tunable capacitor providing an LC resonant enhancement. This technique builds on the work of the \linebreak[4] \abra broadband axion search \cite{Ouellet:2018beu,Salemi:2021gck}, \DMRP resonant dark photon experiment \cite{DMRadio} and techniques originally proposed in \cite{Kahn2016,Chaudhuri15,Sikivie2014b}.

\DMRm will consist of a series of interchangeable coaxial inductive pickups with volume up to $\sim 1.2\,\mathrm{m}^3$ inside of a 4\,T solenoidal magnet.  The experiment will scan the frequency range from 5\,MHz to 200\,MHz. Since the pickup sits inside a large magnetic field, it will be constructed from normal conducting OFHC copper. The resistive losses in the copper will dominate the loss in the circuit. However, this is mitigated by a relatively low anomalous surface resistance at frequencies and temperatures relevant for \DMRm as well as a relatively large volume to surface ratio. Approaches to decrease this loss using superconducting coatings are being pursued by other experiments (see \cite{Romanenko:2014yaa,Ahn:2020qyq} and Sec.~\ref{sec:cavity_rnd}) and may be considered in the future. Multiple coaxial pickups of decreasing size are required to keep the detector in the magneto-quasistatic limit as the detector probes to progressively higher frequencies and the experiment will be designed to allow efficient exchange of the pickup components.

The resonance frequency will be scanned using a tunable capacitor made from superconducting material and low-loss dielectrics like sapphire. The target Q-factor for \DMRm is $Q\sim10^6$ at 30\,MHz and is limited by  loss in the copper pickup. The signal will be read out by low noise SQUID amplifiers with a target noise level of 20$\times$ the standard quantum limit or better. Eventually, the sensitivity could be further improved by incorporating next-generation beyond-SQL sensors, discussed in Sec.~\ref{sec:quantum_rnd}.

The detector will be cooled in a large dilution refrigerator to an operating temperature of 20\,mK to reduce the thermal noise. Cryostats of this scale have been demonstrated before \cite{ALDUINO20199}.

\noindent {\bf Projected sensitivity and timeline:}
The experimental scan rate $\partial \log{\nu_r}/\partial t$ at resonant frequency $\nu_r$ is proportional to the quantity $c_{\rm PU}^4B_0^4V^{10/3}\bar{\mathcal{G}}(\nu_r,k_BT,n_A)$, where $B_0$ and $V$ are the characteristic magnetic field and pickup volume and $c_{\rm PU}$ is a geometric coupling which can be related to a flux coupling in the magneto-quasistatic limit \cite{DMRm3Paper}.
$\bar{\mathcal{G}}$ is a coefficient which relates the thermal noise temperature, $k_BT$, and the number of noise photons added by the amplifier, $n_A$, to the scan rate. In the low frequency range, where the noise budget is dominated by thermal noise photons $n_{\rm th}=(\exp(h\nu/k_BT)-1)^{-1}$, and assuming the readout circuit to be optimally coupled for fastest scan rate, this coefficient is given approximately by $\bar{\mathcal{G}}\approx(12\sqrt(3) n_{\rm th} n_A)^{-1}$. In the frequency range relevant for \DMRm, $\bar{\mathcal{G}}$ takes a more complex form given in \cite{DMRm3Paper} with extended discussion in \cite{Chaudhuri2018}.

The projected \DMRm sensitivity is presented in Fig.~\ref{fig:dmrsensitivity}. \DMRm aims to achieve DFSZ sensitivity over the range from 30\,MHz to 200\,MHz ($120\,\mathrm{neV}\lesssim m_a \lesssim 800\,\mathrm{neV}$) and sensitivity to the QCD axion all the way down to 5\,MHz ($20\,\mathrm{neV}$). \DMRm is designed for a 5 year scan time and targets data taking in 2026.

\subsection{ABRACADABRA Demonstrator}
\label{sec:abracadabra}

\noindent {\bf Physics goals and status:}
\abra aims to search for axion dark matter through the electromagnetic coupling in the sub-$\mu$eV mass range using the lumped element detection technique. 

\noindent {\bf Experimental approach:} \abra uses a toroidal geometry based on the proposal \cite{Kahn2016}. The detector consists of a 10\,cm toroidal magnetic with a maximum field of 1\,T. Axion dark matter converts the DC, toroidal, magnetic field into an AC magnetic field through the center of the toroid -- a region that would otherwise be free of any magnetic field. In the Run 1 configuration \cite{PhysRevD.99.052012}, the central region was read out in a broadband configuration by a superconducting pickup loop connected to a two-stage SQUID readout. The magnet and loop were shielded inside of a superconducting shield to protect them from environmental noise. The detector and SQUIDs were cooled to an operating temperature of 700\,mK in a dilution refrigerator. Vibrational noise was mitigated using a vibration isolation system, and the detector was shielded with a simple MuMetal shielding. A calibration procedure injected a current into a dedicated loop inside the magnet to generate an axion-like magnetic field allowing a direct reconstruction of the detector gain. This procedure indicated parasitic inductance in the readout scheme which decreased the sensitivity by a factor of $\sim$6.5.

The SQUID voltage was sampled continuously at a rate of 10\,MS/s for weeks-long runs. The timestream was FFT'd in real time, downsampled, and FFT'd again, building multiple spectra capable of searching the full axion mass range from 50\,kHz to 5\,MHz with sufficient spectral resolution. Data were collected in magnet-on and magnet-off configurations, the former being used for the axion dark matter search, and the latter to veto spurious signals from other sources. Run 1 consisted of \mbox{$2.6\times10^6$\,s} of magnet-on data, plus another 2 weeks worth of magnet-off data.

\begin{figure}[!ht]
  \centering
  \begin{minipage}[t]{.45\textwidth}
    \centering
    \includegraphics[width=.5\textwidth]{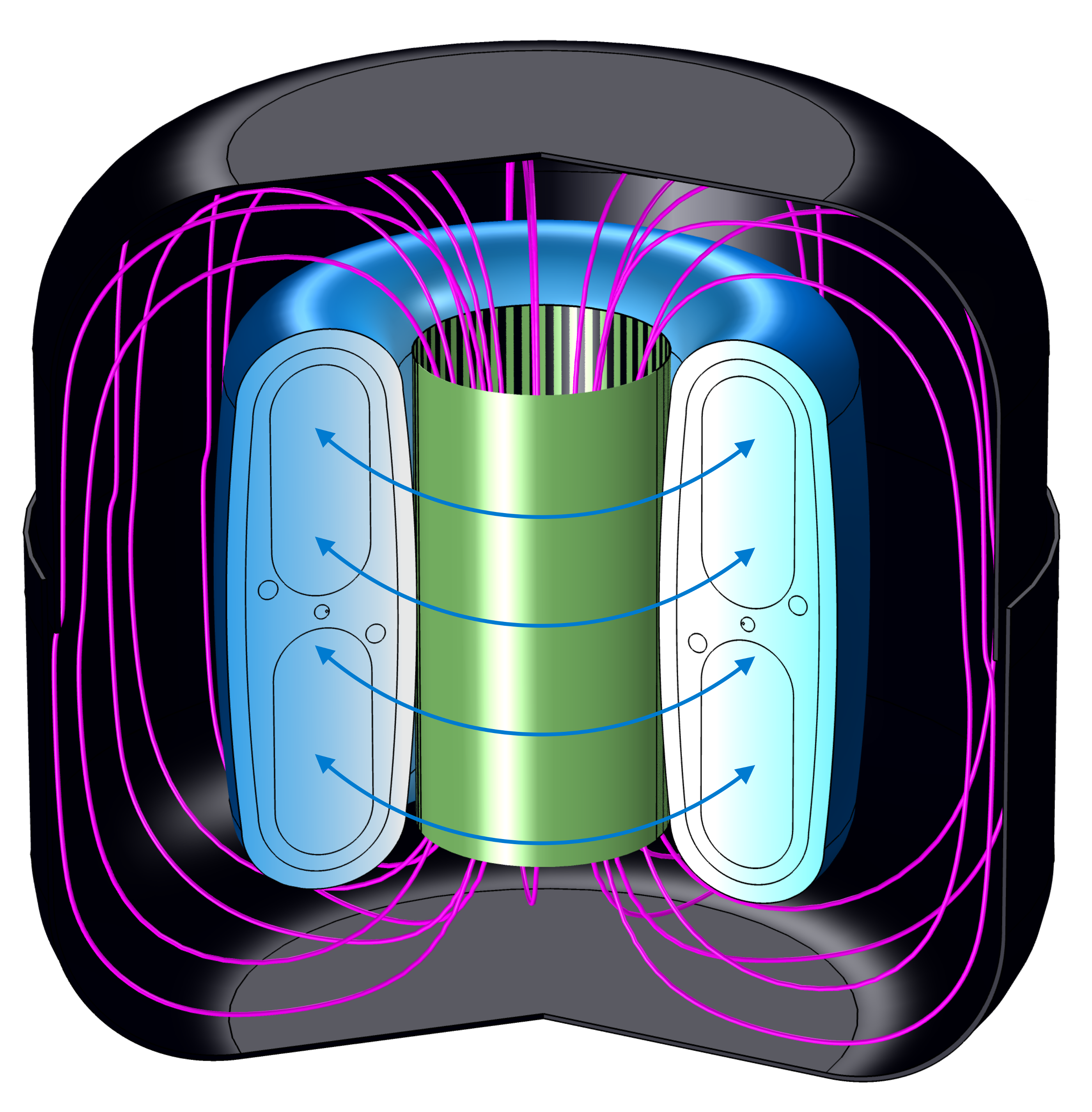}\\
    \includegraphics[width=\textwidth]{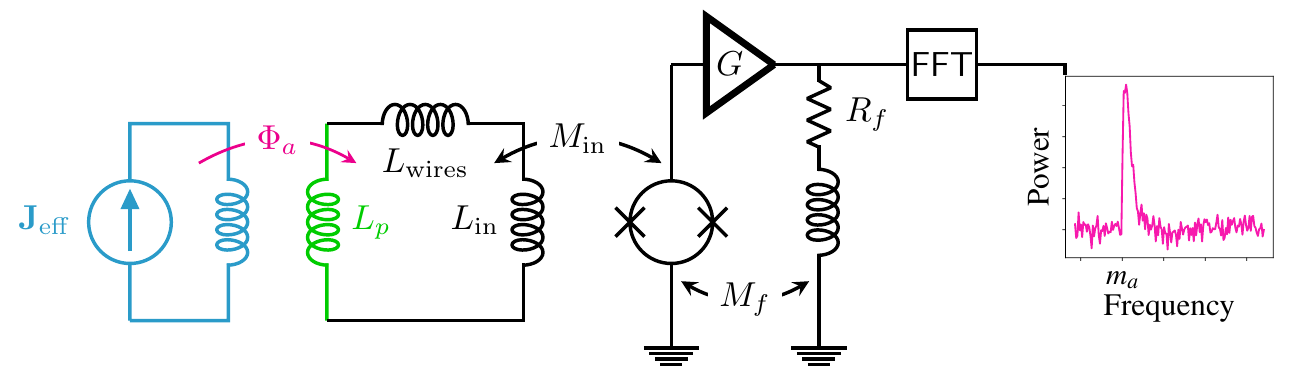}
  \end{minipage}
  \includegraphics[width=.5\textwidth]{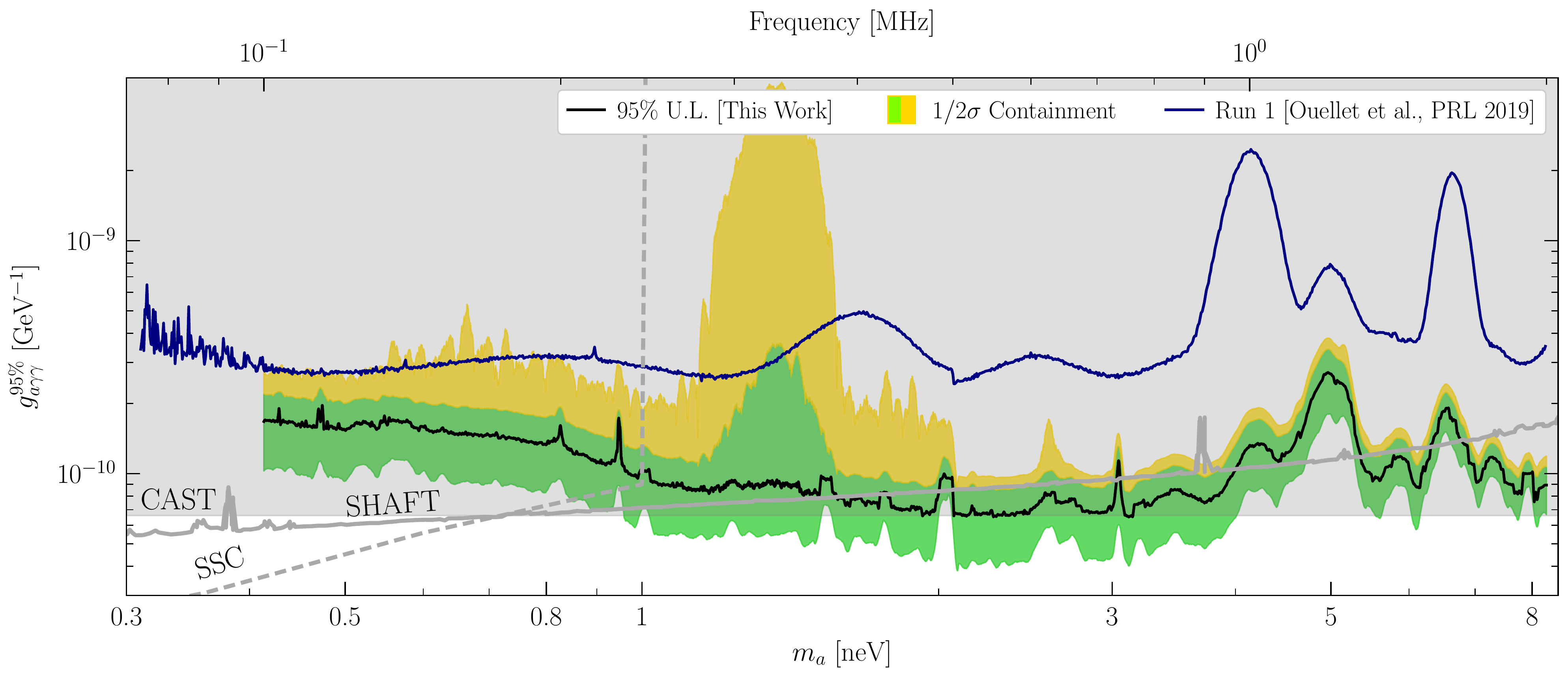}
  \caption{Left Top: Diagram of the \abra detector used for the Run 3
    results. A $\sim$ 10\,cm-scale 1\,T toroidal magnet (blue) drives the axion induced magnetic field (purple lines) which is screened by the cylindrical pickup (green). The detector is enclosed within a superconducting shield. Left Bottom: Schematic of the broadband readout. The axion effective current (blue) drives a flux through the inductive pickup (green), which is inductively coupled to a SQUID readout in flux lock mode and monitored.  Right: 95\% CL exclusion limits from Runs 1 and 3 of the \abra prototype. Figures from~\cite{Salemi2021a}.}
  \label{fig:abra_limits}
\end{figure}

After the first run, the detector underwent an upgrade to increase its sensitivity to axion dark matter. The pickup loop was replaced by a pickup cylinder, which improved the coupling to the axion induced magnetic-field -- i.e. driving more current for the same size axion induced mangetic field (see Fig.~\ref{fig:abra_limits}). The readout wiring was improved to reduce the parasitic inductance. The SQUIDs were moved to the coldest stage of the cryostat and cooled to an operating temperature of $\sim$150\,mK, while the magnet stayed at $\sim$700\,mK. Runs 2 and 3 were collected in this configuration, but Run 2 data were eventually discarded due to excess noise. Run 3 consisted of $\sim2$ weeks of magnet-on data, and $\sim1$ week's worth of magnet-off data.

\noindent {\bf Results and Future Outlook:}
The \abra experiment collected Run 1 data in Summer 2018 and Run 3 data in Summer 2020. The \abra collaboration developed an analysis procedure to search the millions of mass points collected simultaneously in a broadband search -- which does not have rescan capabilities. This procedure based on \cite{Safdi} correctly accounts for the ``look-elsewhere effect.''  

\abra placed the first laboratory based limits on axion dark matter below $1\,\mu\mathrm{eV}$ and was the first to demonstrate the lumped element approach to axion dark matter detection. The Run 1 results probed the mass range \mbox{$0.31\,\mathrm{neV} < m_a < 8.3\,\mathrm{neV}$} (75\,kHz -- 2\,MHz), setting 95\% C.L. limits down to \mbox{$g_{a\gamma\gamma} < 1.4\times10^{-10}\,\mathrm{GeV}^{-1}$} \cite{ABRA,PhysRevD.99.052012}. The Run 3 results improved on this, setting 95\% C.L. limits down to \linebreak[4] \mbox{$g_{a\gamma\gamma}<3.2\times10^{-11}\,\mathrm{GeV}^{-1}$}~\cite{Salemi2021a}.

At low axion masses, the sensitivity was limited by vibrational noise coming from stray DC fields from the magnetic being transduced by vibrational noise from the pulse tube cooling system. In the axion search region, the sensitivity was limited by the SQUID flux noise floor. Further upgrades to \abra could continue to improve sensitivity by factors of a few and perhaps extended in mass range, however, the detector is transitioning to a test bench for future R\&D. 

\subsection{Search for Halo Axions with Ferromagnetic Toroids (SHAFT)}
\label{sec:shaft}
\begin{figure}[ht]
    \centering
    \includegraphics[width=0.6\textwidth]{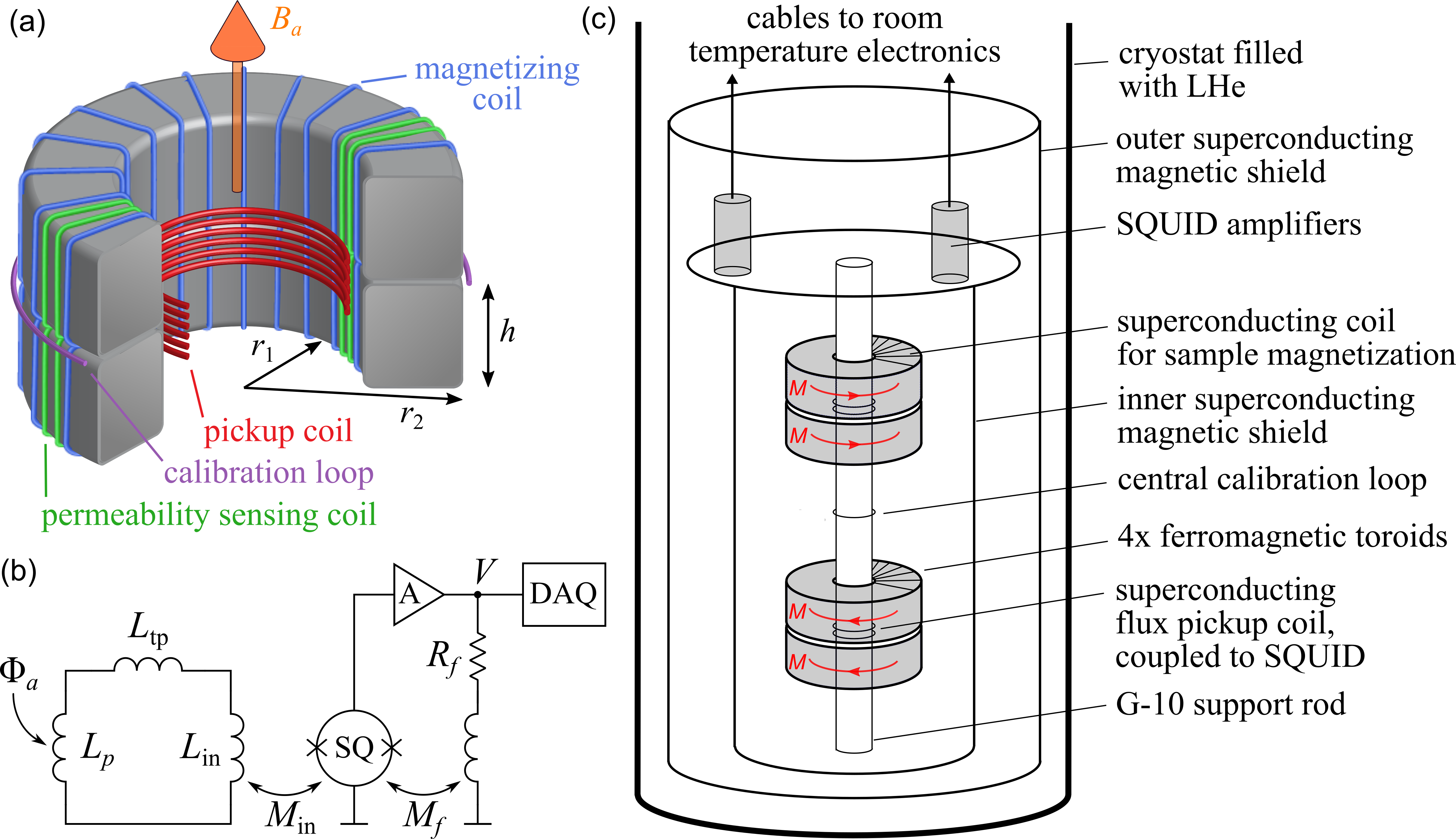}%
    \includegraphics[width=0.4\textwidth]{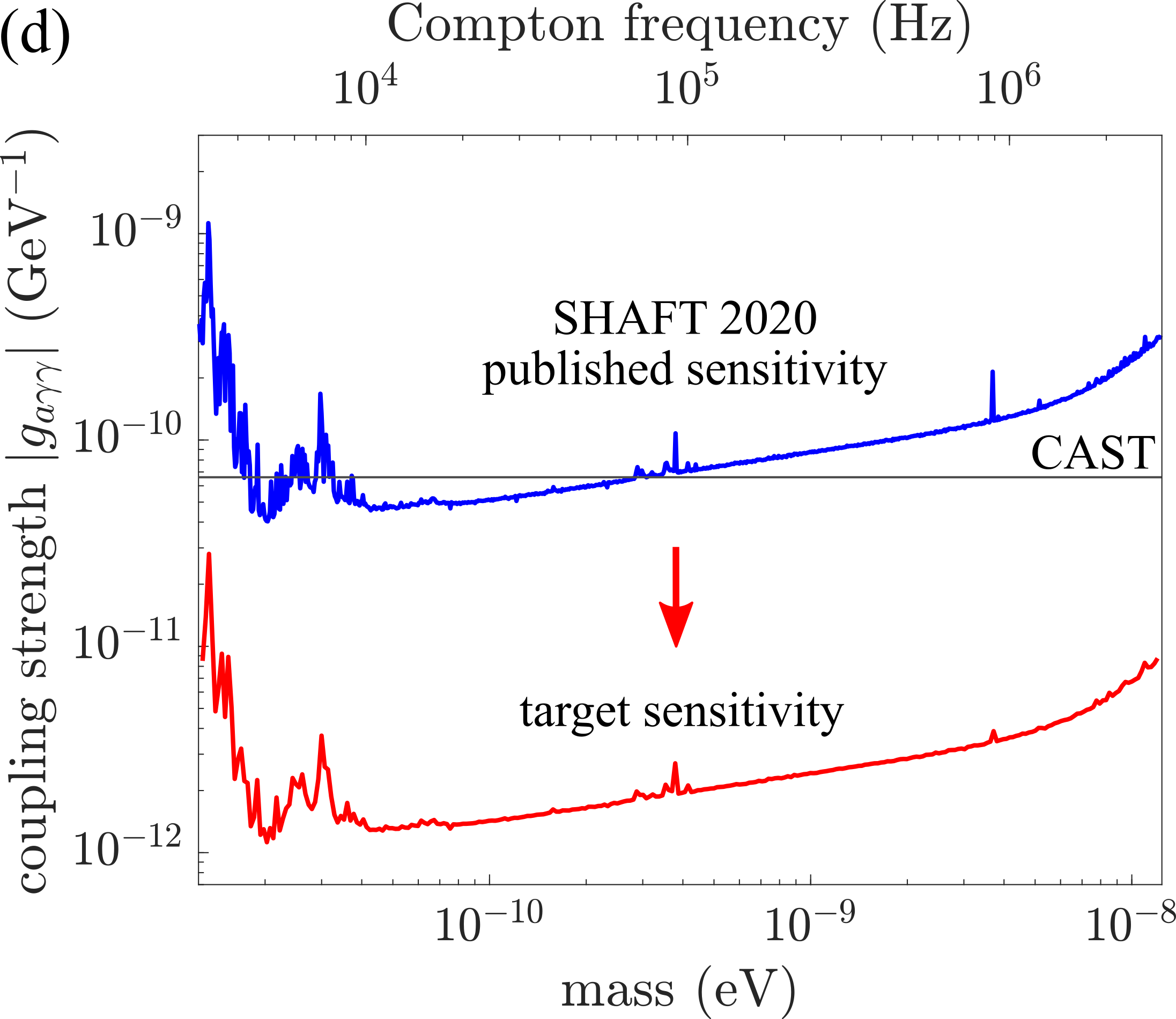}%
    \caption{
    The SHAFT experiment and projected sensitivity~\cite{Gramolin2021}.
    (a) Schematic of a single detection channel. Two permeable toroids are independently magnetized by injecting current into a magnetizing coil wrapped around each toroid. 
    (b) The circuit model that shows the axion-induced flux~$\Phi_a$ coupling into the pickup coil (inductance $L_p$). The pickup coil is coupled to the SQUID magnetic flux sensor (SQ) via twisted pair leads (inductance $L_{\rm tp}$) and input coil (inductance $L_{\rm in}$). The SQUID is operated in the flux-locked-loop mode, with feedback resistance $R_f$. The feedback voltage $V$ was digitized and recorded by the data acquisition system (DAQ).
	(c) The SHAFT experimental schematic. The apparatus contains four permeable toroids, shown such that the top and bottom channels are magnetized in opposite directions, to make use of phase-sensitive systematic rejection. The experiment operates at 4.2~K in a liquid helium bath cryostat. 
    (d) Published limits set by the first-generation SHAFT search are shown in blue~\cite{Gramolin2021}. Target sensitivity of next-generation SHAFT search is shown in red.
    }
    \label{fig:shaft_fig}
\end{figure}

\noindent {\bf Physics goals and status:}
The SHAFT experiment is sensitive to the electromagnetic coupling $g_{a\gamma\gamma}$ of axion-like particles. The goal is to search for axion-like dark matter in the broad mass range between $\approx10^{-12}$~eV and $\approx 10^{-8}$~eV, with coupling strength sensitivity down to $10^{-12}$~GeV$^{-1}$. SHAFT aims to explore the region of parameter space where ALP-photon mixing could explain the anomalous transparency of the universe to TeV gamma-rays~\cite{Matsuura2017,Kohri2017,PDG2019}. The first-generation search has reported limits on the electromagnetic interaction of axion-like dark matter in the mass range that spans three decades from 12~peV to 12~neV~\cite{Gramolin2021}. Over part of this mass range it improved the CAST limits, at 20~peV reaching $4.0 \times 10^{-11}~\text{GeV}^{-1}$ (95\% confidence level), Fig.~\ref{fig:shaft_fig}. 

\noindent {\bf Experimental approach:}
SHAFT takes the lumped-circuit approach, based on a modification of Maxwell's equations in the presence of axion-like dark matter, which mixes with a static magnetic field $B_0$ to produce an oscillating magnetic field~\cite{Sikivie2014b, Chaudhuri15, Kahn2016, Chaudhuri2018, Ouellet2019,Salemi2021a}.
An inductively-coupled SQUID is the detector of this oscillating magnetic field, with sensitivity of 150~$\rm{aT}/\sqrt{\rm Hz}$, which is at the level of the most sensitive magnetic field measurements demonstrated with any broadband sensor~\cite{Gramolin2021}.
A toroidally-shaped permeable iron-nickel alloy material enhances the magnitude of the static magnetic field~$B_0$ by a factor of 24, and thus improves sensitivity to axion-like dark matter. Systematic rejection is especially important to ensure discovery potential of broadband ultralight dark matter searches. The unique systematic rejection mechanism in SHAFT is the design that incorporates two independent detection channels, allowing discrimination between an axion-like dark mater signal and electromagnetic interference systematics. For collecting data sensitive to axion-like dark matter, the top channel toroids are magnetized counter-clockwise, and the bottom channel toroids clockwise, viewed from the top, Fig.~\ref{fig:shaft_fig}(c). An axion-like dark matter signal would then appear in the two detector channels with opposite sign ($180^{\circ}$ out of phase). Electromagnetic interference, uncorrelated with toroid magnetization, would not have such a phase relationship and can be distinguished from an axion-like dark matter signal by forming symmetric and antisymmetric combinations of the two detector channels.

\noindent {\bf Projected sensitivity and timeline:}
The first-generation SHAFT sensitivity was limited by SQUID thermal noise at 4.2~K, apparatus vibrations, and the data taking timescale restricted to 41~h by liquid helium boil-off.
As of 2022, the second-generation search is ongoing, aiming to improve sensitivity by a factor of 12 by mounting the setup in a cryogen-free dilution refrigerator, which will cool the SQUID sensors to millikelvin temperature and allow longer data-taking runs. We project that by the end of 2023 SHAFT will achieve sensitivity to $g_{a\gamma\gamma}$ at the level of $3\times10^{-12}~\text{GeV}^{-1}$ in the mass range $\approx10^{-12}$~eV to $\approx 10^{-8}$~eV. Scaling up the toroid volume by a factor of 3 will improve sensitivity to $10^{-12}~\text{GeV}^{-1}$ in 2024, Fig.~\ref{fig:shaft_fig}(d).

\subsection{DMRadio-GUT}
\label{sec:DMRadio-GUT}

\noindent {\bf Physics goals:}
DMRadio-GUT is a search for QCD axions at the GUT scale \cite{Brouwer:2022bwo}.  As shown in Fig.~\ref{fig:dmrsensitivity}, the experiment would reach DFSZ sensitivities to masses from $0.4\,\textrm{neV}$ to $120\,\textrm{neV}$, the bottom of DMRadio-m$^3$'s range (Sec.~\ref{sec:DMRadio-m3}).  At these low masses, the QCD axion--photon coupling is very small, $g_{a\gamma\gamma}\propto m_a$, necessitating technological advancements outlined here.

\noindent {\bf Experimental approach:}
DMRadio-GUT is based on the lumped element technique pioneered in ABRACADABRA-10\,cm (see Sec.~\ref{sec:abracadabra}) that uses an LC circuit to couple to the axion field.  For optimal sensitivity, the GUT experiment will have a resonant readout, with scan rate (in the relevant thermal noise-dominated regime)
\begin{equation}\label{eq:dmrGUT_scanRate}
    \frac{d\nu_r}{dt}\sim\frac{10^6}{\textsc{SNR}^2}\left(g_{a\gamma\gamma}^4\rho_{DM}^2\nu_r\right)\left(c_{PU}^4\frac{QB_0^4V^{10/3}}{\eta_Ak_BT}\right) \,,
\end{equation}
where the final term contains all of the experimental parameters \cite{Brouwer:2022bwo}.  Our goal is to have a scan time $\sim6\,\textrm{years}$, which we can accomplish by tuning the parameters shown in Eq.~\eqref{eq:dmrGUT_scanRate}.  Here we lay out a baseline scenario and R\&D path.  However, there are a variety of combinations of these parameters which can achieve our goal sensitivity, and so if some R\&D thrusts prove challenging, success improving other parameters can compensate.

The first design decision is the magnet.  DMRadio-GUT will likely be a toroid, which keeps magnetic fields away from the sensitive readout electronics.  The fastest way to increase the scan rate is to improve the strength of the magnetic field, $B_0$, and experimental volume, $V$.  In order to obtain DFSZ sensitivities throughout the mass range, our baseline goal is a 10\,m$^3$ magnet with 12\,T RMS field ($\sim$16\,T peak field for a rectangular toroid with $r_{in}/r_{out}=1/2$).  This is achievable with Nb$_3$Sn or REBCO, as demonstrated by several recent and upcoming projects such as the High-Luminosity LHC~\cite{bottura2012advanced,todesco2018progress}, and the SPARC tokomak~\cite{Creely2020}.

The scan rate also scales with the fourth power of $c_{PU}$, a factor similar to the overlap factor in cavities, that describes how well coupled the energy from the axion current is into the pickup structure.  Based on the calculations in \cite{Chaudhuri:2019ntz}, we can expect $c_{PU}\approx0.1$.

The quality factor of the LC resonator, $Q$, is $20\times10^6$ in our baseline scenario.  This number is twenty times the goal $Q$ value for DMRadio-50\,L.  To reach this goal, we have started R\&D on both superconducting resonators and active feedback at low frequencies.  Either or both of these may be used to improve the readout circuit.

Finally, $k_BT$ and $\eta_A$ represent the thermal and amplifier-added noise, respectively.  By running at 10\,mK, the GUT experiment will minimize thermal backgrounds.  We hope to reduce $\eta_A$ to -20\,dB below the standard quantum limit through techniques like backaction evasion with devices such as Radiofrequency Quantum Upconverters \cite{Kuenstner2022,Kuenstner2022b}.  Although DMRadio-GUT is thermally limited, reductions in amplifier noise can still greatly improve the experiment's scan rate by increasing the sensitivity bandwidth, the bandwidth over which a given configuration can maintain an approximately constant SNR.

\noindent {\bf Projected sensitivity:}
With these baseline parameters, DMRadio-GUT can achieve DFSZ sensitivities to axions over almost two decades of axion masses in $\sim6\,\textrm{years}$ of scan time, as shown in Fig.~\ref{fig:dmrsensitivity}.  Although some of the technologies needed to achieve this reach are still in development, there are several R\&D paths that would be able to maintain the goal sensitivity and mass range with scan times under ten years.  Building off of the successes of predecessor lumped element experiments with a multi-thrust R\&D campaign, DMRadio-GUT will be able to definitively probe low-mass QCD axions down to below 1\,neV.

\subsection{SRF Cavities and Related Techniques}
\label{sec:heterodyne}
\begin{figure}[t]
    \centering
    \includegraphics[width=0.3\textwidth]{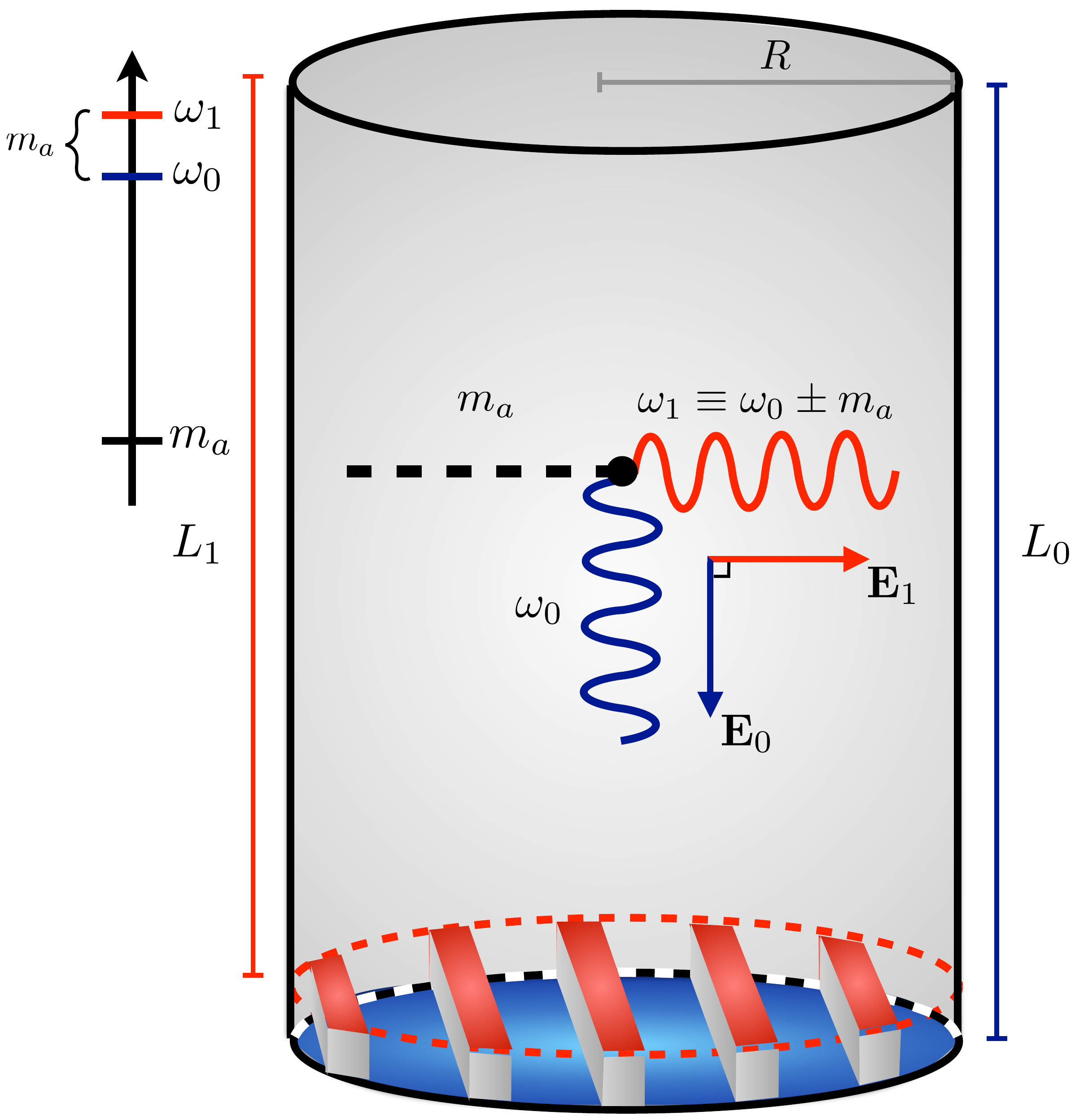}
    \includegraphics[width=0.65\textwidth]{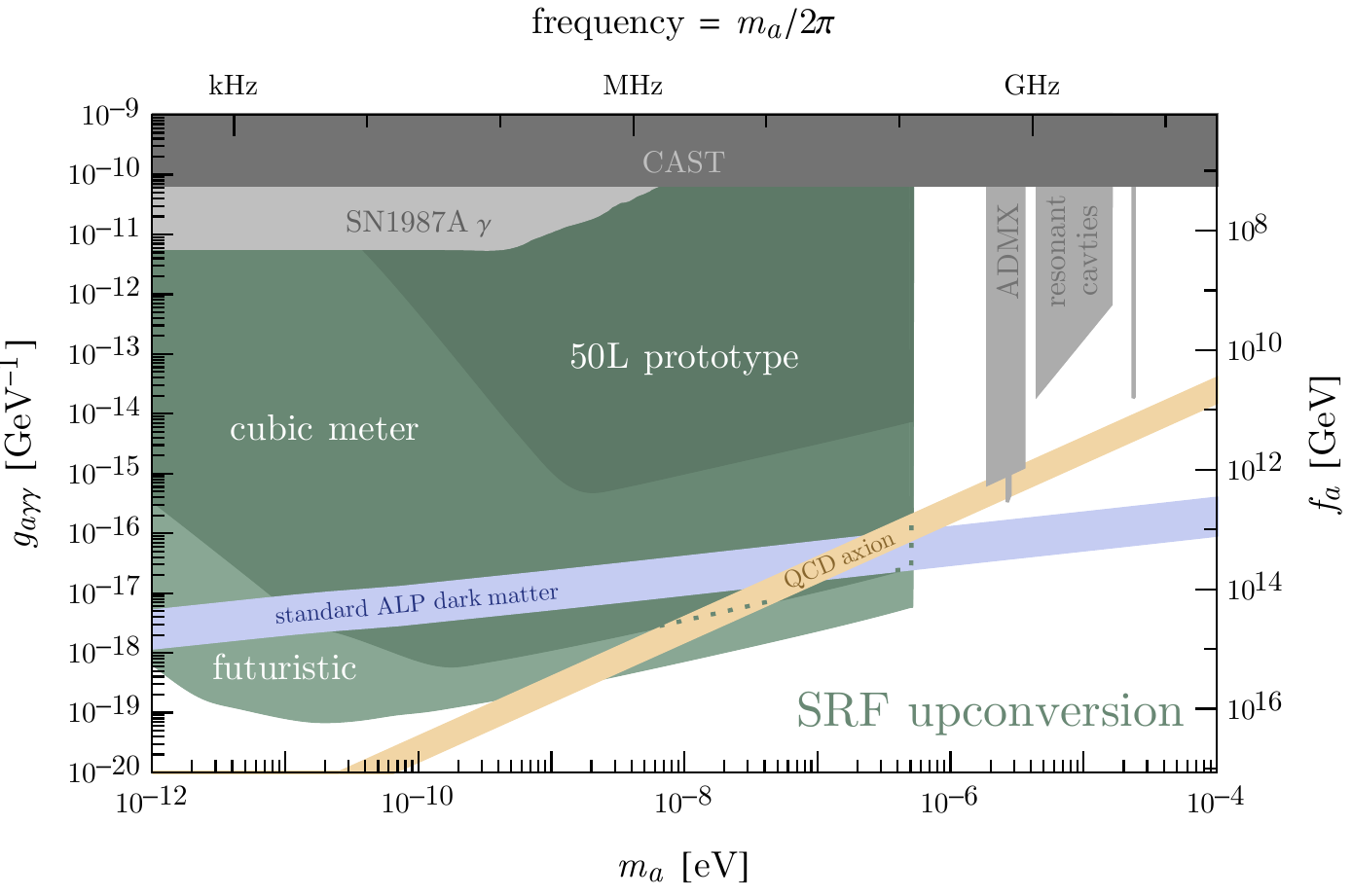}
    \caption{(Left) A schematic of the heterodyne/upconversion experimental setup of Sec.~\ref{sec:heterodyne} (taken from~\cite{Berlin:2019ahk}). In an SRF cavity, a pump mode photon of frequency $\omega_0 \sim \text{GHz}$ is converted by axion dark matter into a nearly degenerate photon of frequency $\omega_1 = \omega_0 + m_a$. (Right) The projected sensitivity of a superconducting upconversion setup. As three representative examples, we show the projected sensitivity of a 50L prototype, a $1 \ \text{m}^3$ setup, as well as a futuristic experiment with a total instrumented volume of $5 \ \text{m}^3$. See text for additional details. Regions motivated by the QCD axion and dark matter produced by the misalignment mechanism are also shown as orange and blue bands, respectively.
    }
    \label{fig:SRF_upconversion}
\end{figure}

The previous subsections describe experiments with static magnetic fields, in which axion dark matter sources an effective current oscillating with frequency $m_a$. By contrast, in the recently proposed heterodyne approach~\cite{Berlin:2019ahk,Berlin:2020vrk,Lasenby:2019prg}, a microwave cavity is prepared by driving a pump mode with nonzero frequency $\omega_0 \sim 1 \ \text{GHz} \gg m_a$. The dark matter axion field causes frequency upconversion, driving power into a signal mode with nearly degenerate frequency $\omega_1 \simeq \omega_0 + m_a$ and distinct spatial geometry. This process is efficient provided that the pump $B$-field and signal $E$-field have a large overlap $B_0 \cdot E_1$ when integrated over the volume of the cavity. This is the case when the pump and signal modes are, e.g., low-lying TE and TM modes. A scan over axion masses can be performed over an extremely large range by only slightly perturbing the cavity geometry, which alters the splitting $\omega_1 - \omega_0$. A schematic of the setup is shown in the left panel of Fig.~\ref{fig:SRF_upconversion}.

The upconversion process leads to a GHz frequency axion signal, even for axion frequencies/masses much smaller than $1 \ \text{GHz} \sim 1 \ \mu \text{eV}$. Compared to other static field approaches targeting long wavelength axions, upconverting the signal to GHz frequencies is advantageous. This is because: 1) the signal power is parametrically enhanced by $\sim \text{GHz}/m_a \gg 1$ and 2) high-frequency GHz resonators can couple to very slowly oscillating axion signals, opening up sensitivity to axion masses below a kHz, a region typically inaccessible with static field resonators. This signal can be efficiently read out with well-developed experimental tools, without requiring quantum metrology techniques, cooling beyond liquid helium temperatures, or magnetic fields in excess of $\mathcal{O}(100) \ \text{mT}$. Furthermore, since the axion signal is extremely narrow, $\Delta \omega / \omega \sim 10^{-6} \, m_a / \omega_1 \ll 10^{-6}$, it benefits from the high quality factors $Q_{\text{int}} \sim \text{few} \times 10^{11}$~\cite{Romanenko:2014yaa} achieved by superconducting radiofrequency (SRF) cavities. The top curve in the right panel of Fig.~\ref{fig:SRF_upconversion} shows the reach of a 50L prototype using existing SRF cavities and commercially available technology, assuming a modest quality factor of $Q = 10^{10}$. The middle curve shows the potential of a dedicated experimental design of a cubic meter cavity and $Q = 10^{12}$. The bottom curve shows the reach of a more futuristic setup, still assuming $Q = 10^{12}$, but a larger volume of $5 \ \text{m}^3$; this corresponds to a total instrumented volume approximately 100 times larger than today's individual SRF cavities (although note that this is only a fraction of the total instrumented volume of SRF-based accelerators, such as PIP II, LCLS 2, and XFEL). In the prototype and cubic meter setups, we assume a peak magnetic field strength of $B_\text{pump} = 0.2 \ \text{T}$ (no larger than that employed in modern SRF cavities), while for the futuristic setup we assume $B_\text{pump} = 0.4 \ \text{T}$. In the above projections, we fix the integration time to correspond to $\sim 10^7 \ \text{sec}$ to cover an $e$-fold in axion mass. Instead, for a search specifically optimized for the QCD axion, we find that the cubic meter setup could probe the DFSZ parameter space for axion masses ranging $9 \times 10^{-9} \ \text{eV}$ to $10^{-6} \ \text{eV}$, while the futuristic setup would be sensitive to the DFSZ region for axion masses of $1 \times 10^{-10} \ \text{eV}$ to $10^{-6} \ \text{eV}$, both assuming a total integration time of $\sim 6 \ \text{years}$. A cubic meter setup could also cover dark matter motivated parameter space for $m_a \gtrsim 10^{-11} \ \text{eV}$, as well as unexplored parameter space at ultralow ($\lesssim \text{kHz} \sim 10^{-11} \ \text{eV}$) frequencies/masses.

The most important determinant of the sensitivity is the ``leakage'' of noise power from the pump mode to the readout, such as through imperfections of the cavity geometry and vibrations of the walls. These effects were previously studied by the MAGO~\cite{Ballantini:2005am} and DarkSRF~\cite{DarkSRF} collaborations, which precisely controlled separate aspects of SRF cavities. In Fig.~\ref{fig:SRF_upconversion}, the prototype sensitivity assumes no attenuation of mechanical vibrations, whereas the projections for the cubic meter and futuristic setups assume 100 dB and 130 dB of vibrational attentuation, respectively. This is no larger than the vibrational attenuation employed in current resonant bar experiments searching for gravitational waves. Currently, potential cavity geometries and tuning mechanisms are being studied at SLAC (funded by an LDRD grant), CERN, and the SQMS center at Fermilab. Ongoing developments at SQMS are aimed at performing in-situ measurements of leakage noise arising from, e.g., microphonics and surface imperfections, as well as the design and construction of a prototype using state of the art cavities. Such developments in SRF technology will also benefit related experiments searching for dark photons, millicharged particles, and gravitational waves, as discussed in Ref.~\cite{Snowmass2021:SRF}.

A close relative of this approach was first proposed by Refs.~\cite{GORYACHEV2019,THOMSON2021}. In that setup, the signal mode is excited as well, and the axion causes its frequency to periodically shift, so that its effect can be detected by precision frequency metrology. A prototype built at the University of Western Australia has already demonstrated the principle of operation~\cite{upload21,uploadcorrect21}. Future work will improve the sensitivity by better stabilizing the oscillators, which will also be relevant for the frequency conversion approach.

\section{\label{sec:ultralight}
Alternative Couplings}
Due to the exquisite control achievable for electromagnetic fields, many axion searches are focused on the interaction with two photons. However, as already mentioned above, establishing the nature of a QCD axion requires confirmation of the coupling to gluons.
Moreover, the need to explore a mass range spanning many orders of magnitude also encourages the exploration of new methods based on other couplings of the axion. This section highlights the developments in this direction.

\subsection{CASPEr-electric}
\label{sec:casper-e}
\noindent {\bf Physics goals and status:}
The Cosmic Axion Spin Precession Experiments (CASPEr) use the nuclear magnetic resonance approach to search for axion-like dark matter.
The CASPEr-electric experiment at Boston University searches for the defining \linebreak[4] model-independent interaction of the QCD axion, which solves the strong-CP \linebreak[4] problem~\cite{Budker2014}. Any claim of QCD axion detection will need to verify the existence of this EDM coupling $g_d$, which is equivalent to the coupling constants given in section~\ref{sec:qcd_axion}: $g_d=|g_{ap\gamma}|=|g_{an\gamma}|$, see eq.~\eqref{eq:gd}. 
CASPEr-e is also sensitive to the nucleon spin gradient coupling $g_{aNN}$, which is equivalent to the coupling constant given in eq.~\eqref{eq:gann}. 
The goal of CASPEr-e is to search for axion-like dark matter in the mass range between $10^{-12}$~eV and $10^{-6}$~eV, with sufficient sensitivity to detect the QCD axion with mass between $10^{-12}$~eV and $5\times10^{-9}$~eV. This corresponds to the theoretically-favored range of axion decay constants $f_a$ near the grand unified and Planck scales.
The first-generation search in the mass range 162~neV to 166~neV has reported limits on the EDM and gradient interactions of axion-like dark matter~\cite{Aybas2021a}. Measurements with a 5~mm sample place the upper bounds
$|g_d|<9.5\times10^{-4}\,\text{GeV}^{-2}$ and 
$|g_{aNN}|<2.8\times10^{-1}\,\text{GeV}^{-1}$
(95\% confidence level) in this mass range. The constraint on $g_d$ corresponds to an upper bound of 
$1.0\times 10^{-21}\,\text{e}\cdot\text{cm}$
on the amplitude of oscillations of the neutron electric dipole moment, and 
$4.3\times 10^{-6}$
on the amplitude of oscillations of CP-violating $\theta$ parameter of quantum chromodynamics.

\begin{figure}[t]
    \centering
    \includegraphics[width=\textwidth]{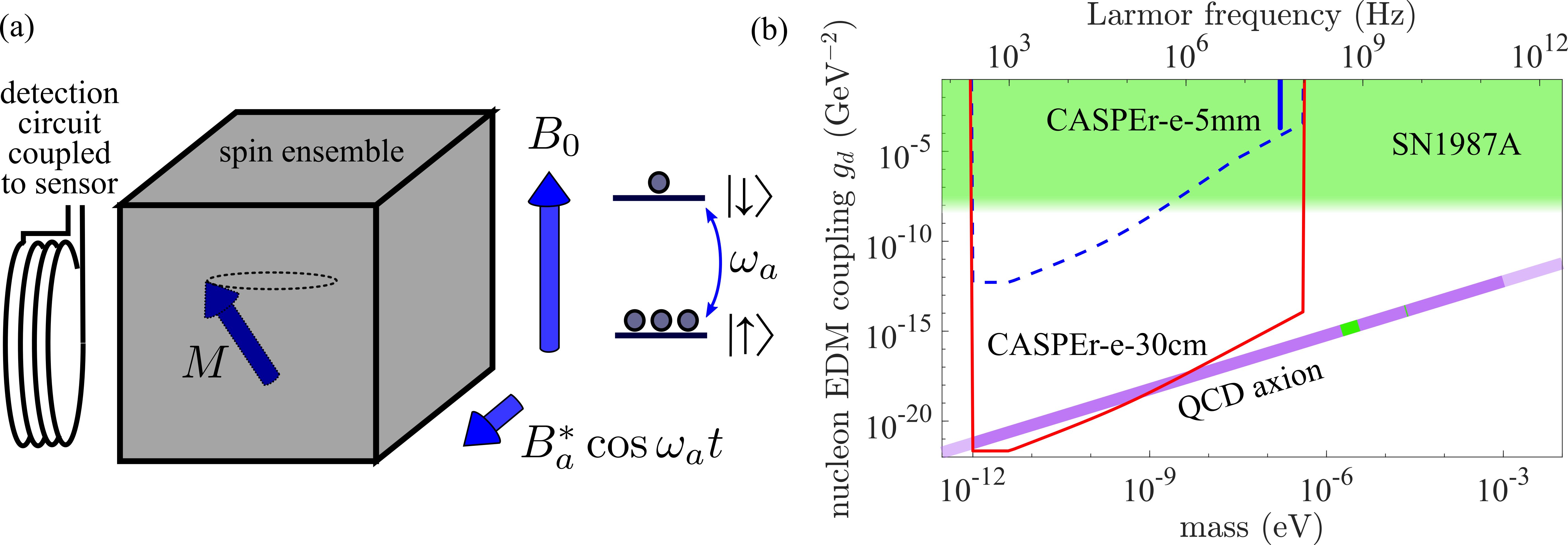}%
    \caption{
    The CASPEr experimental schematic and CASPEr-e projected sensitivity.
    (a) The CASPEr experimental schematic, showing the nuclear spin ensemble whose spin states are split by the applied bias field $B_0$. When this splitting is resonant with the axion-like dark matter Compton frequency $\omega_a$, the ensemble magnetization $M$ is tilted and undergoes precession that is detected by an inductively-coupled sensor.
    (b) CASPEr-e projected sensitivity, showing the published limits (blue) and the sensitivity of the search with a 5~mm sample (blue dashed line)~\cite{Aybas2021a}. The red line shows the quantum spin projection noise-limited sensitivity of a CASPEr-e search with a 30~cm sample. The green region is excluded by constraints on excess cooling of SN1987A~\cite{Graham:2013AxionDM,Chang2018a,PDG2019}. The QCD axion is in the purple band, whose width shows theoretical uncertainty~\cite{Graham:2013AxionDM}.
    }
    \label{fig:casper_fig}
\end{figure}

\noindent {\bf Experimental approach:}
CASPEr-electric uses precision solid-state magnetic resonance to search for an oscillating spin torque induced by interaction of axion-like dark matter with nuclear spins, Fig.~\ref{fig:casper_fig}(a).
Such an interaction can be represented by a pseudo-magnetic field $B_a^*\cos{(\omega_at)}$, which oscillates at the axion Compton frequency $\omega_a=m_ac^2/\hbar$, with the amplitude $B_a^*$ proportional to the axion-spin coupling strength: $B_a^*\propto g_d$ or $B_a^*\propto g_{aNN}$~\cite{Graham:2013AxionDM,Budker2014}. A solid sample is chosen to maximize the magnitude of this interaction. CASPEr-e uses poled ferroelectric crystals, such as PbTiO$_3$ or PMN-PT, leveraging the enhanced effective electric field $E^*\approx340$~kV/cm~\cite{Mukhamedjanov2005a,Ludlow2013,Budker2014,Skripnikov2016,Aybas2021a}. The $^{207}$Pb nuclear spins inside the crystal are pre-polarized in a 9~T applied magnetic field; ongoing work explores how to speed up and/or increase this polarization using light-induced paramagnetic centers. The long spin relaxation time $T_1=26$~min allows the bias magnetic field $B_0$ to be swept to lower values without losing polarization. If $B_0$ is chosen so that the spin Larmor frequency matches the axion Compton frequency $\omega_a$, then the pseudo-magnetic field drives the spins on resonance, the magnetization vector tilts and precesses around the bias field. This precession is detected by an inductively-coupled sensor, such as a SQUID. The search for axion-like dark matter is performed by sweeping the bias field $B_0$~\cite{Aybas2021a}.

\noindent {\bf Sensitivity and timeline:}
The first-generation experiments have demonstrated how CASPEr-e can search for the defining EDM interaction of the QCD axion~\cite{Aybas2021a}, Fig.~\ref{fig:casper_fig}(b). In order to achieve QCD axion sensitivity, the experiment will be scaled up in volume, due to the limits imposed by the quantum spin projection noise~\cite{Aybas2021b}. 
An experiment with a 60~cm sample is projected to reach the QCD axion sensitivity in the mass range $10^{-12}$~eV to $5\times10^{-9}$~eV. In parallel with R\&D efforts on a scaled-up experiment, work will proceed on the following activities: (1) integration with quantum electromagnetic sensors that offer superior performance compared to SQUIDs, (2) characterization of optimal material to host the solid-state spin ensemble, (3) application of quantum control tools to optimize spin ensemble coherence time and polarization.
An experiment limited only by spin projection noise (standard quantum limit) is projected to reach QCD axion sensitivity up to masses $\approx 5\times10^{-8}$~eV. The timeline for CASPEr-e aims to demonstrate sensitivity at the QCD axion level on a 5-7 year timescale.

\subsection{CASPEr-gradient}
\label{sec:casper-g}
\noindent {\bf Physics goals and Status:}
The CASPEr-gradient setups at the Johannes Gutenberg University Mainz, Germany, are dedicated experiments to search for the gradient coupling $g_{aNN}$ of axion-like particle dark matter to  liquid hyperpolarized nuclear spin samples.
First experimental results were published in 2019 excluding ultralight ALP dark matter in the mass ranges $10^{-22}$~eV to $1.3\times 10^{-17}$~eV with coupling constants $g_{aNN}>6\times 10^{-5}$ (95\% confidence level)~\cite{Wu:2019AxionDM} and $1.8\times 10^{-16}$~eV to $7.8\times 10^{-14}$~eV 
corresponding to Compton frequencies ranging from ~45~mHz to 19~Hz \cite{Garcon:2019AxionDM} with coupling constants $g_{aNN}>5\times 10^{-5}$.
The existing CASPEr-gradient low field (LF) apparatus will search for ALP masses up to $\approx 1.6\times10^{-8}$~eV corresponding to 4\,MHz ALP Compton frequency with sensitivities to find ALP dark matter with nucleon couplings at the level of $g_{aNN}>10^{-14}$.
A second apparatus, CASPEr gradient high field (HF), currently under construction will be capable of searching for ALPs with masses up to $\approx 2.4\times10^{-6}$~eV corresponding to 600\,MHz ALP Compton frequency with sample volumes of ~1\,ml.

\noindent {\bf Experimental approach:}
The approach is based on magnetic resonance, in the same way as described in previous section~\ref{sec:casper-e} for CASPER-electric, Fig.~\ref{fig:casper_fig}(a).
CASPEr-gradient is only sensitive to the nucleon spin gradient coupling $g_{aNN}$, which is equivalent to the coupling constant given in eq.~\eqref{eq:gann}.
CASPEr-gradient LF and HF use liquid xenon-129 hyperpolarized via spin-exchange optical pumping with optically polarized rubidium vapor, or protons hyperpolarized by e.g. parahydrogen-induced polarization. The liquid state allows for long coherence time (compared to solid state NMR) matching or exceeding the coherence time of virialized cold dark matter. This allows for a given sample density an optimum search sensitivity. The search for axion-like dark matter is performed by sweeping the bias field $B_0$ and observing non-trivial transverse magnetization or unexplained modulations of the nuclear magnetic resonance frequency. The detection modality is matched to the maximum Compton frequency in the respective search range: CASPEr-gradient LF featuring a superconducting magnet with a maximum field of 100\,mT uses superconducting pick-up coils in a gradiometer configuration attached to a super-conducting-quantum-interference device (SQUID), CASPEr-gradient HF is housed in a superconducting magnet with a maximum field of 14.1\,T deploying a tunable inductive pick-up circuit. 
Ongoing work addresses the production and optimization of hyperpolarized xenon-129 and proton samples as well as characterization of the detection system and optimization of the magnetic field homogeneity.

\noindent {\bf Projected sensitivity:}
As for CASPEr-E the magnitude of the signal detected by the sensor is proportional to the number of spins in the ensemble, the degree of polarization, the coherence time of the spin ensemble, and the magnitude of the pseudo-magnetic field $B_a^*$. Electronic noise in the detection circuit, by thermal noise in the circuit that couples to this amplifier, or by fundamental quantum spin projection noise.

\noindent {\bf Projected timeline:}
The CASPEr Gradient LF experimental measurements are ongoing with a spherical 8~mm-diameter sample of thermally polarized protons. Furthermore upgrades to hyperpolarized liquid xenon-129 are actively pursued. The sensitivity of the apparatus is limited by SQUID detection noise floor at $1\,\mu\Phi_0/\sqrt{\rm Hz}$ and able to search for ALP dark matter well beyond the astrophysical SN1987A constraints. First magnetic field scans are projected to be completed on a 1-year timeline.
CASPEr Gradient HF is currently being build. The collaboration envisions the experiment to be operational within the next 2 years.

\subsection{ARIADNE}
\label{sec:ariadne}
Axions can couple to fundamental fermions through a scalar vertex and a pseudoscalar vertex. Depending on the model, the coupling can occur for either electrons or nuclei, with possible allowed interactions being monopole-monopole, monopole-dipole, or dipole-dipole.  Monopole-dipole interactions include a scalar coupling, $g_{s}$, and a pseudo-scalar coupling, $g_{p}$, thereby violating $\mathrm{P}$ and $\mathrm{T}$ symmetry.  In the non-relativistic limit, this interaction is proportional to $g_{s}g_{p}\vec{\sigma}\cdot\vec{r}$ where $\vec{\sigma}$ is the spin of one particle, and $\vec{r}$ is the distance between two particles \cite{Fadeev2019spinforces}. 
Although the couplings are extremely small for the interaction between single particles, a macroscopic object with $10^{22} \sim 10^{23}$ components would produce a coherent field which can be detectable with a sensitive laboratory experiment, with an interaction range set by the Compton wavelength of the axion, which is inversely proportional to its mass~\cite{1984PhRvD..30..130M}. Many of the experimental tests of this interaction have been done with polarized gases~\cite{1996PhRvL..77.2170Y}. 
In many cases, combined laboratory measurements taken in parallel with astrophysical data have produced the most stringent constraints on the products of the coupling constants $g_s$ and $g_p$ \cite{Raffeltgsgp,OHare:2020wah}.
Nonetheless, existing constraints are insufficient by several orders of magnitude in order to probe the range of interaction strength expected for the QCD axion.

ARIADNE aims to detect QCD-axion-mediated spin-dependent interactions between an unpolarized source mass and a spin-polarized $^3$He low-temperature gas \cite{arvanitaki2014resonantly,progressariadne}.  Unlike other direct axion search experiments which depend on the local dark matter density at the Earth, with the axion being the particle that constitutes dark matter, the signal can be modulated in a controlled way.  The experiment probes QCD axion masses in the higher end of the traditionally allowed axion window, up to $6$ meV, which are not accessible by any existing experiment including dark matter haloscopes such as ADMX.  Thus ARIADNE fills an important gap in the search for the QCD axion in this important region of parameter space.
The axion can mediate an interaction between fermions (e.g. nucleons) with a potential given by $
 U_{sp}(r)=\frac{\hbar^2 g_s^N g_p^N}{8 \pi m_f}\left( \frac{1}{r \lambda_a}+\frac{1}{r^2}\right) e^{-\frac{r}{\lambda_a}} \left(\hat \sigma \cdot \hat r \right)$, where $m_f$ is their mass, $\hat{\sigma}$ is the Pauli spin matrix, $\vec{r}$ is the vector between them, and $\lambda_a = h/m_A c$ is the axion Compton wavelength. For the QCD axion the scalar and dipole coupling constants $g_s^N$ and $g_p^N$ are correlated to the axion mass. Since it couples to $\hat{\sigma}$ which is proportional to the nuclear magnetic moment, the axion coupling can be treated as a fictitious magnetic field~ $B_{\mathrm{eff}}$.  This fictitious field is used to resonantly drive spin precession in a laser-polarized cold $^3$He gas.  This is accomplished by spinning an unpolarized tungsten mass sprocket near the $^3$He vessel. As the teeth of the sprocket pass by the sample at the nuclear larmor precession frequency, the magnetization in the longitudinally polarized He gas begins to precess about the axis of an applied field. This precessing transverse magnetization is detected with a superconducting quantum interference device (SQUID). The $^3$He sample acts as an amplifier to transduce the small fictitious magnetic field into a larger real magnetic field detectable by the SQUID. Superconducting shielding is needed around the sample to screen it from ordinary magnetic field noise which would otherwise limit the sensitivity of the measurement. 

The experiment sources the axion in the lab, and can explore all mass ranges in our sensitivity band simultaneously, unlike experiments which must scan over the allowed axion oscillation frequencies (masses) by tuning a cavity or magnetic field.

\begin{figure}[t!]
\begin{center}
\includegraphics[width=1.0\columnwidth]{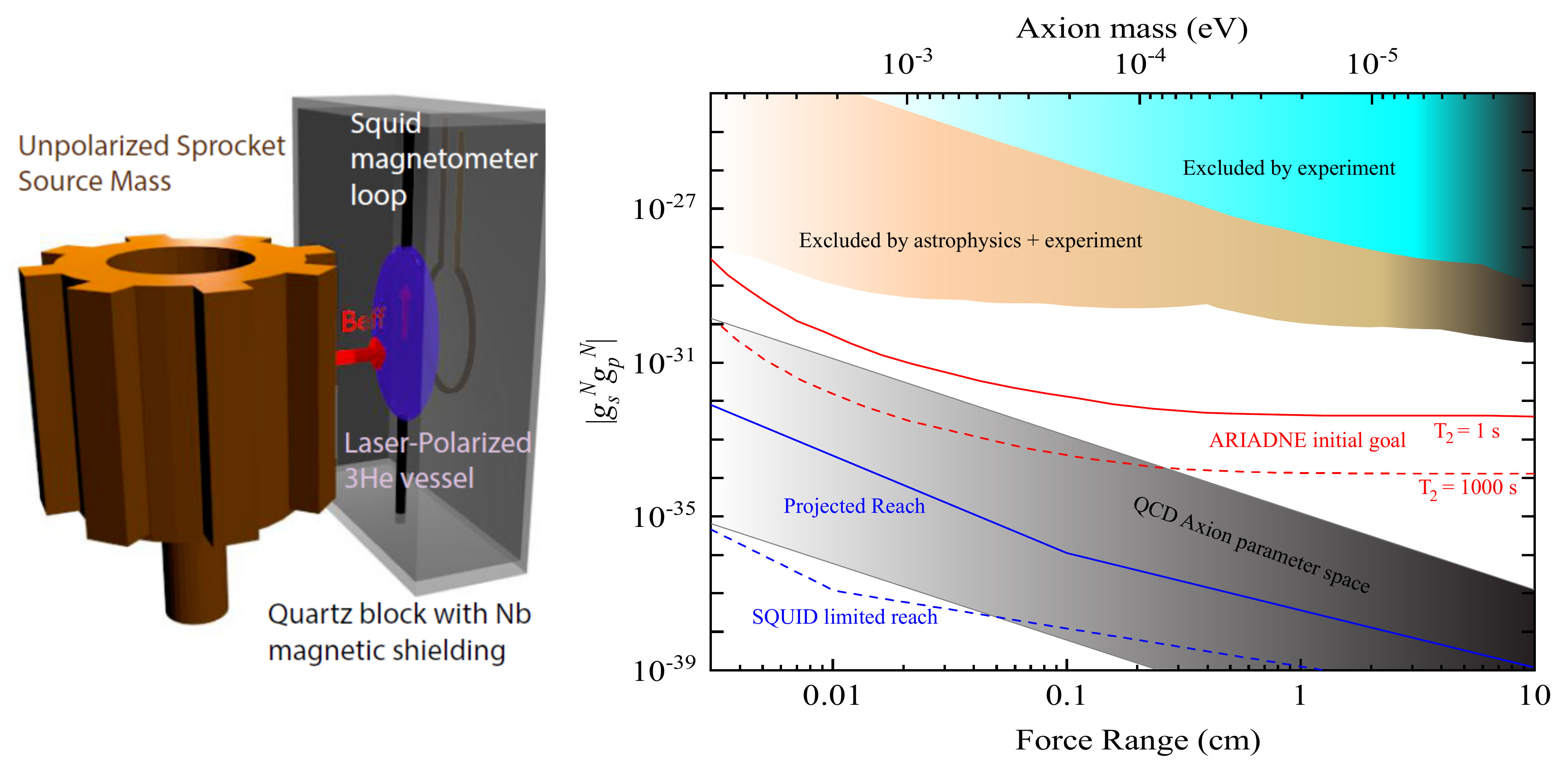}
\caption{(left) Setup: a sprocket-shaped source mass is rotated so its ``teeth'' pass near an NMR sample at its resonant frequency. (right) Projected reach for monopole-dipole axion mediated interactions. The band bounded by the red (dark) solid line and dashed line denotes the limit set by transverse magnetization noise, depending on achieved $T_2$. Current constraints and expectations for the QCD axion also are shown, adapted from Refs. \cite{arvanitaki2014resonantly,romalis2018,OHare:2020wah}.\label{fig:ariadne}}
\end{center}
\end{figure}
Distinct from other magnetometry experiments \cite{chuexperiment, romalis2018,northropgrummanpaper}, the experiment uses a  resonant enhancement technique. Assuming sources of systematic error and noise can be mitigated, the approach is expected to be spin-projection noise limited \cite{arvanitaki2014resonantly}, and in principle allows several orders of magnitude improvement, yielding sufficient sensitivity to detect the QCD axion (Fig. \ref{fig:ariadne}).

\noindent\textbf{Projected timeline:} Several critical components have been tested for the experiment including the superconducting shielding, magnetic characterization of the source mass, and rotary stage mechanism. Final assembly and commissioning of the experiment is currently underway with first results expected in the next 1-2 years. Assuming systematic uncertainties and background noise can be controlled at the level anticipated, QCD-axion sensitivity is expected within a 3-5 year timescale. 

Future prospects for improvements in the search for novel spin dependent interactions could include investigations with a spin polarized source mass, or improved sensitivity with new cryogenic or quantum technologies.
Spin squeezing or coherent collective modes in $^3$He could offer prospects for improved sensitivity beyond the Standard quantum limit of spin projection noise in experiments such as ARIADNE, potentially allowing sensitivity all the way down to the SQUID-limited sensitivity (dashed-dotted line in Fig. \ref{fig:ariadne}). This would allow one to rule out the axion over a wide range of masses, and when combined with other promising techniques \cite{CASPERPRX,ABRA,DMRadio}, and existing experiments \cite{ADMX,HAYSTACK} already at QCD axion sensitivity, could allow in principle the QCD axion to be searched for over its entire allowed mass range.  

\subsection{Magnetometer Networks}
\label{sec:ultralight_ideas}
\noindent {\bf GNOME:} Due to topology or self-interactions, ultralight bosonic fields such as ALPs can form stable, macroscopic field configurations in the form of boson stars \cite{Kolb:1993zz,Bra16,Kim18AxionStars,eby2019global} or topological defects (e.g., domain walls \cite{sikivie1982axions,Vil85,kawasaki2015axion,Pos13}). 
Even in the absence of topological defects or self-interactions, bosonic dark matter fields exhibit stochastic fluctuations \cite{centers2021stochastic}. 
Additionally, it is possible that high-energy astrophysical events could produce intense bursts of exotic ultralight bosonic fields \cite{dailey2021quantum}. 
In any of these scenarios, instead of being bathed in a uniform flux, terrestrial detectors will witness transient events when ultralight bosonic fields pass through Earth. 
The Global Network of Optical Magnetometers for Exotic Physics searches (GNOME) is a network of more than a dozen time-synchronized optical atomic magnetometers with stations in North America, Europe, Asia, Australia, and the Middle East.
GNOME is designed to search for correlated signals heralding beyond-the-Standard-Model physics \cite{Pus13,afach2018characterization,afach2021search}. 
There are a limited number of ``portals'' through which ultralight bosonic fields such as ALPs can couple to Standard Model particles \cite{Saf18RMP}, prominent among them spin-dependent interactions to which GNOME is particularly sensitive. 
GNOME has the potential to search a wide range of unexplored parameter space for such ALP fields \cite{Pos13,Pus13,Kim18AxionStars,dailey2021quantum,mas2022intensity}.

GNOME uses precision optical atomic magnetometers \cite{budker2007optical,Bud13} to search for spin torques induced by interaction of axion-like fields with nuclear spins, such as that described by the nonrelativistic Hamiltonian \cite{Saf18RMP}
\begin{align}
\hat{H}_{aNN} = - 2 g_{aNN} \left( \hbar c \right)^{3/2} \boldsymbol{S} \cdot {\boldsymbol{\nabla}} \phi(\boldsymbol{r},t)~,
\label{Eq:ALP-spin-Hamiltonian}
\end{align}
where $g_{aNN}$ is the ALP-nucleon coupling constant, $\boldsymbol{S}$ is the nuclear spin, and $\phi(\boldsymbol{r},t)$ is the ALP field.
The form of the Hamiltonian in Eq.~\eqref{Eq:ALP-spin-Hamiltonian} is analogous to the Zeeman Hamiltonian, so such effects manifest as pseudo-magnetic fields that can be detected using spin-based magnetometers.
GNOME magnetometers are located within multi-layer magnetic shields to reduce the influence of external magnetic noise and perturbations, while still maintaining sensitivity to spin-dependent ALP interactions with nuclei \cite{Kim16}.
While a single magnetometer would measure transient and stochastic events associated with ALP fields, in practice it is difficult to confidently distinguish a true ALP signal from false positives induced by changes of magnetometer operational conditions (e.g., magnetic-field spikes, laser mode hops, electronic noise, etc.).  
Effective vetoing of prosaic noise and transient events (false positives) requires an array of individual, spatially distributed magnetometers to eliminate spurious local effects.  
Furthermore, a global distribution of sensors is beneficial for event characterization, providing the ability to resolve ALP field properties by observing the relative timing and amplitude of transient events at different sensors \cite{Mas20,kim2021machine}.
New data analysis efforts and upgrades of GNOME magnetometers to noble gas comagnetometers \cite{Kor02,Kor05} are underway.

\smallskip

\noindent {\bf Unshielded magnetometer networks:} Recently, it was proposed that the Earth itself could act as a transducer for ultralight dark-matter detection, both for dark (hidden) photons \cite{Fedderke:2021aqo,Fedderke:2021rrm} and ALPs \cite{Arza:2021rrm}.
Either via kinetic mixing in the case of hidden photons or via coupling to photons in the case of ALPs, the ultralight dark matter field induces a monochromatic oscillating magnetic field with a particular global vectorial pattern.
Such a global magnetic field pattern can be searched for using a network of unshielded magnetometers (as opposed to the shielded magnetometers used in GNOME).
Searches for hidden photons and ALPs using a publicly available dataset from the SuperMAG Collaboration \cite{Fedderke:2021rrm,Arza:2021rrm} established experimental constraints on such scenarios that are competitive with astrophysical limits \cite{Payez:2014xsa,McDermott:2019lch,wadwan21} and the CAST experiment \cite{CAST:2017uph} in the probed mass ranges (from around $10^{-18}$~eV to $10^{-16}$~eV).

\subsection{Broadband Axion Dark Matter Searches via Non-Electromagnetic Couplings}
\label{sec:broadband_non-em}

\textbf{Fermion couplings:}
Non-electromagnetic interactions of an axion or axion-like dark-matter field with standard-model fields can induce time-varying spin-dependent effects. 
In particular, an axion dark-matter field $a \approx a_0 \cos(m_a t - \boldsymbol{p}_a \cdot \boldsymbol{x})$ can couple to the axial-vector currents of fermions via the derivative interaction,
\begin{equation}
    \label{derivative_coupling_fermions}
    \mathcal{L} = - \frac{C_\psi}{2 f_a} \partial_\mu a \, \bar{\psi} \gamma^\mu \gamma^5 \psi \, , 
\end{equation}
where $\psi$ and $\bar{\psi} = \psi^\dagger \gamma^0$ denote a fermion field and its Dirac adjoint, respectively, $f_a$ is the axion decay constant, and $C_\psi$ are model-dependent dimensionless parameters. 
In the non-relativistic limit, the spatial components of the derivative interaction (\ref{derivative_coupling_fermions}) give rise to time-varying energy shifts of spin-polarised fermions according to the Hamiltonian \cite{Flambaum:2013Axion_Patras,Graham:2013AxionDM,Stadnik:2014AxionDM},
\begin{equation}
    \label{axion-wind_Hamiltonian}
    H (t) \approx \frac{C_\psi a_0}{2 f_a} \sin(m_a t) \, \boldsymbol{\sigma}_\psi \cdot \boldsymbol{p}_a \, , 
\end{equation}
where $\boldsymbol{\sigma}_\psi$ denotes the fermion's Pauli spin vector and $\boldsymbol{p}_a \approx m_a \boldsymbol{v}_a$ denotes the axion momentum relative to the fermion. 
It has been proposed to search for the ``axion wind'' spin-precession effect described by Eq.~(\ref{axion-wind_Hamiltonian}) using co-magnetometry \cite{Stadnik:2017Thesis,Graham:2018AxionDM}, spin-polarised torsion pendula \cite{Stadnik:2017Thesis,Graham:2018AxionDM}, Penning traps \cite{Smorra:2019AxionDM} and storage rings \cite{Smorra:2019AxionDM,Graham:2021AxionDM}. 

A co-magnetometry search using mercury atoms and ultracold neutrons constrained the nucleon interaction parameters up to $f_a/C_N \approx 4 \times 10^5~\textrm{GeV}$ over the mass range $10^{-22}~\textrm{eV} \lesssim m_a \lesssim 10^{-17}~\textrm{eV}$ \cite{Abel:2017AxionDM}. 
A co-magnetometry search using two different transitions of the acetonitrile molecule constrained the nucleon interaction parameters up to $f_a/C_N \approx 2 \times 10^4~\textrm{GeV}$ over the mass range $10^{-22}~\textrm{eV} \lesssim m_a \lesssim 10^{-17}~\textrm{eV}$ \cite{Wu:2019AxionDM}. 
An ultra-low-field nuclear magnetic resonance search constrained the nucleon interaction parameters up to $f_a/C_N \approx 2 \times 10^4~\textrm{GeV}$ over the mass range $10^{-16}~\textrm{eV} \lesssim m_a \lesssim 10^{-13}~\textrm{eV}$ \cite{Garcon:2019AxionDM}. 
A spin-polarised torsion pendulum search constrained the electron interaction parameter up to $f_a/C_e \approx 2 \times 10^6~\textrm{GeV}$ over the mass range $10^{-23}~\textrm{eV} \lesssim m_a \lesssim 10^{-18}~\textrm{eV}$ \cite{Terrano:2019AxionDM}. 
A Penning trap search with antiprotons constrained the antiproton interaction parameter up to $f_a/C_{\bar{p}} \approx 0.6~\textrm{GeV}$ over the mass range $10^{-23}~\textrm{eV} \lesssim m_a \lesssim 10^{-16}~\textrm{eV}$ \cite{Smorra:2019AxionDM}. 
Existing co-magnetometers using noble-gas species are expected to already be able to probe nucleon interaction parameters up to $f_a/C_N \sim 10^9~\textrm{GeV}$ over the mass range $10^{-24}~\textrm{eV} \lesssim m_a \lesssim 10^{-16}~\textrm{eV}$ \cite{Stadnik:2017Thesis}, while future co-magnetometers may probe nucleon interaction parameters up to $f_a/C_N \sim 10^{12}~\textrm{GeV}$ over the mass range $10^{-24}~\textrm{eV} \lesssim m_a \lesssim 10^{-14}~\textrm{eV}$ \cite{Graham:2018AxionDM,Bloch:2020AxionDM}. 
Future spin-polarised torsion pendula experiments may probe the electron interaction parameter up to $f_a/C_e \sim 10^{11}~\textrm{GeV}$ over the mass range $10^{-22}~\textrm{eV} \lesssim m_a \lesssim 10^{-14}~\textrm{eV}$ \cite{Graham:2018AxionDM}. 
Future storage ring experiments using protons may probe the proton interaction parameter up to $f_a/C_p \sim 10^{12}~\textrm{GeV}$ over the mass range $10^{-22}~\textrm{eV} \lesssim m_a \lesssim 10^{-18}~\textrm{eV}$, while future storage ring experiments using muons may probe the muon interaction parameter up to $f_a/C_\mu \sim 10^6~\textrm{GeV}$ over the mass range $10^{-22}~\textrm{eV} \lesssim m_a \lesssim 10^{-11}~\textrm{eV}$ \cite{Graham:2021AxionDM}.

\noindent\textbf{Gluon coupling:}
An axion dark-matter field can also couple to the gluons via the non-derivative interaction,
\begin{equation}
    \label{nonderivative_coupling_gluons}
    \mathcal{L} = \frac{C_G}{f_a} \frac{g^2}{32 \pi^2} a G_{\mu \nu}^{b} \tilde{G}^{b \mu \nu}  \, , 
\end{equation}
where $G$ and $\tilde{G}$ are the gluonic field tensor and its dual, respectively, $b=1,2,...,8$ is the colour index, $g^2/(4\pi)$ is the colour (strong) coupling constant, and $C_G$ is a model-dependent dimensionless parameter. 
The axion-gluon coupling (\ref{nonderivative_coupling_gluons}) induces an oscillating electric dipole moment (EDM) of the neutron \cite{Graham:2011AxionDM} via a chirally enhanced one-loop process \cite{Crewther:1979nEDMchiral,Pospelov:1999Theta}. 
Additionally, the axion-gluon coupling (\ref{nonderivative_coupling_gluons}) induces oscillating EDMs of atoms and molecules via the one-loop-level oscillating nucleon EDMs and tree-level oscillating P,T-violating intranuclear forces (which generally give the dominant contribution in diamagnetic species) \cite{Stadnik:2014AxionDM}, as well as via two-photon exchange processes between electrons and the nucleus that induce oscillating CP-odd semileptonic interactions (which can give the dominant contribution in paramagnetic species) \cite{Flambaum:2020EDM}. 
A search for an oscillating EDM of the neutron constrained the gluon interaction parameter up to $f_a/C_G \approx 10^{21}~\textrm{GeV}$ over the mass range $10^{-24}~\textrm{eV} \lesssim m_a \lesssim 10^{-17}~\textrm{eV}$ \cite{Abel:2017AxionDM}, while a search for an oscillating EDM of the HfF$^+$ molecule constrained the gluon interaction parameter up to $f_a/C_G \approx 10^{15}~\textrm{GeV}$ over the mass range $10^{-22}~\textrm{eV} \lesssim m_a \lesssim 10^{-15}~\textrm{eV}$ \cite{Roussy:2021AxionDM}.

\section{Solar Axions}
\label{sec:solar_axions}
Although this white paper is mainly focused on searches for dark matter axions, experiments that can detect axions independent of its role as dark matter can provide important information on the existence of the axion as well as its properties. 
One of the most powerful techniques is to look for axions coming from the strongest astrophysical source of axions visible from Earth: the Sun~\cite[e.g.][]{Raffelt:1996wa}.
\subsection{IAXO and BabyIAXO}

\noindent{\bf Physics goals and status:}
The International Axion Observatory (IAXO~\cite{Irastorza:2011gs,BabyIAXOCDR,IAXO:2019mpb}) is a next-generation axion helioscope, designed to search for axions from the Sun that are reconverted into X-ray photons via a strong laboratory magnetic field. 
IAXO will deliver a factor of 20 improvement in sensitivity to the axion-photon coupling $g_{a\gamma}$ over the current leading helioscope,  the CERN Axion Solar Telescope (CAST~\cite{Arik:2015cjv}), across a wide mass range extending up to $\sim0.25$~eV. 
BabyIAXO, a preliminary experiment scheduled for 2025, will validate all key technologies and deliver a factor of $\sim5$ improvement in sensitivity, already opening novel discovery space for axions and ALPs. 

\begin{figure}
\vspace{-0.in}
\includegraphics[width=1\linewidth]{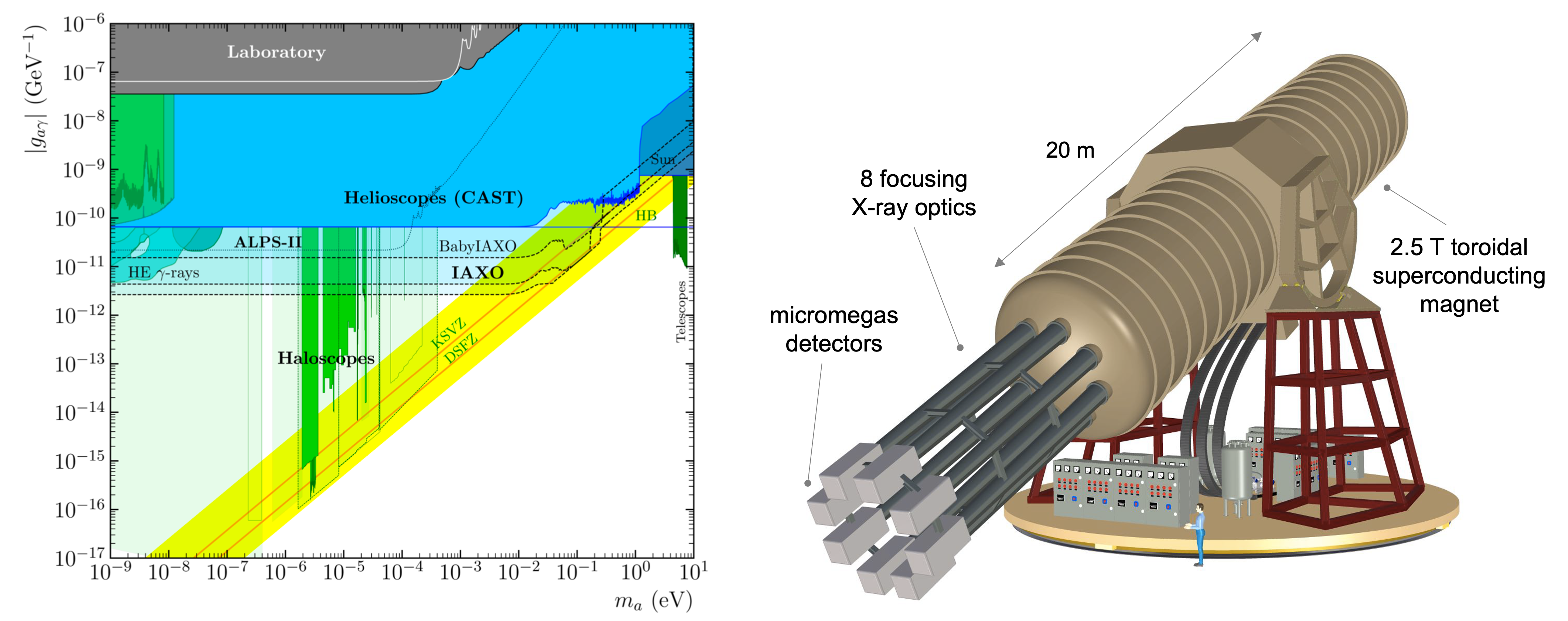}
\vspace{-0.2in}
\caption{\label{fig:iaxo_design} {\it (left:)} BabyIAXO and IAXO sensitivity~\cite{IAXO:2019mpb}, compared with the QCD axion (yellow) band and other current (solid) and future (shaded) experimental limits. BabyIAXO (IAXO) will improve sensitivity to $g_{a\gamma}$ by a factor of 5 (20) for the wide mass range up to 0.25\,eV, and are the only experiments with sensitivity to high-mass ($ > 10^{-3}$\,eV) QCD axions. The lower dotted line refers to an optimized IAXO+ configuration. {\it (right:)} Illustration of the IAXO instrument~\cite{IAXOLoI2013,BabyIAXOCDR}. IAXO will consist of a superconducting toroid magnet with eight custom X-ray telescopes that focus the reconverted photons onto ultra-low background detectors.}
\vspace{-0in}
\label{fig:IAXO}
\end{figure}

Helioscopes such as BabyIAXO and IAXO are the only experiments sensitive to QCD axions  in the $m_a > 10^{-3}$\,eV region, which is motivated by several dark matter scenarios~\cite{IAXO:2019mpb}, as well as to a broad mass range of ALPs that could constitute all of dark matter~\cite{Arias:2012az}. 
The IAXO program is thus a necessary complement to other methods that are sensitive to lower-mass axions, as shown in Fig.~\ref{fig:iaxo_design}.

A thorough review of the implications of BabyIAXO and IAXO sensitivity on various physics models has recently been published~\cite{IAXO:2019mpb}.
Many models permit dark matter axions up to the meV masses that will be uniquely probed by IAXO and BabyIAXO. 
These include models in which axion strings and domain-wall decays contribute to QCD axion dark matter~\cite{Gorghetto:2018myk,Gorghetto:2020qws} or in which there is fine-tuning of the initial axion field value. Models with long-lived domain walls in fact favor QCD axion dark matter in the meV mass range~\cite{Ringwald:2015dsf}. 
Generic ALPs have an even larger mass range in which they could constitute all of dark matter, including a large section of the parameter space accessible by IAXO~\cite{Arias:2012az}. 
BabyIAXO and IAXO will resolve if axions are the origin of several long-standing astrophysical anomalies, including the anomalous transparency of the universe to high-energy gamma-rays~\cite{Meyer:2013pny,Kohri:2017ljt} and the anomalously large cooling rates observed in diverse stellar systems~\cite{Ayala:2014pea,OscarPatras,DiLuzio:2021ysg,Giannotti:2017hny}.  
BabyIAXO and IAXO will also investigate non-axion particles at the low-energy frontier of physics, such as hidden photons~\cite{Schwarz:2011gu,Redondo:2013lna}, and possible particle physics solutions to the dark energy problem. 
These dark energy candidates include new particles, such as chameleons~\cite{Brax:2010xq}, as well as regions of axion parameter space -- the so-called ``ALP miracle" region at $m_a \approx 0.01$-$1$\,eV and $g_{a\gamma} = 10^{-11}$-$10^{-10}$\,GeV$^{-1}$ -- where ALPs would have an adequate potential and coupling to photons to be responsible for both cosmic inflation and dark matter~\cite{Daido:2017tbr,Daido:2017wwb}.
In the case of axions or ALPs coupled not just with photons but also with electrons or nucleons, IAXO will have the potential to probe the products of couplings $g_{ae}\cdot g_{a\gamma}$~\cite{IAXO:2019mpb} and $g_{ae}\cdot g_{aN}$~\cite{DiLuzio:2021qct}.
In particular,
IAXO will also supersede the astrophysical limits on the $g_{ae}$ coupling, opening investigations of a new set of axion models~\cite{DiLuzio:2021ysg}.
Furthermore, if the ALP couplings are large enough to produce a sizeable statistics, the IAXO signal might provide information about the ALP mass~\cite{Dafni:2018tvj} 
and even help distinguish among different axion models~\cite{Jaeckel:2018mbn}.
Finally, axions/ALPs could provide the most efficient messenger to probe the magnetic field in the interior of the sun. 
In fact, due to the coupling to photons, axions can be produced in the solar magnetic field~\cite{Caputo:2020quz,Guarini:2020hps} and generate a flux potentially observable with IAXO~\cite{OHare:2020wum}.

Although this generation of solar axion searches will be sensitive to a wide range of dark matter models, the helioscope technique, in contrast to haloscopes, is  independent of the assumption that axions are all or part of dark matter. Thus any exclusions are robust to uncertainties in the local dark matter density or specific production mechanisms. 

\noindent{\bf Experimental approach:}
X-ray photons will convert into axions in the strong fields of ions and electrons in stellar cores, such as our own Sun, via the Primakoff effect.
Axion helioscopes are designed to search for axions from the Sun that are reconverted into detectable photons via a strong laboratory magnetic field.
The reconverted photons then have energy equal to that of the incident axion, $\sim$1-10~keV~\cite{Sikivie:1983ip,Zioutas:2004hi,Andriamonje:2007ew}.
For low masses, the axion momenta is dominated by this kinematic contribution, and thus the helioscope technique offers sensitivity to a broad mass range.
By changing the pressure of a buffer gas in the magnet beam pipes, coherence can be maintained, allowing sensitivity at higher axion masses~\cite{vanBibber:1988ge, Arik:2008mq}.
If axions exist, such a signal is a robust prediction depending on well-studied solar physics.

Improving on the successful design of CAST, IAXO will consist of a superconducting toroid magnet with eight custom X-ray telescopes that focus the reconverted photons onto ultra-low background detectors (Fig.~\ref{fig:iaxo_design}).
As a preliminary stage of this program, BabyIAXO will consist of a superconducting magnet with two bores, each instrumented with X-ray optics and detectors with dimensions similar to those foreseen for IAXO.

BabyIAXO will advance technology for all IAXO subsystems at relevant scale while also offering novel physics reach and discovery potential.
Full design details and construction status were recently published in the BabyIAXO CDR~\cite{BabyIAXOCDR}.
DESY has agreed to provide general infrastructure, with experiment operation scheduled to begin in early 2025.
The magnet will feature a common-coil ``racetrack" design, with two 10\,m-long, 70\,cm-diameter bores, demonstrating similar engineering parameters (winding, geometry, etc.)  as are necessary for the IAXO toroidal magnet. 
The detectors are baselined to be small (6\,cm wide and 3\,cm thick) Time Projection Chambers with pixelated Micromegas readout, as have been used in CAST~\cite{Irastorza:2011hh, Dafni:2012fi, Dafni:2012zz}.
Surface tests have already demonstrated the improved shielding and veto necessary to achieve the required background levels of $10^{-7}$\,keV$^{-1}$cm$^{-2}$s$^{-1}$, and underground tests are underway. 
The X-ray optics are necessary to focus the X-ray signal, reducing detector size and background. 
 One bore will be instrumented with a custom X-ray optic with the same design and performance requirements of the final IAXO optic.  
The existing flight-spare telescope from the X-ray Multi-mirror Mission (XMM~\cite{10.1117/1.OE.51.1.011009}) is baselined for the second bore, although this may be replaced with another custom optic if funds become available.
Support and drive structure will be provided by a modified prototype of the support system for the Cherenkov Telescope Array (CTA) Mid-Size Telescope.

\noindent{\bf Projected sensitivity and timeline:}
IAXO will improve helioscope signal-to-noise by a factor of $>$10$^4$, improving sensitivity to $g_{a\gamma}$ by a factor of $\sim20$. BabyIAXO will already improve the signal-to-noise and $g_{a\gamma}$ sensitivity by factors of $>$10$^2$ and $\sim5$, respectively. 

The current BabyIAXO schedule anticipates data taking beginning in early 2025. 
The main physics program will proceed in two phases, each lasting approximately two years. 
In Run-I, the magnet volume will be in vacuum, providing sensitivity to the wide mass range up to $\sim0.02$\,eV. 
In the subsequent Run-II, the magnet volume will be filled with buffer gas, improving sensitivity to QCD axions up to $\sim0.25$\,eV.
Construction of full-scale IAXO could begin as soon as BabyIAXO is commissioned and operational.

The IAXO program builds on the success of the CAST.
CAST~\cite{CAST:2017uph}, the most powerful axion helioscope thus far, was the first experiment to surpass the astrophysical limits on $g_{a\gamma}$ for ALPs and placed the leading limits on QCD axions with masses $\sim0.01-1$\,eV.
In its final phase of operation, CAST hosted a ``pathfinder" detection line, including a custom X-ray telescope and a low-background Micromegas detector, both utilizing the same technologies proposed for IAXO.

The IAXO program will also constitute a generic infrastructure, in particular the large magnetic field volume, for axion haloscopes. The BabyIAXO magnet will already surpass the figure of merit ($B^2V$) of current haloscope magnets, while the IAXO magnet figure of merit one will be an order of magnitude larger. The RADES collaboration is performing R\&D on a new RF-cavity haloscope concept, in which sub-cavities coupled by inductive irises are used to instrument large volumes~\cite{Melcon:2018dba}. The RADES proof-of-concept has already been successfully demonstrated at small-scale in CAST~\cite{CAST:2020rlf}, as part of a seed program to implement axion haloscopes in BabyIAXO. 

Looking toward the future, there is further room for helioscope technology to expand beyond IAXO.
Since the conversion probability of axions back into photons increases as $B^2$, a stronger magnetic field would substantially increase sensitivity.
Though such a strong, large-area magnet is not currently cost-effective, the success of IAXO is a critical step towards allowing helioscopes to take advantage of ongoing advances in magnet technology.

\section{Synergies with astrophysical searches}
\label{sec:synergies}



\begin{figure}[ht!]
\begin{center}
\includegraphics[width=\textwidth]{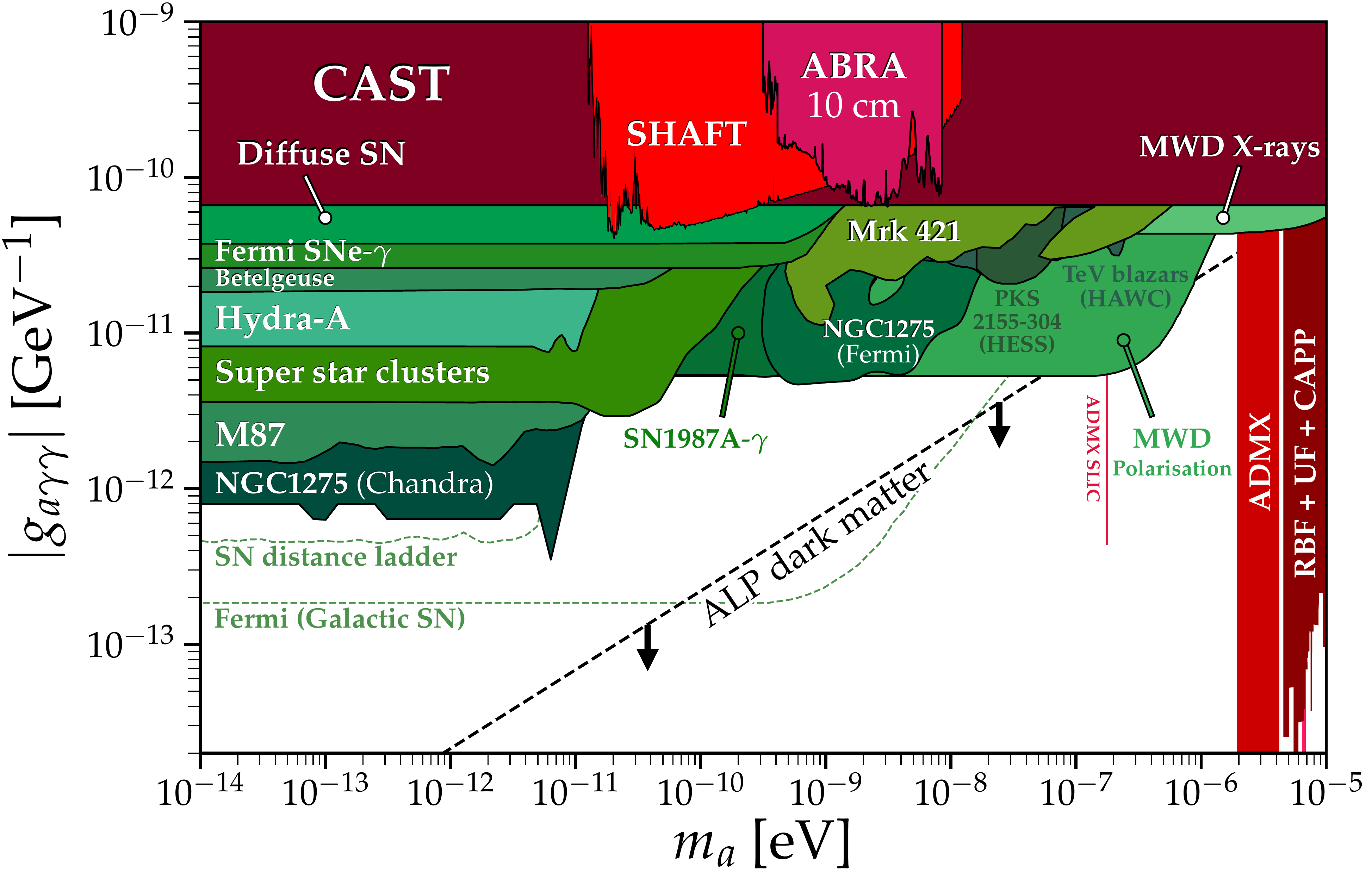}
\end{center}
\caption{Constraints on low-mass axions, primarily from high-energy astrophysical probes using X-ray and $\gamma$-ray data. Astrophysical bounds are shown in shades of green whereas experimental (haloscope/helioscope) bounds are shown in shades of red. We also mark the upper bound on ALP dark matter from Ref.~\cite{Arias:2012az} with a dashed black line. The experimental bounds shown in this range are from CAST~\cite{Andriamonje:2007ew,CAST:2017uph}, SHAFT~\cite{Gramolin2021}, ABRA~\cite{Ouellet:2018beu,Salemi:2021gck}, ADMX~\cite{Asztalos2010,Du:2018uak,ADMX:2019uok,ADMX:2021abc}, ADMX SLIC~\cite{Crisosto:2019fcj}, RBF~\cite{DePanfilis}, UF~\cite{Hagmann}, and CAPP~\cite{Lee:2020cfj,Jeong:2020cwz,CAPP:2020utb}. The astrophysical bounds shown are from: searches using Fermi-LAT data on extragalactic supernovae $\gamma$-rays~\cite{Meyer:2020vzy}, NGC1275~\cite{Fermi-LAT:2016nkz}, and for the decay of the diffuse supernova ALP background~\cite{Calore:2021hhn,Calore:2020tjw}. NuSTAR observations of super star clusters~\cite{Dessert:2020lil} and Betelgeuse~\cite{Xiao:2020pra}. Chandra observations of M87~\cite{Marsh:2017yvc}, NGC1275~\cite{Reynolds:2019uqt}, and Hydra-A~\cite{Wouters:2013hua}. HESS analysis of PKS 2155-304~\cite{HESS:2013udx}. HAWC observations of TeV blazars~\cite{Jacobsen:2022swa}. Constraints from nearby magnetic white dwarfs observed via X-ray (Chandra)~\cite{Dessert:2021bkv} and polarisation signals~\cite{Dessert:2022yqq}. Gamma rays from SN1987A~\cite{Payez:2014xsa}, and ARGO-YBJ+Fermi observations of the cluster Mrk 421~\cite{Li:2020pcn}. Finally we also show projections for constraints (green dashed lines) from the supernova distance ladder~\cite{Buen-Abad:2020zbd} and if a 10~$M_\odot$ SN occurred around the galactic centre and was observed by Fermi~\cite{Meyer:2016wrm}. A word of caution should be made about the differing statistical methodologies, and levels of assumption in deriving these bounds---original references should always be consulted when attempting a detailed comparison. Plotting scripts and limit data available at Ref.~\cite{ciaran_o_hare_2020_3932430}.}\label{fig:LowMassAstroBounds}
\end{figure}

\begin{figure}[ht!]
\begin{center}
\includegraphics[width=0.95\textwidth]{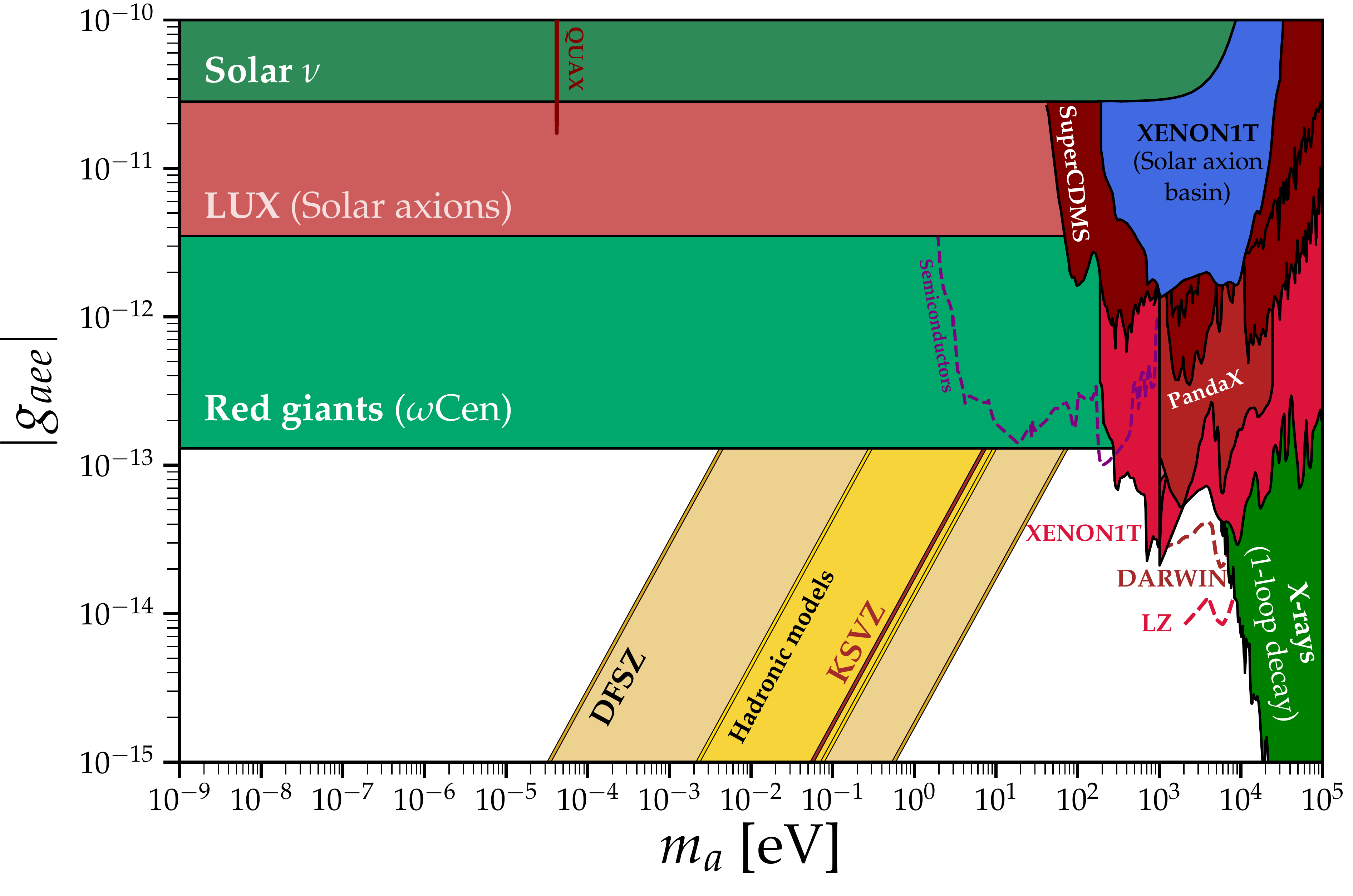}
\end{center}
\caption{Current and projected constraints on the axion-electron coupling $g_{aee}$ demonstrating the landscape is dominated currently by the stellar cooling bounds across the majority of the mass range, the bound shown here is from the red giants (RG) in the globular cluster $\omega$Centauri~\cite{Capozzi:2020cbu} (though other derived bounds are comparable). We also show the LUX bound on solar axions~\cite{LUX:2017glr} and the solar luminosity bound~\cite{Gondolo:2008dd}. The only competitive constraints on DM axions exist for $m_a\gtrsim 100$~eV, where underground recoil searches like XENON1T~\cite{XENON:2019gfn,XENON:2020rca,VanTilburg:2020jvl}, PandaX~\cite{PandaX:2017ock}, DARWIN~\cite{DARWIN:2016hyl}, LZ~\cite{LZ:2021xov}, and semiconductors~\cite{EDELWEISS:2018tde,Bloch:2016sjj}, can set bounds, before they are superseded by the non-observation of X-rays from the one-electron-loop decay to two photons~\cite{Ferreira:2022egk}. Plotting scripts and limit data available at Ref.~\cite{ciaran_o_hare_2020_3932430}.}\label{fig:AxionElectron}
\end{figure}

Important constraints on axion/ALPs couplings with the visible sector come from astrophysical probes (see Fig.~\ref{fig:LowMassAstroBounds}, \ref{fig:AxionElectron}, and \ref{fig:AxionPhoton_withProjections} for summaries). In many cases astrophysical considerations are still the leading limits thereby setting the benchmark for further exploration in experiments.

Taking the axion-photon coupling as an example, 
it can be constrained by a large number of astrophysical probes. 
These include: radio searches for the conversion of axions to photons in neutron star magnetospheres \cite{Pshirkov:2007st, Huang:2018lxq, Hook:2018iia, Safdi:2018oeu, Leroy:2019ghm,Foster:2020pgt, Darling:2020plz,Darling:2020uyo, Witte:2021arp, Battye:2021xvt, Millar:2021gzs, Battye:2021yue, Foster:2022fxn}, radio searches for the ratio of stars in the horizontal over the RG branch~\cite{Ayala:2014pea,Straniero:2015nvc,Carenza:2020zil,Lucente:2022wai},model-independent searches for axion-like signals public datasets of radio spectra~\cite{Keller2022}, 
the evolution of intermediate mass stars~\cite{Friedland:2012hj,Carosi:2013rla},
the white dwarf initial-final mass relation~\cite{Dolan:2021rya},
the non-observation of $\gamma$-rays associated with SN 1987A~\cite{Payez:2014xsa},
the diffuse $\gamma$-rays from all past supernovae~\cite{Calore:2020tjw,Calore:2021hhn}, the energy deposit due to ALPs decay in core-collapse supernovae with low explosion energies~\cite{Caputo:2021rux,Caputo:2022mah}, 
X-ray searches from super star clusters~\cite{Dessert:2020lil} and from Betelgeuse~\cite{Xiao:2020pra}, 
X-ray spectroscopy from AGN NGC 1275~\cite{Reynolds:2019uqt}, 
heating of interstellar medium~\cite{wadwan21, wadwan22}, 
and HE and VHE gamma-ray searches for spectral irregularities and EBL transparency in AGN sources~\cite{ajello2016PhRvL.116p1101A, abramowski2013PhRvD..88j2003A, buehler2020JCAP...09..027B, cheng2021PhLB..82136611C, guo2021ChPhC..45b5105G, horns2012PhRvD..86g5024H, zhang2018PhRvD..97f3009Z}.

Astrophysical means offer strategies to probe also the other axion couplings. 
In particular, RG branch stars and white dwarfs are powerful laboratories to constraint the axion electron coupling. 
Currently, the most stringent constraint on this coupling, $g_{ae}\lesssim 1.5\times 10^{-13}$, was derived in in 2020 in two independent analyses~\cite{Straniero:2020iyi,Capozzi:2020cbu} of the RG branch tip in several globular clusters, taking advantage of the newly determined cluster distances from the \emph{Gaia} DR2 data~\cite{chen2018}. Leading limits on the axion-muon coupling have also been set using the muons in SN1987A~\cite{Bollig:2020xdr, Croon:2020lrf, Caputo:2021rux}.

The axion-nucleon coupling, on the other hand, can be effectively constrained by observations of neutron stars cooling~\cite{Buschmann:2021juv} and by the analysis of the neutrino signal from SN 1987A~\cite{Carenza:2019pxu}.
Supernovae and neutron stars are difficult objects to model. 
Nevertheless, the progress in recent years has been substantial. 
A significant step forward was the recent realization that the pion abundance in supernovae may be considerably larger than expected~\cite{Fore:2019wib}, which led to a substantial revision of the axion production rates~\cite{Carenza:2020cis} and of the expected SN axion spectrum~\cite{Fischer:2021jfm}.
Finally, the phenomenon of black hole superradiance offers a possibility to probe the axions and ALPs coupled just to gravity~\cite{Arvanitaki:2014wva}, allowing to derive very general constraints on the axion mass and its self-coupling strength.

Note that due to the complexity of astrophysical systems, there are discussions regarding alternative scenarios that could potentially loosen up some of the constraints.
For example, Ref.~\cite{Bar:2019ifz} proposes a looser bound from SN 1987A cooling due to an alternative modeling of the neutrino emission, while the ICM magnetic field modeling for NGC 1275 bound is questioned in Ref.~\cite{Libanov:2019fzq,Matthews:2022gqi}. 
 
Future studies by next-generation very-high energy gamma-ray observatories such as LHAASO and CTA may well contribute further constraints. CTA's improved energy resolution among IACTs will provide greater sensitivity to the spectral irregularities expected in VHE AGN spectra as a result of ALP-photon oscillations in the magnetic fields of sources and the Milky Way ~\cite{abdalla2021ctaprop}. Longitudinal data collected by LHAASO will be able to probe heavier ($\mu$eV scale) ALP masses via the transparency of VHE gamma rays to the EBL across cosmological distances~\cite{long2021lhaasoPhRvD.104h3014L}.
The better understanding of the stellar population, expected in the coming years, will also guarantee a deeper comprehension of the axion/ALPs properties. 
As discussed above, data from the \emph{Gaia} survey have already allowed to revise the analysis of the impact of ALPs on RG stars and will likely lead to well needed revisions of the white dwarf bound (see discussion in Ref.~\cite{DiLuzio:2021ysg} for further details).
Further improvements are likely to follow, in the near future, thanks to the next-generation space-based missions, such as JWST~\cite{JWST:Gardner:2006ky}, 
which will enlarge the statistical sample of RG branch members near the cores of GCs, and the Vera Rubin Observatory, expected to detect WDs much fainter than those detected by Gaia and to increase the census of WDs to tens of millions~\cite{LSSTDarkMatterGroup:2019mwo,Mao:2022fyx}.

If the programmed experimental effort increases the optimism in a much deeper understanding of the axion properties,
on the theoretical side a series of original analyses of novel astrophysical observables is enlarging the region of the axion parameter space accessible to astrophysics and, in several cases, reinforcing the constrains on the ALP couplings with SM fields. Among those are recent studies regarding the ALP DM stimulated decay~\cite{Caputo:2018vmy,Arza:2019nta,Ghosh:2020hgd,Buen-Abad:2021qvj,Sun:2021oqp} which amplifies the ALPs decay signal through Bose enhancement and can set constraints with SKA and FAST. Observations of the neutron star population around the Galactic Center with SKA, FAST and other radio telescopes can also detect or set strong constraints on axion dark matter which converts to radio photons\footnote{Prospects for constraints are described at greater length in CF3 ``Dark Matter In Extreme Astrophysical Environments.''~\cite{Baryakhtar:2022obg}}, providing sensitivity even in DM substructure scenarios that render direct detection experiments insensitive \cite{Vaquero:2018tib, Buschmann:2019icd, Eggemeier:2019khm, Kavanagh:2020gcy, Foster:2022fxn}.

Let us report a few more significant examples.
The investigation of exotic contributions to the 511 keV line has recently allowed a new strategy to probe the parameter space of ALPs~\cite{Calore:2021klc} coupled with multiple SM fields.
Another class of observables root from the chiral nature of the Chern-Simons term, which induces photon polarization when photons propagate through ALPs DM or large magnetic fields. This leads to constraints across different bands using AGN~\cite{Ivanov:2018byi}, protoplanetary disk polarimetry~\cite{Fujita:2018zaj}, CMB birefringence~\cite{Fedderke:2019ajk}, X-ray polarization~\cite{Day:2018ckv}, polarization measurement of supermassive black holes~\cite{Chen:2019fsq}, radio waves from galactic pulsars~\cite{Castillo:2022zfl}, to polarization of magnetic white dwarfs~\cite{Dessert:2022yqq}.
In addition, photon lines from two-body decay of ALPs can be searched for directly~\cite{Blout:2000uc,Grin:2006aw,Regis:2020fhw,Keller:2021zbl,Foster:2021ngm}, or they might show up as signals in line intensity mapping~\cite{Bauer:2020zsj,Bernal:2020lkd}, or as spectral distortions to the CMB~\cite{Bolliet:2020ofj}. Lastly, it is shown that with the help of ultralight DM simulations, constraints of energy density fraction can be derived with disk galaxy rotational curves~\cite{Bar:2021kti,Bar:2019ifz,Bar:2018acw}.

On cosmological scales, ALPs-induced CMB spectral distortion places strong bounds at $m_a > 10^{-14}\; \mathrm{eV}$ \cite{Mirizzi:2005ng, Mirizzi:2009nq}. The coupling between ALPs and photons could lead to an apparent dimming of the photon sources, the effect of which is redshift dependent. This leads to a modification of the inferred distance of the source, which can in turn be constrained using cosmic distance measurement datasets~\cite{Avgoustidis:2010ju, Liao:2015ccl, Tiwari:2016cps,Buen-Abad:2020zbd}. When cosmological history is considered, the suppression of structure formation due to the quantum pressure associated to the light pseudo scalar can be constrained to \(m_a > 2 \times 10^{-20}\,\mathrm{eV}\) by Lyman-$\alpha$ forest 1D flux power spectrum~\cite{2021PhRvL.126g1302R, 2020RogersPRD, Kobayashi:2017jcf, 2017PhRvL.119c1302I, Armengaud:2017nkf}. The very lightest pseudo scalars are also probed by, e.g., the cosmic microwave background \cite{Hlozek:2017zzf, Poulin:2018dzj}, galaxy clustering \cite{2022JCAP...01..049L}, galaxy weak lensing \cite{2021arXiv211101199D} and the Milky Way sub-halo mass function \cite{Nadler200800022}.

Axions/ALPs contributing to the DM are typically considered as a non-relativistic oscillating field. However, bursts of relativistic axions from transient astrophysical sources allow for unique probes of axion physics and distinct classes of signatures~\cite{Eby:2021ece}. After formation in the early universe, axions could
form gravitationally-bound compact objects such as axion miniclusters and axion stars (see e.g.~\cite{Colpi:1986ye,Seidel:1993zk,Eggemeier:2019jsu,Kolb:1993zz,Xiao:2021nkb}).
For the QCD axion, axion bursts from axion star explosions~\cite{Levkov:2016rkk,Eby:2016cnq} (or other possible transients) can lead to detectable signals~\cite{Eby:2021ece} over a wide range of axion masses $10^{-15}~{\rm eV} \lesssim m_a \lesssim 10^{-7}$~eV in experiments such as ABRACADABRA, DMRadio and SHAFT discussed in the previous sections. Since the sensitivity to axion bursts is not strictly suppressed by large decay constant $f$, such signatures could probe a different parameter space window to conventional searches. Due to axion emission spectrum being intimately related to the fundamental axion potential, axion burst signals can yield new insights into the basic properties of axions that are challenging to probe otherwise~\cite{Eby:2021ece}. 
Relativistic axion signals contribute an additional instrument to the multimessenger toolkit for testing physics beyond the Standard Model. Besides, the formation of compact objects also opens up the potential to search for their gravitational and electromagnetic signals~\cite{Dror:2019twh,Dai:2019lud,Fairbairn:2017sil,Croon:2020ouk,Hertzberg:2020dbk,Amin:2020vja}.

\section{R\&D needs for future axion experiments}
\label{sec:randd}
As can be easily gleaned from the preceding sections there are a number of common themes in the challenges facing axion dark matter experiments. 
In particular, we can identify three R\&D streams -- Quantum sensors, magnets and cavities -- that will be important to fully realize the potential of the next generation of experiments. Collaborations on these topics is already underway, both within the axion community and beyond, offering also interesting opportunities for technological spin-offs. Expanding the breadth and scope of these collaborations will be key to fully realizing the existing synergies and to explore the full axion parameter space fast and efficiently.

\subsection{Quantum Sensor R\&D}
\label{sec:quantum_rnd}
Quantum techniques can potentially relax the volume requirements if successfully employed in future haloscopes by suppressing the noise. Quantum-limited amplifiers enabled haloscopes to dig deeper, reaching a few hundred millikelvins of system noise temperature. However, at high frequencies the noise will be dominated by quantum rather than thermal noise. Quantum squeezing and photon-counting are two techniques that can be employed to evade the standard quantum limit, speed up the scan rate, and help in closing the detection volume gap discussed in Section~\ref{sec:above_microeV}. Quantum squeezing was successfully used by the HAYSTAC collaboration to achieve a quantum advantage gain of about 2.5. Single photon counting promises an even larger quantum advantage \cite{lamoreaux2013analysis} and a superconducting qubit-based single photon detector was recently used in a dark photon search using a fixed frequency cavity \cite{Dixit:2020ymh}. High Q cavities ($Q > 5\times 10^{5}$) are needed for photon counters to allow the quantum sensors employed in the photon counter to do repeatedly enough parity measurements for noise suppression. 
Recently a magnetic-field-compatible cavity using nested shells of low loss sapphire~\cite{DiVora:2022tro} exhibited $Q=9\times 10^6$, thus boosting the signal rate and enabling quantum measurements in axion searches.  At higher signal frequencies, superconducting qubits can be used as Cooper-pair breaking sensors rather than as atomic clocks.  Although detectors like the Quantum Capacitance Detector~\cite{Echternach:2018} have demonstrated the lowest noise equivalent power of any THz photon detection technology, the corresponding dark count rate of 1~Hz is still orders of magnitude too high to be useful in dark matter searches.  Ongoing research targets the origins of low energy disturbances that create this dark rate, and results will benefit both dark matter and quantum computing applications.

Quantum technologies have a crucial role in engineering spin ensembles that are used in experiments, such as CASPEr, to search for the EDM and the gradient interactions of ultra-light dark matter~\cite{Aybas2021a}. Optical techniques can be used to prepare spin ensembles in low-entropy high-polarization states that have enhanced sensitivity to these interactions. Coherent control and decoupling can extend spin coherence time, also enhancing sensitivity. The goal is to achieve sensitivity to spin ensemble dynamics at the level of quantum spin projection noise. Approaches that make use of spin squeezing and other spin ensemble correlations may ultimately make it possible to extend sensitivity even further. In addition, it is necessary to optimized coupling between a spin ensemble and the electromagnetic sensor used to detects its dynamics, in order to avoid back-action effects~\cite{Aybas2021b}. An important part of these activities is to identify and synthesize optimal materials that host spin ensembles with enhanced sensitivity to the EDM and the gradient interactions of ultra-light dark matter~\cite{Budker2014}.

Superconducting-nanowire single-photon detectors are ideally suited for sensing low-count-rate signals due to their high internal efficiency and low dark-count rates.  Recent proposals for axion search \cite{Baryakhtar2018, BREAD:2021tpx} either require SNSPDs that can operate in the presence of large magnetic fields, or require some  means of carrying the light generated by the haloscope from the high-field region to a low-field region where the detectors can operate.  The recently established robustness of SNSPDs to operation in high fields~\cite{Lawrie:2021, Polakovic2020} and their ability to operate at elevated temperatures (relative to alternative superconducting detector technologies) make them well-suited for photon detection in the mid-infrared (meV) to visible (eV) energy range.  The suitability of SNSPDs to applications requiring low dark-count rates is illustrated by recent progress in the LAMPOST prototype search for dark photon dark-matter using these devices~\cite{Chiles:2021gxk}.

\subsection{Magnet R\&D}
\label{sec:magnet_rnd}
To maximize axion sensitivity, most axion experiments require optimized magnetic fields. At the same time, these experiments also have to balance the designs against practical engineering considerations such as cooling requirements and quench protection. As cavity based haloscopes~\cite{Backes2021,Braine2020} push towards higher masses/frequencies the optimal cavity size in order to fully trap the axion energy (proportional to $\approx \lambda$/2) gets smaller. Since the axion signal sensitivity scales as $\ensuremath{g_{a\gamma\gamma}}\xspace^{-1} \propto B_0 V^{\alpha} $,\footnote{The actual value of $\alpha$ depends on the specifics of the axion search and the frequency range, but ranges between $\frac{1}{2}$ and $\frac{5}{6}$.} this means that in order to maintain the same level of sensitivity down to the DFSZ~\cite{Braine2020} coupling, a dedicated magnet R\&D program for higher strength (up to 45+ Tesla) and optimized magnet designs will be needed for the upcoming experiments. While lumped element based experiments will also require high magnetic fields, they place additional requirements on their overall field profiling necessitating a careful optimization program to meet their sensitivity targets.

\begin{figure}[ht!]
	\begin{center}
		\includegraphics[width=.40\textwidth]{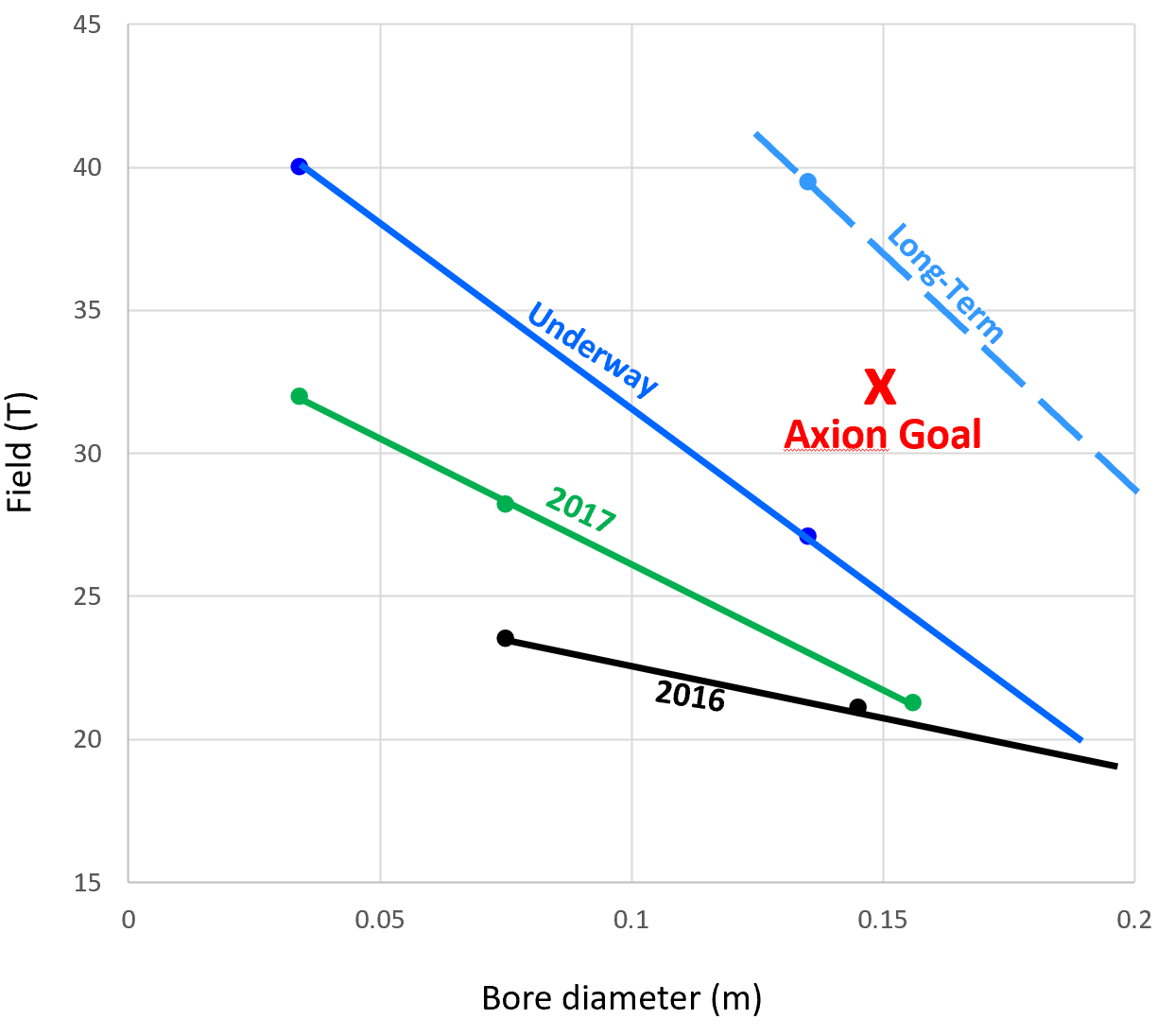}
		\hspace*{1.0cm}
		\includegraphics[width=.45\textwidth]{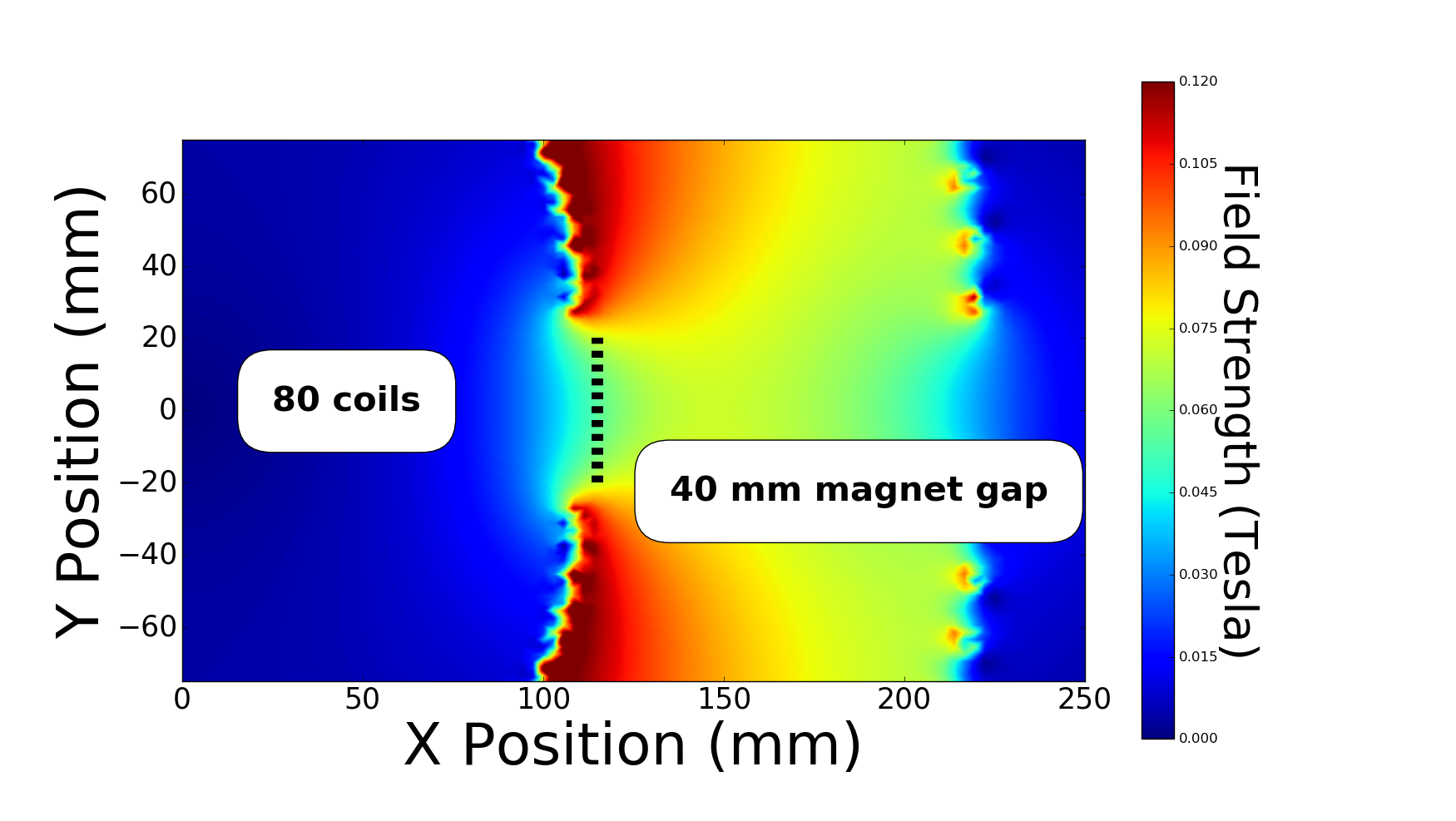}   
		\caption{(Left) Evolution of peak fields and bore sizes available from superconducting magnets in recent years. (Right) Magnetic field profile for a 1 Tesla peak spilt toroidal magnet design. The strongest field leakage occurs not only in the gap region, but also in between individual magnet coils. The superconducting elements of this experiment will be placed within 10 mm of the coils/gap region while the field strength falls below 100 mT}
		\label{fig:fieldbores}
	\end{center}
\end{figure}

For increasing the maximum field strength a sequential two pronged approach is proposed, each targeting a different axion mass region for cavity based axion searches. For the 8–30 $\mu$eV range, we seek to design and build a conventional large magnet based on Nb$_{3}$Sn and/or NbTi technology with the aim of producing a 15 T field in a 25 cm clear bore and with an overall length of 60 cm. For experiments searching in the 30–50 $\mu$eV range, we seek to design a high-temperature superconducting (HTS) insert to allow for a maximum central magnetic field strength of 32 T over a 15 cm diameter. Production and quality of rare-earth barium copper oxide (REBCO) tapes, a vital component of HTS inserts, has risen steadily in recent years due to numerous commercial suppliers worldwide and a strong demand. Over the past six years, the National High Magnetic Field Laboratory (MagLab) has been developing higher field REBCO inserts with current designs reaching a maximum field of 45 T~\cite{Bai2020} and future designs aiming for even higher field strengths across smaller bore diameters as shown in Figure~\ref{fig:fieldbores}. At the same time as field strengths continue to grow, engineering R\&D must be undertaken in order to provide quench management/protection and thermalization management to allow for safe operations of these magnets in cryogenic environments.~\cite{Gavrilin2021,Kolb-Bond2021}

Upcoming lumped element axion experiments, looking at axion masses below $\approx$ 1~$\mu eV$ will have unique field requirements that require precise magnet profiling to minimize signal leakage and detrimental effects on sensitive readout electronics - see Figure~\ref{fig:fieldbores}. These specific axion sensitivity requirements may dictate non-standard toroidal or solenoidal magnet designs in order to optimize the coupled energy between the detector and the axion signal in order to cover multiple mass decades of axion mass parameter space~\cite{Chaudhuri15,Chaudhuri2019,Chaudhuri:2019ntz}. These upcoming experiments will build on the experience of previous toroidal DM searches \cite{Ouellet2019} as well as the expertise of the LBL, SLAC and the MagLab magnet design groups. In addition to the engineering R\&D work discussed above additional work aimed at reducing environmental backgrounds such as mechanical vibrations and electromagnetic interference will be vital to ensure the success of these experiments.

All these R\&D efforts will allow for the development of cost-effective large-volume high-field magnet designs that will benefit many axion detection experiments. We seek to establish a framework by which experiments can create optimized magnetic field profiles based on the individual experimental needs, minimizing signal losses/leakage while still maximizing the science potential of axion searches in the highest possible magnetic field. We are interested in partnering with experts at national labs and industry in order to design, construct and then successfully implement these next generation of magnets for use in axion experiments.

\subsection{Cavity R\&D}
\label{sec:cavity_rnd}
\begin{figure}[ht!]
\centering
\includegraphics[width=\linewidth]{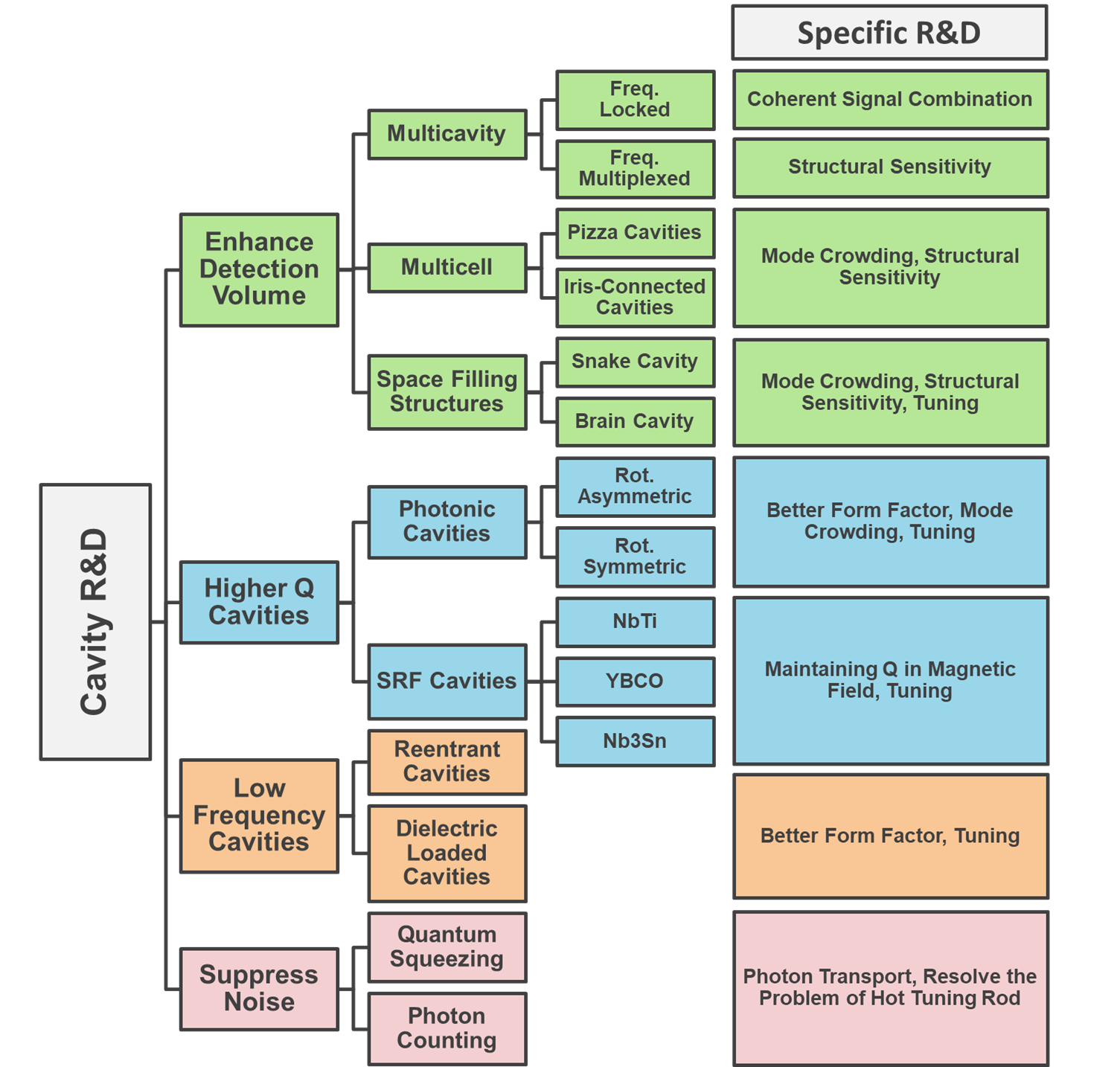}
\caption{\label{fig:cavity_RnD} Layout of design challenges and potential R\&D avenues to pursue. Enhancing the detection volume can take several general design paths. Multicell cavities can be used in a frequency locked configuration (such as what is being pursed by ADMX-G2 and ADMX-EFR) or in a frequency multiplexed configuration. Multicell cavities consist of geometries in which effective cavities communicate through irises.}
\end{figure}

Cavities with sophisticated geometry and tuning mechanisms are required to allow large volume experiments to be operated at high frequencies~\cite{Kuo_2021}.
Microwave cavities for axion searches leverage new ideas from both accelerators and quantum information sciences. Moving to higher frequencies cavity volumes run into rapidly degrading volumes and lower quality factors for ordinary metals. There are several methods to recover this:
\begin{itemize}
    \item Increase volume by coherently adding multiple cavities in phase at the same frequency to take advantage of the axion coherence
    \item Increase search scan rate but building an array of cavities with complementary frequencies and small tuning range to perform a frequency comb search. Does not take advantage of axion coherence but may be simpler to implement 
    \item Take advantage of higher order modes in cavities. Clever use of dielectrics in photonic configurations can increase the form-factor and quality factor for such cavities.
    \item Increase the quality factor of cavities by developing magnetic field tolerant superconducting surfaces that maintain a higher Q than copper in high magnetic fields.
\end{itemize}

Going in the opposite direction also lower frequency cavities below 500 MHz present their own unique challenges.

Figure \ref{fig:cavity_RnD} outlines the various design paths to tackle the related issues of scalability to higher and lower masses

\section{Conclusion -- Towards the Ultimate Axion Facility}
\label{sec:conclusion}

\begin{figure}[ht!]
\begin{center}
\includegraphics[width=\textwidth]{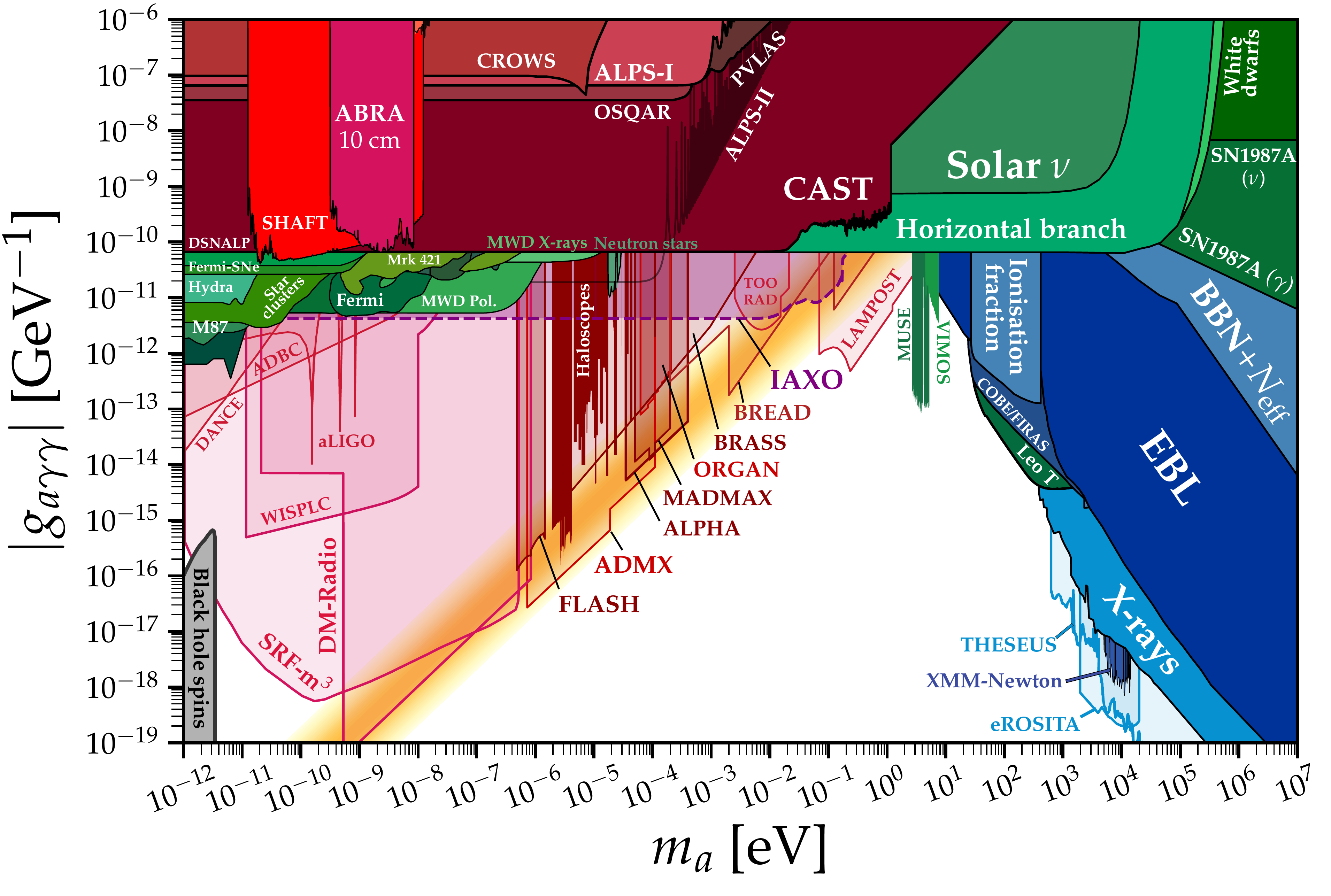}
\end{center}
\caption{Current (opaque) and projected (transparent) constraints on the axion-photon coupling. All existing constraints are described in the preceeding sections. The estimated sensitivities of several proposed and in-construction haloscopes are shown including (from low to high masses, roughly): DANCE~\cite{Michimura:2019qxr}, ADBC~\cite{Liu:2018icu}, aLIGO~\cite{Nagano:2019rbw}, SRF cavity~\cite{Berlin:2020vrk}, WISPLC~\cite{Zhang:2021bpa}, DM-Radio~\cite{DMRadio}, FLASH~\cite{Alesini:2017ifp}, ADMX~\cite{Stern:2016bbw}, ALPHA~\cite{Lawson:2019brd}, MADMAX~\cite{Beurthey:2020yuq}, ORGAN~\cite{McAllister:2017lkb}, BRASS~\cite{BRASS}, BREAD~\cite{BREAD:2021tpx}, TOORAD~\cite{Schutte-Engel:2021bqm}, and LAMPOST~\cite{Baryakhtar:2018doz}. We also show projections for IAXO~\cite{Shilon:2013xma} and ALPS-II~\cite{Ortiz:2020tgs}. Finally, we have displayed forecasted sensitivity to heavy dark matter ALP decays to X-rays using eROSITA~\cite{Dekker:2021bos} and THESEUS~\cite{Thorpe-Morgan:2020rwc}, as well as a projection for a Fermi-LAT observation of a galactic supernova~\cite{Meyer:2016wrm}. Plotting scripts and data available at Ref.~\cite{ciaran_o_hare_2020_3932430}.}\label{fig:AxionPhoton_withProjections}
\end{figure}

Axion physics and in particular the search for axion dark matter is at the dawn of a golden era. In the time since the last Snowmass, the number of experiments but perhaps more importantly the range of axion parameter space they are sensitive to has grown by order(s) of magnitude. This becomes most evident, when looking at Fig.~\ref{fig:AxionPhoton_withProjections}, showing the present status (solid) as well as the projections for the future (transparent; many of which are discussed in this Snowmass white paper). From a single experiment sensitive sensitive to axion dark matter in a narrow range of masses (solid ``Haloscopes'') we can now see the road to a nearly full coverage of the favored axion dark matter parameter space (various transparent) that spans nine, perhaps more, orders of magnitude in mass.  

However, realizing these ambitions is not automatic. The sensitivities indicated in Fig.~\ref{fig:AxionPhoton_withProjections} still require significant developments and even breakthroughs in technology. Achieving this in a timely and efficient will require a concerted effort.
Importantly, certain technologies are ubiquitous among axion experiments: superconducting magnets, dilution refrigerators, and the requisite cooling infrastructure. There is therefore strong motivation as well as potential to share resources between and amongst various groups. 
This could greatly benefit from a shared facility dedicated to axion physics. 
By investing in several strong, large bore magnets and associated cryogenics that would be able to accommodate multiple experiments, the science reach per experiment could be extended and the search could be conducted more efficiently with shared infrastructure.

\bibliographystyle{jhep}
\bibliography{main.bib}

\vspace{1em}
\section*{Note}
Some of the text may be based on the respective LOIs on \url{https://snowmass21.org/cosmic/dm_wave}. 
\end{document}